\documentstyle[epsf,wrapfig,12pt,twoside]{report}
\def\insertplot#1#2#3{\par \hbox{ \hskip #3
              \vbox to #2{ \special{em:graph #1} \vfil } } }

\begin{document}
\setcounter{footnote}{0}
\setcounter{figure}{0}
\thispagestyle{empty}
\pagestyle{myheadings}

\title{
\LARGE \bf Disentangling of spectra\\ -- theory and practice}
\author{P. Hadrava,\\
Astronomical Institute, Academy of Sciences of the Czech Republic,\\
Bo\v{c}n\'{\i} II 1401, 141 31 Praha 4, Czech Republic\\
e-mail: had@sunstel.asu.cas.cz\\
http://www.asu.cas.cz/\~{}had/korel.html}
\date{Release September 19, 2008\\
 modified August 31, 2009}
\markboth{Summer school on disentangling, Ond\v{r}ejov, 15. -- 19. 9. 2008}
 {P. Hadrava, Disentangling of spectra -- theory and practice}

\maketitle

\begin{abstract}
In this document a review of the author's method of Fourier
disentangling of spectra of binary and multiple stars is
presented for the purpose of the summer school organized at
Ond\v{r}ejov observatory in September 2008.
Related methods are also discussed and some practical
hints for the use of the author's code KOREL and related
auxiliary codes with examples are given.
\\
keywords: 
Spectroscopic binaries and multiple stars -- orbital elements --
line profiles -- 
 spectra disentangling 
\end{abstract}



\onecolumn

\tableofcontents

\clearpage

\noindent{\Large \bf Introduction}\\[3mm]
\addcontentsline{toc}{section}{Introduction}
\noindent{}Observations of binary and multiple stars enable us to determine
their basic physical parameters which are not so easily reachable
for single stars. This is why, in spite of their more complicated
physics, the binaries are a clue to understanding the physics,
structure and evolution of stars in general. This knowledge is then
a basis for understanding the physics of higher systems (clusters
of stars and nebulae of interstellar matter, galaxies, as well as
the Universe as a whole). Observations of binaries can also be used
as a primary method of distance determination.

 Different methods of observations of binaries are complementary
and should be used simultaneously to combine and exploit optimally
their sensitivities to different parameters of the observed systems.
The spectroscopic observations reveal the variations of Doppler-shifts
of spectral lines of the component stars caused by changes of radial
velocities in the course of the orbital motion. (The conception of
spectroscopic radial velocity only seems to be straightforward --
for its definition and discussion see, e.g., Lindegren and Dravins 2003).
In combination with inclination of the orbital plane which can be found
e.g. from a light-curve solution
for eclipsing binaries, radial velocities yield the orbital velocity
and thus also the information on absolute dimensions of the orbit.
Using the third Kepler law also the masses of the stars can be
determined and attributed to the spectral types of the stars.
This, however, requires to distinguish spectral features belonging
to the individual component stars, i.e. to decompose the observed
superposition of the light of the components into the individual
spectra of the component stars. This decomposition can be performed
by comparison of spectra taken at different known radial velocities
of the individual components or at phases with different known light
ratios of the components (e.g. during eclipses) or with some other
well established variations of the component spectra. Because the
information on the component spectra as well as on the radial
velocities or the orbital and other parameters of the observed
system is entangled in the observed spectra, the procedure of
solution for this information is referred to as the ``disentangling"
 -- cf. Fig.~\ref{disch}.
Such a method of solving for orbital parameters and simultaneous
decomposition of spectra of binaries was first published (and named)
by Simon and Sturm (1994) who used the method of singular-value decomposition
(SVD) in the wavelength domain to separate the component spectra.

\setlength{\unitlength}{1mm}
\begin{figure}[hbt]
\noindent\begin{picture}(140,100)
 \put(0,0){\epsfxsize=100mm \epsfbox{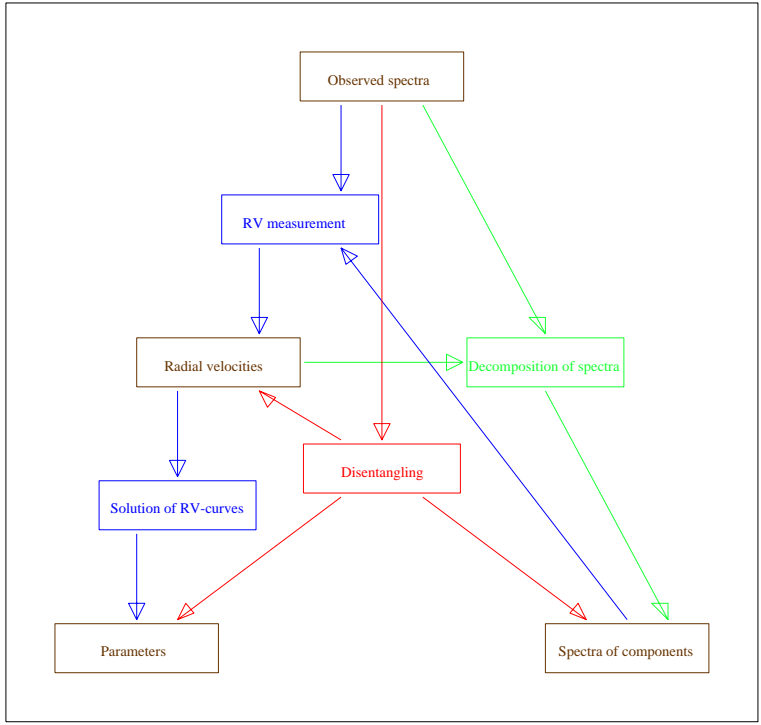}}
\put(110,95){\parbox[t]{50mm}{\caption{
 \label{disch} A scheme of disentangling method compared with
the classical processing of spectra of binary stars: orbital
parameters are determined from the observed spectra with an
intermediate step of radial-velocity measurement (blue),
spectra of components may be separated using known radial
velocities (green), while in disentangling all unknown quantities
are determined together as the best fit of the observations}}}
\end{picture}
\end{figure}

 The method by Simon and Sturm turned out to be in principle (by the
formulation of the problem and by its properties) equivalent to the method
which I developed from the method of cross-correlation for radial velocity
measurements between 1990 and 1993. The aim of this my work was originally
to enable measurement of radial velocities of highly blended lines
of the components (i.e. at phases close to conjunctions, but also
practically at all phases for wide lines of early type stars with
small amplitudes of radial velocities). I realized soon that it would
be advantageous to skip the step of determination of radial velocities
and to include instead of it directly the observed spectra. These spectra
could thus be another possible
input for the code FOTEL which I designed originally for solution of
light-curves (i.e. determination orbital ELements from the FOTometry),
and which I enriched also for a simultaneous solution of radial velocity
curves (for which also the code SPE(ctroscopic)EL(ements) was written by
Ji\v{r}\'{\i} Horn) and later also for astrometric/interferometric data.
The incorporation of cross-correlation should avoid measurements of radial
velocities and it is this relation with the cross-correlation why I renamed
the corresponding code KOR(elation)EL(ements) from its original name
BA\v{Z}ANT (which means pheasant whose colours were resembled by the curves on
the screen) given to it by my son (who used then to help me at his age
of three years).
My additional intention was to improve the method of cross-correlation
by cross-correlating mutually spectra of the same system obtained at
different phases instead of correlating each exposure with an arbitrarily
chosen template. To distinguish which peak in cross-correlation is due
to which combination of components, it turned out to be preferable to
decompose the spectra. The formulation of the problem in this complexity
was later defined (which means that it was developed practically at
the same time) by Simon and Sturm (1994) as the disentagling. Such a
decomposition of spectra is relatively easy in Fourier representation,
which is commonly used in the cross-correlation, but it qualitatively
surpasses the original cross-correlation in the sense explained by
Rucinski (2002).
At the same time, the Fourier disentangling is numerically much more
efficient than the wavelength-domain disentangling. This efficiency enables
a further generalization of the method (e.g. to include easily more
components).

 Similarly as before with the code FOTEL, I made from the very beginning
the source-file of the KOREL-code publicly available first on FTP and then
on my web-pages. For a convenience of users, I published in an electronic
form on the web also reviews of the methods and manuals for the use of
FOTEL and KOREL, which I upgraded and supplemented simultaneously (or with
some delay or advance) with the development of the codes and publications
of the news in the literature. While skilled users appreciated this
practice, some other disabled it by misusing the texts and spreading
of various desinformations. The versions of the codes FOTEL and KOREL
from the year 2004, when also their descriptions were printed are thus
the last made available. Because these versions of 2004 are already
obsolete it is no more worth to support them.

 In the mean time, several users in the world reached very good results
applying the method of Fourier disentangling and the KOREL code for
particular stellar systems. There is an increasing audience of
astrophysicists willing to use the method, but some of them meet
problems in the beginning or they do not use it in an optimal way.
Frequent requests for a help with the use of disentangling gave rise
to the idea to organize a summer school, in which the theory of the
method as well as the practice with the code KOREL will be explained.

 In comparison with the version of KOREL of 2004, the new version of
2008 is much more powerful in its precision and efficiency. This is
why I prefer to deal with the new one, in spite of suggestions that
even the old one is not yet properly understood. The new version
also enables some new options. Fortunately, the previously available
possibilities are still controlled basically in the same way as in
the version of 2004, despite it will be probably necessary to change
in some way in near future. Until such a version will be prepared,
verified and properly described, a technical solution is to provide
users with a remote access to computer of the Astronomical Institute
with a compiled version of KOREL08, where they could disentangle
their spectra in the way which will be shown in this summer school.

 The present document is a preprint of textbook on the method
of Fourier disentangling and related topics from physics of binaries.
It is provided for a personal use of participants of the school
with a warning that it is still a very preliminary version.
It explains in details some theoretical aspects of the method in
Chapter~\ref{Theory} and provides a practical experience in the use
of its implementation in the code KOREL in Chapter~\ref{Practice}.
It is based on the previous review of the method (Hadrava 2004c)
but it includes some later results including some quite recent,
which will be published elsewhere. Partly updated manual on KOREL
from the same publication is included in Appendix together with
some other auxiliary resources.

 To save the time and to enable practical exercises from the very
beginning of the course, I shall explain first the simplest version
of the method of Fourier disentangling in Section~\ref{entry} and
the use of the KOREL code in Section~\ref{start}.
However, a thorough understanding of the theoretical background
is needed for a safe use of the method, not to say for its further
development. We shall thus return in subsequent Sections to different,
in some cases even very general topics, which, from the purely logical
point of view should precede.

\clearpage
\chapter{Theory of spectra disentangling}\label{Theory}

\section{A swift entry into the theory} \label{entry}

Let us suppose that a multiple stellar system consists of $n$
stars and that the intrinsic spectrum $I_{j}(x)|_{j=1}^{n}$ of each
component is constant in time (i.e. it has no intrinsic physical
or geometric variability) apart of being Doppler-shifted according
to the instantaneous radial velocity $v_{j}(t)$ of the star $j$
at the time $t$.
As it is customary, here we use $x$ as an independent variable of
the spectrum, the logarithmic wavelength scale
\begin{equation}\label{lnlamb}
 x=c\,{\rm ln}\lambda\; ,
\end{equation}
in which the Doppler shift is the same for all frequencies. For instance,
for a radial velocity $v<<c$ much smaller than the speed of light
the non-relativistic approximation gives
\begin{equation}\label{Doppler}
 v=c\Delta\lambda/\lambda=\Delta x\; ,
\end{equation}
while the relativistic formula
\begin{equation}
\Delta x (v)=c\,{\rm ln}\frac{\sqrt{c^{2}-v^{2}-v_{\perp}^{2}}}{c-v}
\end{equation}
(where $v_{\perp}$ is the tangential component of the velocity) is
non-linear in $v$ but still independent on $\lambda$ equally as the
general-relativistic red-shifts.\footnote{ This could be violated
 by a dispersive medium between the source and observer.}

 According to Eq.~(\ref{Doppler}), the observable spectrum of the star
$j$ is given by convolution with shifted Dirac delta-function
$\delta(x-v_{j}(t))$ and the composite spectrum of the whole stellar
system observed at time $t$ is the superposition of these
convolutions\footnote{ In fact the Doppler shift modifies also the
 value of intentensity depending on the way of its normalization,
 however, we shall mostly use the rectified spectra, i.e. ratios of
 the actual spectra and their continua.}
\begin{equation}\label{Kor1}
 I(x,t)= \sum_{j=1}^{n}I_{j}(x)\ast\delta(x-v_{j}(t))\; .
\end{equation}
Comparing such spectra obtained at different times, we would like
to find what is common for all of them, i.e. the spectra
$I_{j}(x)$ of the components, and what is changing, i.e. the
instantaneous radial velocities $v_{j}(t)$. Obviously, without
any additional conditions (e.g. on smoothness of $I_{j}(x)$ or
its shape) such a solution is in principle possible if we have
$N$ ($N>n$) spectra $I(x,t_{l})$ taken in suitable times
$t_{l}|_{l=1}^{N}$ at different radial velocities.
Typically, the observed spectra are sampled in number of bins
of the order of thousands. With a few unknown radial velocities
(or orbital parameters by which the radial velocities are bound),
the solving for all the unknown variables is overdetermined already
if $n+1$ observed spectra are taken into account. However, because
each exposure includes also some noise, it is desirable to treat even
more observations at once and to fit them simultaneously by all the
unknowns -- both the component spectra as well as the radial velocities
or some other relevant parameters.

 Assuming first the radial velocities to be known, Eq.~(\ref{Kor1})
is a huge set of linear equations coupled for all wavelengths
(provided differences between $v_{j}(t)$ vary with $t$). The inversion
of the corresponding large (but sparse) matrix can be simplified in
a suitable representation, in which the operator separates into smaller
submatrices (cf. the experience from quantum mechanics). Really,
Fourier transform provides such a representation, because it changes
the convolution of functions to a simple product of their Fourier
modes.
Fourier transform ($x\rightarrow y$) of Eq.~(\ref{Kor1}) reads
\begin{equation}\label{FKor1}
 \tilde{I}(y,t)=
  \sum_{j=1}^{n}\tilde{I}_{j}(y)\exp(iyv_{j}(t)) \; ,
\end{equation}
which really simplifies the task essentially, because the huge set of
linear equations splits now into many simple systems of dimension $n$,
i.e. the number of components in the observed system. In this set of
equations, independent for each Fourier mode and labeled by the new independent
variable $y$, the expressions $\exp(iyv_{j}(t))$ are multiplicators
(known as simple functions of the fixed radial velocities $v_{j}(t)$)
of the component spectra's unknown modes $\tilde{I}_{j}(y)$, which
are to be solved to fit the Fourier modes $\tilde{I}(y,t)$ of the observed
spectra. For the simple case of a double-lined spectroscopic binary $n=2$
the decomposition of spectra simplifies to the solution of $N/2$ ($\frac{1}{2}$
is due to the symmetry) sets of complex linear equations for two unknowns,
one in each Fourier mode. It is immediately obvious from Eq.~(\ref{FKor1})
that for $y=0$ this set of equations is singular ($\tilde{I}(y,t)=
\sum_{j=1}^{n}\tilde{I}_{j}(y)$), it means that in practice the ratio
of continua cannot be determined from the Doppler shifts only.

 Owing to the Parseval theorem (\ref{Parseval}), the squares of
residuals in the wavelength domain $x$ are equal (or proportional
depending on the chosen normalization of the Fourier transform) to
the residual in the domain of Fourier frequencies $y$. These residuals
($O-C$) of the observed spectra can thus be minimized by the
least-squares method also with respect to the unknown radial velocities
$v_{j}(t_{l})$ or directly with respect to the orbital parameters
which determine these velocities. This optimization with respect to
the parameters is the procedure which complements the above described
method of decomposition to become a method of disentangling of the
spectra.

 Generally, the dependence of the squared residuals of spectra on these
parameters is not quadratic (unlike the dependence on the component
spectra), so the minimization of the residuals with respect to these
parameters must be performed by some general method (e.g. the simplex
method used in the KOREL code). This also implies directly, that there
may exist some local minima and hence multiple solutions. Some of
these solutions may be fully or nearly equivalent (e.g. solutions
which refer to different epochs of the periastron passage or with
different combinations of periastron longitude and epoch for very small
eccentricities), but some may be artifacts of incorrect minimization
or of errors or gaps in the data.

\clearpage
\section{Background thoughts}\label{mysl}
A short cut leading readers directly {\it in medias res} of the
method of disentangling has been described in the Sections~\ref{entry}
and \ref{start}. The purpose of the present Section is to warn the
readers against taking such a way. Instead, let us return first to
a very general, rather philosophical principles of the methodology
of science. It is commonly accepted, that a progress in development
of science is usually achieved by combining experimental (in astronomy
it means observational) and theoretical approaches. A routine
application of an experimental method without taking into account
its theoretical background threatens with an incorrect interpretation
of the data.

\setlength{\unitlength}{1mm}
\begin{figure}[hbt]
\noindent\begin{picture}(140,80)
 \put(5,0){\epsfxsize=80mm \epsfbox{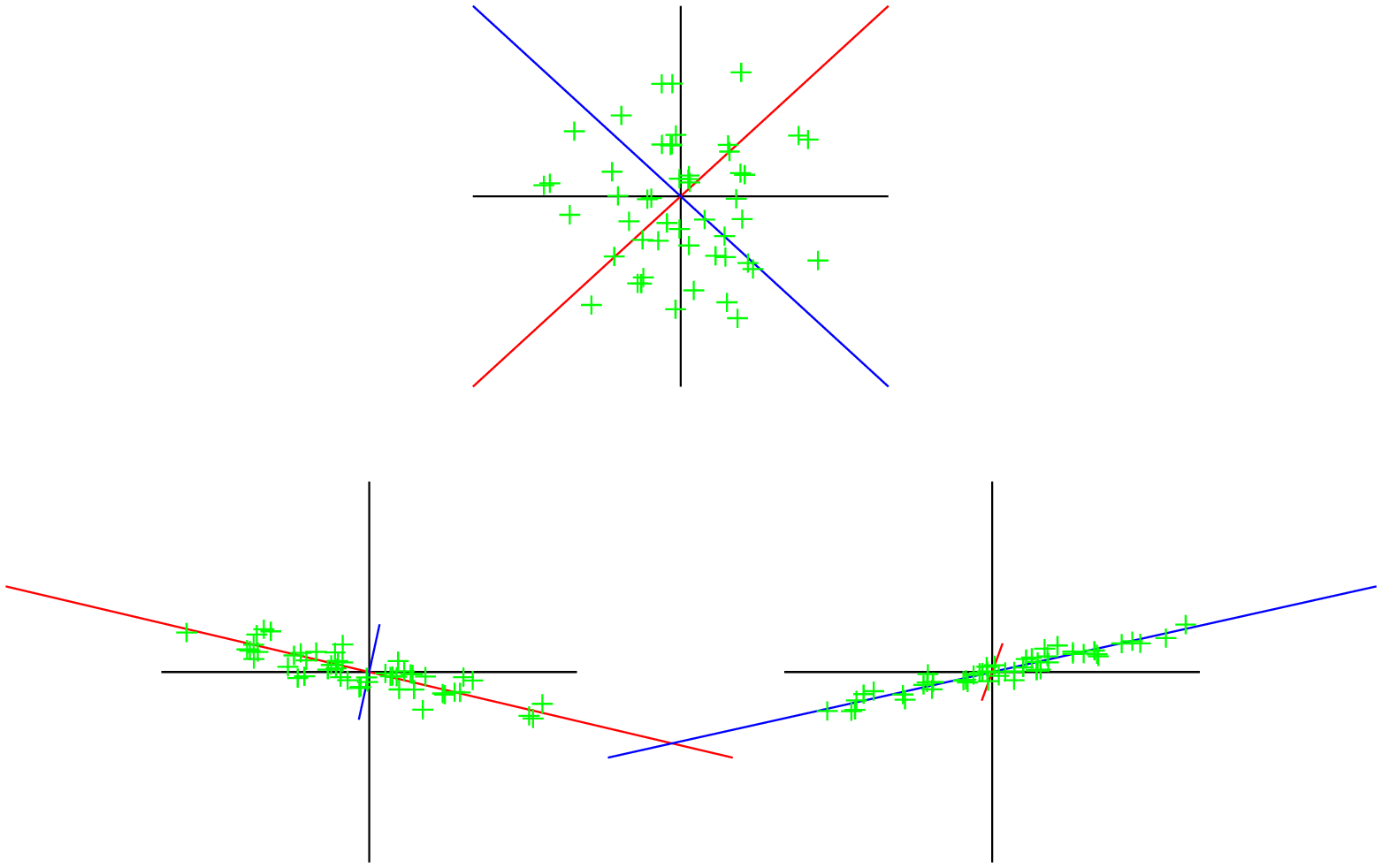}}
 \put(60.,55){$p^{1}$}
 \put(46.,70){$p^{2}$}
 \put(60,70){A}
 \put(60,42){B}
 \put(39.,22.){$x^{1}$}
 \put(24.,35.){$x^{2}$}
 \put(42,19.3){A}
 \put(24.5,25.5){B}
 \put(83.,22.){$y^{1}$}
 \put(68.,35.){$y^{2}$}
 \put(85.,28){B}
 \put(68.5,25.5){A}
\put(90,77){\parbox[t]{65mm}{\caption{
 \label{modely} Variable physical quantities $p^{i}$ of a source
lead to variations of observable quantities $x^{k}$, $y^{k}$. Differences
between $x$ and $y$ in sensitivity with respect to subspaces of $p$ may
cause that $x$ seems to confirm validity of model A and exclude model B,
while $y$ confirms B and excludes A
 }}}
\end{picture}
\end{figure}


 To illustrate the danger of pure observational approach
which is seemingly independent of theory or even capable of deciding
which theoretical model is correct, let us assume an object, the state
of which is described by a set of parameters $p^{i}$. In our case the
object may be a star and the parameters quantities like density, velocity,
pressure, chemical composition, ionization state, magnetic field etc.
in different parts of the star. Our observations of the object usually
do not allow us to measure directly $p^{i}$, but some observable quantities
$x^{k}=x^{k}(p^{i})$ which are their functions. In astrophysics, these
quantities are mostly intensities of light in different wavelength bands,
polarization states etc. and they reflect the parameters owing to a
complex process of radiative transfer. The parameters $p^{i}$ are
mutually linked by physical laws, so that only a few of them are free
parameters and have to be determined from the observation. The dependence
of the observed quantities $x^{k}$ on $p^{i}$ is also a matter of
corresponding physics of the observed object, but also of the detecting
instrument and the interlying medium, which usually contributes by some
noise. The theory of the observed object consists first in the proper choice
of its model and the proper physics governing relations of $p^{i}$ as well
as $x^{k}(p^{i})$ and than in a technical problem to calculate $p^{i}$
and $x^{k}$ for chosen values of the free parameters, eventually to
solve the inverse problem and determine the free parameters from the
observed quantities. The observation of the object consists first in real
getting of the values $x^{k}$ and its aim is to interpret the observations
by comparing the observed values with the theoretical predictions and
hopefully to decide which theoretical model with which values of
free parameters corresponds better to the observational results.
However, even a very simplified model illustrated in Fig.~\ref{modely}
reveals, that the commonly used practice of best fit (which is
also basis of the disentangling) need not be at all a good fit,
if the theoretical model is not appropriate or the data are not
sufficiently sensitive to parameters to be determined. The upper
diagram shows a two-dimensional subspace of parameters $p^{i}$
centered at their mean values for the object. Assume the object to
be perturbed by some process (e.g. a star oscillates around its
equilibrium state, or it is affected by magnetic field) and the
parameters change for values $\delta p^{i}$. Different theoretical
models (e.g. model A and B) of the perturbations may predict different
bounds $F(\delta p^{i})=0$ of the perturbations (drawn in the Figure
by the skew lines $\frac{\partial F}{\partial\delta p^{i}}\delta p^{i}=0$).
These perturbations of the object result in perturbations of
observed quantities
$\delta x^{k}=\frac{\partial x^{k}}{\partial p^{i}}\delta p^{i}$
and the models are represented by (generally) different lines also
in the space of the observed quantities $\delta x^{k}$.
In Fig.~\ref{modely}, random uncorrelated perturbations of
$\delta p^{i}$ (drawn by crosses) are chosen, which suggest that neither
model A nor B is a sufficient explanation of the process of perturbations
(or that both processes described by these models take place simultaneously).
However, different observed quantities $x^{k}(p^{i})$ and $y^{k}(p^{i})$
(in the left and right bottom diagrams, resp.) may be sensitive to
different subspaces of $\delta p^{i}$ and each one of them seems to be
a convincing observational proof of one of the models and an elimination
of the other, but their evidence is mutually opposite. The problem seems
to be trivial, if the quantities $p^{i}$ or $x^{k}$ are of similar
nature, so that their scaling could be compared. However, if $p^{1}$
and $p^{2}$ are, e.g., density and temperature, resp., and $x^{1}$ and
$x^{2}$ magnitude and radial velocity, there is no way to judge if the
transform $x^{k}(p^{i})$ is distorted similarly as the left or right
diagram.

 It follows from this example that the value of observational evidence
depends on the class of tested theoretical models as well as on the
quality of the data and the method of data processing. It is preferable
to take into account a wider variety of data, but even then a good
agreement with a theoretical model is not a guaranty that another model
is not better. Let us note that in stellar spectroscopy the directly
observed quantities are usually the fluxes in different wavelength
channels, while radial velocities are already an intermediate result of
an interpretation of the spectra, which is biased by several assumptions
on the observed source and the instrumentation. For instance, it is obvious
that a solution of radial velocity curve will lead to an underestimate of
$K$-velocity if it is smaller than the width of line, which is blended by
an unresolved weak line of secondary component, so that the radial
velocities measured classically (e.g. by means of moments of line-profile)
are in fact a weighted mean between primary and secondary component.
It is thus preferable to fit directly the spectra using the disentangling
instead of fitting radial-velocity curve in which the reliability of
individual data points is questionable. It is also known that the
measured radial velocity curves may be defleted from the simple Keplerian
form due to phase-dependent line-profile asymetries caused by the proximity
effects (reflection, tidal distortions etc.). Theoretically estimated
corrections are used for the radial-velocity curves, however, for the
disentangling it is needed to model directly the line-profile distortions.

\setlength{\unitlength}{1mm}
\begin{figure}[hbt]
\noindent\begin{picture}(140,60)
 \put(4.5,4.5){\epsfxsize=50mm
  \epsfbox{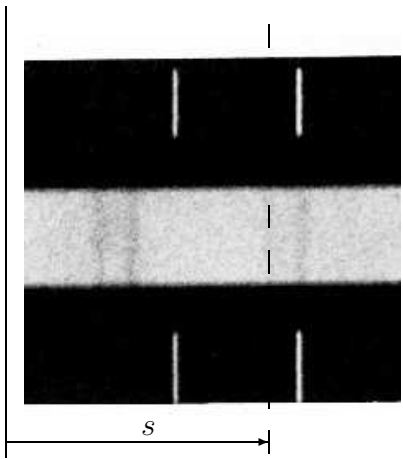}}
 \put(2.,2.){\vector(1,0){35}}
 \put(2.,0.){\line(0,1){60}}
 \put(37.,0.5){\line(0,1){3}}
 \put(37.,6.5){\line(0,1){3}}
 \put(37.,12.5){\line(0,1){3}}
 \put(37.,18.5){\line(0,1){3}}
 \put(37.,24.5){\line(0,1){3}}
 \put(37.,30.5){\line(0,1){3}}
 \put(37.,36.5){\line(0,1){3}}
 \put(37.,42.5){\line(0,1){3}}
 \put(37.,48.5){\line(0,1){3}}
 \put(37.,54.5){\line(0,1){3}}
 \put(20.,3.){$s$}
\put(70,45){\parbox[t]{85mm}{\caption{
 \label{spek} Part of Ond\v{r}ejov spectrogram No. 5197 of double-line binary
 96 Her. Wavelength scale is defined by (emission) comparison spectrum
 exposed on both sides of the stellar (absorption) spectrum in the central
 horizontal belt
 }}}
\end{picture}
\end{figure}
 Let me illustrate this problem by a practical experience which is
behind the development of my codes FOTEL and KOREL. In 1969 I was asked
by my supervisor S. K\v{r}\'{\i}\v{z} to investigate the precision
of measurement of radial velocities from photographic spectrograms
taken with the (then new) Ond\v{r}ejov 2m-telescope and to elaborate an
optimal method. The classical method was the measurement of positions
$s$ of stellar and comparison lines fitted visually by a line in microscope
of Abbe comparator -- cf. Fig.~\ref{spek}. (The measurement was very
laborious and for tens of lines measured in two directions and two
orientations of the plate it took usually a whole day for one plate.)
The wavelength scale $\lambda{s}$ was then calculated by least-squares
fit to the positions $s_{i}$ of all measured comparison lines with
known wavelengths $\lambda_{i}$ to minimize the infuence of errors
in measurement of individual lines. For the fit a polynomial approximation
\begin{equation}\label{polynom}
 \lambda(s)\simeq\sum_{k=0}^{m}a_{k}s^{k}
\end{equation}
was usually accepted, because
it was computationally straightforward. The problem was that for a small
degree of the polynome (e.g. $m=3$) the curve was not elastic enough
to follow the dispersion curve $\lambda{s}$ with sufficient precission,
and for higher degrees it was too much elastic and the new degrres of
freedom were misused by the code to fit individual errors in some lines,
which caused large defletions of the curve in regions, where suitable
comparison lines were missing. This is also an example, where a more
sophisticated theoretical model is needed for correct interpretation
of the measured data to restrict unrealistic degrees of freedom. Such
a model is given by the grating angular dispersion
\begin{equation}\label{grating}
 \lambda=\frac{d}{n}(\sin\alpha+\sin\beta)\; ,
\end{equation}
where $d$ is the constant of grating, $n$ is the order of spectrum,
$\alpha$ the angle of incidence (fixed by configuration of the spectrograph)
and $\beta=\beta(s)$ is the angle of difraction joined with the measured
position $s$ of a line by the projection properties of the spectrograph
camera. In the case of Ond\v{r}ejov spectrograph (with bent spectrogram
photoplates) it could be well approximated as $\beta=(s-s_{0})/f$, where
$f$ is the focal length of the camera, the desired model is thus
\begin{equation}\label{grating1}
 \lambda=b_{0}+b_{1}\sin(b_{3}s-b_{2})\; ,
\end{equation}
which should fit the relation $\lambda(s)$ with high precision with
just four free parameters (like the polynome of the third degree).
A disadvantage with the then available computer (type MINSK with
about 20kB of memory but extending in our present lecture-room) was
that unlike Eq.~(\ref{polynom}) linear in $a_{k}$, Eq.~(\ref{grating1})
is non-linear with respect to coefficients $b_{2}$ and $b_{3}$.
Generally, the equations
\begin{equation}\label{LSF}
 0=\frac{\partial}{\partial p_{k}}S(p)=
 \frac{\partial}{\partial p_{k}}\sum_{i}\left(y_{i}-F(x_{i},p)\right)^{2}\; ,
\end{equation}
which give the condition of minimum with respect to parameters $p$
of summed squares of residuals of measurements $\{(x^{i},y^{i})|_{i=1}^{n}\}$
are linear in $p$ if the function $F$ of the model $y=F(x,p)$ is
linear in $p$, i.e. if $F$ can be expressed as a linear combination
of some functions $f_{k}$, i.e. $F(x,p)=\sum_{k}p_{k}f_{k}(x)$.
It is thus advantageous to rewrite Eq.~(\ref{grating1}) in the form
\begin{equation}\label{grating2}
 \lambda=c_{0}+c_{1}\sin(c_{3}s)+c_{2}\cos(c_{3}s)\; ,
\end{equation}
which is linear in $c_{0}$, $c_{1}$ and $c_{2}$ and non-linear in
$c_{3}=1/f$ only. The minimization with respect to the linear
coefficients can thus be performed by solution of the set of linear
equations and by a computationally more expensive sampling with
respect to the non-linear parameter only (which is, moreover, well known
in this case). A similar linearization in $\gamma$ and $K$-velocity
helped to speed up the solution of radial-velocity curve in FOTEL
code and it is also the clue to the efficiency of Fourier disentangling.

\clearpage
\section{Methods of radial-velocity measurements}

\begin{wrapfigure}{r}{60mm}
\begin{picture}(57,35)
 \put(0,0){\epsfxsize=57mm \epsfbox{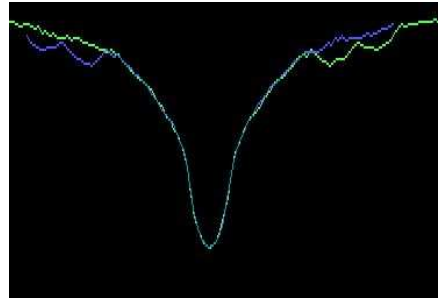}}
\end{picture}
\caption{Radial velocity measurement by oscilloscopic method}\label{Spefo}
\end{wrapfigure}

In the previous Section the determination of radial velocities using
classical Abbe comparator was described as an example of different
possibilities of numerical treatment of the measured data. The method
of measurement and its necessary instrumentation has been improved in
many variants. Especially the problem of determination of suitable
center of line was attacked using different devices and the corresponding
data-processing which enable to display the line-profiles (sometimes
together with their mirror images) -- e.g. osciloscopic machine
(Hossack 1952, cf. Fig.~\ref{Spefo}), tracings (Heinzel and Hadrava 1975),
TV-technique (Minaroviech et al. 1984), scans (\v{S}koda 1996).
In this Section we shall deal with some techniques
of radial-velocity measurement which are related to the method of
disentangling. My intention of this description is not to teach how
to use those methods (readers are refered to the original works in
this respect), but to compare the various approaches and to learn
from them in which respect the disentangling should be improved.


\subsection{Method of cross-correlation}
The idea of this method is based on the fact that the presence
of a weak signal of particular type, like is the spectrum
of a faint secondary component, blended with a stronger signal or
hidden in a noise can be better revealed from overall coincidence
with the observed signal than from some local features (cf.
Fig.~\ref{obr01}). Moreover, this method avoids the need to
identify spectral lines and to determine their centers to be
compared with the laboratory wavelengths. Instead, it simply
assumes that the spectral features in the observed star are
(up to the measured Doppler shift) identical with those in a
comparison star or a synthetic spectrum from a model atmosphere
chosen as a template. The cross-correlation is usually performed
numerically with digitalized observation, however, it is in principle
analogy of the photoelectric measurement of radial velocities
introduced by Griffin (1967) in which the measured spectrum
of single or binary star (Griffin 1975) is matched against an
appropriate diaphragm, e.g. a mask given by a template spectrum
of a suitable comparison star.

\setlength{\unitlength}{1mm}
\begin{figure}[hbt]
\noindent\begin{picture}(160,60)
 \put(5,5){\epsfxsize=80mm
  \epsfbox{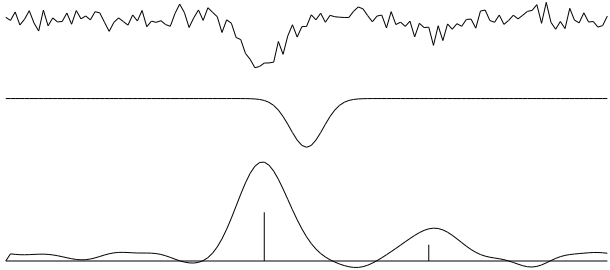}}
\put(90,62){\parbox[t]{65mm}{\caption{
 \label{obr01} Example of cross-correlation -- synthetic
 spectrum constructed as a superposition of lines of
 two components and a noise (the upper curve) correlated with an
 ideal line-profile (middle spectrum) gives a curve with
 amplitudes and positions of local maxima corresponding to the
 strengths and radial velocities of lines of the components
 (marked by vertical abscissae at the bottom curve), while
 the noise is smeared out by the integration.}}}
\end{picture}
\end{figure}

The cross-correlation
\begin{equation}\label{cc}
 F(v)\equiv\int I(x+v)J(x) dx
\end{equation}
of the observed spectrum $I(x)$ with a properly chosen template
spectrum $J(x)$ calculated in the logarithmic wavelength-scale
$x$ defined by (\ref{lnlamb}) indicates by its extremes the velocity
shifts $v$ at which a similar contribution appears in the spectrum
(cf. e.g. Simkin 1974, Hill 1993).\footnote{ In practice only
 the differences of the spectra from their continua are cross-correlated
 to subtract the large slowly varying contributions from the integral.}
If the spectrum $I$ is a superposition of spectra of two or more
components, each of them contributes to $F$ by its peak with maximum
shifted for the instantaneous radial velocity of the component. The
cross-corelation function $F$ calculated for each exposure separately
is thus fitted as a superposition of two or more Gaussian curves, which
approximate the cross-corelations of individual components.\footnote{
 Gaussian curves correspond to the thermal broadening of lines. However,
 the corresponding Maxwellian distribution is wider for light ions than
 for the heavier ones and it varies also with the temperature throughout
 the atmosphere. The Lorentzian wings and other individualities of
 different line profiles make the Gaussian approximation even more
 questionable. On the other hand, the turbulent broadening is often
 approximated by a Gaussian, the width of which equal for all particles
 is added (in squares -- cf. Eq.~(\ref{mom7})) to their thermal widths.
 Moreover, the profile of cross-corelation function is closer to Gaussian
 than the profiles of cross-corelated functions -- cf. Eq.~(\ref{mom9}).}

 An ideal template $J$ should contain the same lines in the same
ratios of their strengths as the spectrum of component to be
measured. In practice an observed spectrum of a star of similar
spectral type or a synthetic spectrum from a model-atmosphere
are used. It should be noted that a misunderstanding is spread
between many users who suppose that also the rotational broadening
of the template $J$ should be similar to that of the measured star
and they even artificially broaden a slowly rotating or synthetic
template $J_{0}$ taking $J=J_{0}\ast R$.\footnote{ It will be shown
 later that this commonly accepted procedure for ``spinning" the
 profiles is a crude approximation inconsistent with the physics
 of stellar atmospheres.}
However, the cross-correlation with such $J$ is identical with
cross-correlation of $I$ and $J_{0}$ additionally cross-correlated
in with $R$ (or equivalently convolved with its reverse). It means
that the final cross-correlation is only smeared out, which makes its
interpretation less precise.\footnote{
 Spectra of stars with small rotational $v\sin i$ are thus
 preferable templates.}

 Some complication in the method may be due to side-peaks which
arise in the cross-correlation function (CCF) from coincidence of lines
in the observed spectrum with different lines in the template.
These aliases are suppressed with increasing number of lines in
the used spectral region (and they are not present at all for just
one line). They are remote from the main peak of CCF if the lines
are sparse in the region, but they may appear in the case of multiplets
and blends of lines.

 Experience with the method shows that the cross-correlation is not
much sensitive with respect to the choice of the spectral type of the
template spectrum. Nevertheless, the template spectrum should be known
a~priori. This can be quite difficult task to estimate for a faint
companion.\footnote{ As already mentioned, an attempt to overcome
 just this problem using a cross-correlation of spectra of the same
 binary obtained at different phases was one of the ideas from
 which the Fourier disentangling has arisen.}
A partial aid in this respect may be the elegant option used by
Bagnuolo et al. (1994) -- to choose as the template a spectrum of
the same binary taken near a conjunction, where both components
have nearly the same radial velocity.

 A generalization of this method is a two-dimensional cross-correlation
\begin{equation}
 F(v_{1},v_{2})=\int I(x)[J_{1}(x-v_{1})+J_{2}(x-v_{2})] dx\; ,
\end{equation}
enabling one to choose templates corresponding to different spectral
types for the primary and secondary component (cf. Zucker and
Mazeh 1994, Zucker et al. 1995).

 The same authors (Zucker and Mazeh 2006) tried to avoid the problem
of finding a proper template in their modification of cross-correlation
method which they designed for single-line binaries (and named TIRAVEL).
Like in the disentangling, they use another exposure of the same star
as a template and calculate a matrix $R(v_{1},... ,v_{N})$ with components
given by cross-correlations between all pairs of normalized exposures
\begin{equation}
 R_{ij}=\int I(x-v_{i},t_{i})I(x-v_{j},t_{j})dx\; ,
\end{equation}
which is a function of radial-velocity shifts $v_{i}$ of the star at
individual exposures. The proper combination of these velocities
giving the best alignment of all observed spectra is then indicated
by maximum of the largest eigenvalue of $R$. The authors suggested a
possibility to generalize their method for double-lined binaries using
a combination with some method of decomposition, which would actually
bring it closer to a method of disentangling. Zucker and Mazeh compared
TIRAVEL with KOREL and they advertised alleged advantages of their
method, which should consist in the fact that their velocities are free
variables not bounded by an orbital motion. This claim reveals that
they are not aware of KOREL's capabilities (for which mere radial
velocities of SB1 are only an extremely trivial application). However,
just in the case of perturbation by a third body, which they mentioned as
an example of the need of free velocities, the possibility to converge
parameters of both the close and the wide orbit is an advantage.

 Even if the need of an external template would be avoided and coincidences
between lines belonging to different components distinguished, there is
still a disadvantage of the cross-correlation technique caused by the fact
that the cross-correlation profile is broadened due to widths of both
the observed spectrum and the template (cf. Eq.~(\ref{mom7})).
This shortcoming of the cross-correlation method is well analyzed
by Rucinski (2002), who developed a method of broadening function
to overcome it.

\subsection{Method of broadening function}
In the method introduced by Rucinski (2002), the broadening function
$B(x)$ is sought from observations to satisfy the relation
\begin{equation}\label{bf}
 I=B\ast J
\end{equation}
between the template $J$ and the observed spectrum $I$. If the observed
spectrum corresponds to a Doppler-shifted template or to a superposition
of such shifted spectra for each component, then $B$ is a shifted
Dirac delta-function, or a sum of them. One or more components
may also correspond to a broadened template spectrum (e.g. due
to a rotation broadening). In favourable cases when lines of $I$
are wide and lines of $J$ are narrow, $B$ will be a Doppler
shifted broadening profile. For a given $I$ and chosen $J$,
Eq.~(\ref{bf}) can be solved with respect to $B$ using e.g.
the SVD method.
Position of peak of $B$ gives again the information about the
instantaneous Doppler-shift of the observed spectrum. The shape
and especially the width of peak of $B$, which unlike $F$ in
Eq.~(\ref{cc}) is not quadratic in the line profile, can yield
also an information about the broadening of lines in $I$.
An example of such a treatment of Schlesinger -- Rossiter --
Mc Laughling rotation effect from the profile of broadening function
(cf. Albrecht et al. 2007).

 In comparison with the method of cross-correlation function,
the method of broadening function is a more direct way of interpreting
the spectroscopic data, but the problem of the proper choice of the template
persists in the later method and, it is at least equally important
as in the former. A progress in detailed analysis of the observed
broadening profile in this method can be implemented into disentangling
and vice versa, but an important difference between these methods remains
in the avoiding the need of template in disentangling.

\clearpage
\section{Methods of decomposition of spectra}
It is important to distinguish contributions of individual
components of a multiple stellar system to its common spectrum
not only to be able to measure the radial velocities and to solve
for orbital parameters, but also to find the physical characteristics
of atmospheres of the stars and to put them into context with their
masses and evolutionary status.

 An exceptional opportunity is yielded by eclipsing binaries for
which spectrum of one component can be obtained during a total
eclipse and the other spectrum can be found as its difference
from spectrum obtained out of eclipse. However, for majority
of stellar systems some more sophisticated methods must be used.

 In principle, any method able to separate the spectra of components
can be used in combination with any method for determination of radial
velocities and orbital parameters as a part of the procedure of
disentangling (i.e. as the right-hand green branch in the scheme
of disentangling in Fig.~\ref{disch} on p.~\pageref{disch}).
On the other hand, the versatility, preciseness and numerical
efficiency of the chosen method of decomposition dominates the
properties and applicability of the overall disentangling. It is
thus worth to study different methods of the decomposition of
spectra in details.

\subsection{Direct subtraction}
The most straightforward method of decomposition of spectra is
the direct subtraction involved by Ferluga et al. (1991, 1997), in
which two spectra $I_{a}, I_{b}$ of a binary obtained at different
phases (best of all at the opposite elongations, at extremes of
radial velocities $v_{a,b}$ of both components) are used.
If the spectra of individual components are $J_{1}, J_{2}$,
the relation
\begin{eqnarray}\label{Fer1}
 I_{a}(x)&=&J_{1}(x-v_{1a})+J_{2}(x-v_{2a})\; ,\\
 I_{b}(x)&=&J_{1}(x-v_{1b})+J_{2}(x-v_{2b})\; ,\label{Fer2}
\end{eqnarray}
should be valid. From here we can calculate both spectra
recurrently pixel by pixel
\begin{eqnarray}
 J_{1}(x)&=&J_{1}(x-v_{1a}+v_{1b}+v_{2a}-v_{2b})
  -I_{a}(x+v_{1b}+v_{2a}-v_{2b})+I_{b}(x+v_{1b})\; ,\\
 J_{2}(x)&=&J_{2}(x-v_{1a}+v_{1b}+v_{2a}-v_{2b})
  +I_{a}(x+v_{2a})-I_{b}(x-v_{1a}+v_{1b}+v_{2a})\; ,\hphantom{m}
\end{eqnarray}
starting from a wavelength, where the spectra of both components
can be approximated by continuum, and proceeding in positive
or negative direction of the logarithmic wavelength variable
$x$ (depending on the sign of the expression
$v_{1a}-v_{1b}-v_{2a}+v_{2b}$) toward groups of spectral lines.

 This straightforward method clearly shows an obstacle, which
is in principle a problem also for later more sophisticated methods.
The region of values of $x$ covered by each exposure contains
information about spectra of individual components in regions
shifted for $v_{1,2}$, so that the information on both spectra
is available only in the overlap of both regions. To be able to
separate both spectra, one must add the missing information about
the spectrum of the other component, usually the assumption that
it is a pure continuum without any line. However, this can be a
source of error, which may then spread also inside the region
where the solution should be well determined.

 The main disadvantage of this method is the fact, that owing
to the recurrent procedure of solution the influence of random
observational noise (which should be added on the right-hand
sides of Eqs. (\ref{Fer1}) and~(\ref{Fer2})) is cumulative, so
that after passing a group of spectral lines the solution can
deflect from the correct value of the continuum. The influence
of the noise can be reduced by averaging results of solutions
obtained from a larger number of pairs of exposures. However,
this suggests to develop a method searching ab initio for
the best fit to a higher number of observed spectra,
so that the solution $J_{1,2}$ of corresponding system of
linear equations of type (\ref{Fer1}) would be overdetermined.

\subsection{Iterative subtraction}
Another similar method of decomposition of spectra based on an iterative
procedure of subtraction has been introduced by Marchenko et al. (1998)
as described by Gonz\'{a}lez and Levato (2006) who suggested to involve
it in a kind of disentangling with radial velocities (and incorrectly
reffered to the full disentangling of orbital parameters by both Simon
and Sturm 1994, and Hadrava 1995, as a decomposition only).
This method overcomes the above mentioned disadvantage of Ferluga's
method. It has been described for double-line binaries only, but it can
easily be generalized to more than two components as follows.

 Suppose we have $N$ exposures taken at times $t_{k}|_{k=1}^{N}$ for
which the radial velocities $v_{j}(t_{k})$ of the components $j=1,2,... ,n$
are known. According to Eq.~(\ref{Kor1})
\begin{equation}\label{Marchen1}
 I(x,t_{k})=\sum_{j}I_{j}(x)\ast\delta(x-v_{j}(t_{k}))\; .
\end{equation}
To get estimate $I_{j}^{(i)}$ of spectrum $I_{j}$ of the component $j$
in the $i$-th step of the iteration (taking e.g. $I_{j}^{(0)}=0$ as the
initial approximation), we can average all observed spectra shifted to
the rest wavelength-scale of the component $j$ after subtracting the
properly shifted estimates of spectra of other components according to
their estimates in the previous step,
\begin{eqnarray}
 I_{j}^{(i+1)}(x)&=&\frac{1}{N}\sum_{k=1}^{N}\left(
  I(x,t_{k})\ast\delta(x+v_{j}(t_{k}))
 -\sum_{l\neq j}^{n}I_{l}^{(i)}(x)\ast\delta(x-v_{l}(t_{k})+v_{j}(t_{k}))
 \right)\nonumber\\
 &=&\frac{1}{N}\sum_{k=1}^{N}I(x,t_{k})\ast\delta(x+v_{j}(t_{k}))\label{Marchen2}\\
 && -\sum_{l\neq j}^{n}I_{l}^{(i)}(x)\ast
 \left(\frac{1}{N}\sum_{k=1}^{N}\delta(x-v_{l}(t_{k})+v_{j}(t_{k}))
 \right)\; . \nonumber
\end{eqnarray}
Substituting here for $I(x,t_{k})$ from Eq.~(\ref{Marchen1}), we
get the difference between $I_{j}(x)$ and its estimate in the $i+1$-step
\begin{equation}\label{Marchen3}
 I_{j}^{(i+1)}(x)-I_{j}(x)=
 \sum_{l\neq j}^{n}(I_{l}(x)-I_{l}^{(i)}(x))\ast
 \left(\frac{1}{N}\sum_{k=1}^{N}\delta(x-v_{l}(t_{k})+v_{j}(t_{k}))
 \right)\; .
\end{equation}
Fourier transform of this equation reads
\begin{equation}\label{Marchen4}
 \tilde{I}_{j}^{(i+1)}(y)-\tilde{I}_{j}(y)=
 \sum_{l\neq j}^{n}(\tilde{I}_{l}(y)-\tilde{I}_{l}^{(i)}(y))
 \left(\frac{1}{N}\sum_{k=1}^{N}\exp(i(v_{l}(t_{k})-v_{j}(t_{k}))y)
 \right)\; .
\end{equation}
It means that provided the radial velocities at individual exposures
are so well randomly spaced over the orbital period that the complex
exponentials in the last sum partly cancel each other for all Fourier
modes $y$ to a quantity smaller in absolute value smaller than 1 (it is
again impossible for $y=0$, but it can happen also for some harmonics
of the orbital frequency), the method converges exponentially to
the true solution of Eq.~(\ref{Marchen1}). As we shall see, the method
is in practice equivalent with the subsequent method introduced
by Bagnuolo and Gies much earlier and it cannot compete with the
methods of disentangling. However, it may provide an alternative
insight into these methods.

\subsection{Method of tomographic separation}
A method for decomposition of a larger number of observed spectra
of a binary (with known radial velocities) was introduced already by
Bagnuolo and Gies (1991). The averaging tendency of their method
diminishes the observational noise in individual exposures.

 This method is based on the mathematical equivalence of the task of
decomposition with the problem of image reconstruction in tomography.
The superposition of Doppler-shifted component
spectra at different orbital phases can be treated as projection
of two parallel linear objects (e.g. photographic spectrograms)
viewed from different angles. Unlike the standard tomography, the
line of a detector is not perpendicular to the direction of projection,
but it remains parallel with the lines of component spectra (which
differs by rescaling of each projection).
Any standard numerical method
of tomographic reconstruction should thus be able to calculate
the distribution of intensities (or opacities) in this object
of dimension $2\times N$ from a higher number of exposures (each
one consisting typically of $N$ pixels) if a sufficient coverage
of viewing angles / orbital phases is available -- cf.
Fig.~\ref{obr02}.\footnote{ The equivalence of both mathematical
 problems can be used also in the opposite direction, it means
 that numerical methods for spectra decomposition could be
 applied for computer tomography as well (cf. Hadrava 2001a).}

\setlength{\unitlength}{1mm}
\begin{figure}[hbt]
\noindent\begin{picture}(140,47)
 \put(20,30){$f_{i_1,2}$}
 \put(70,35){$f_{i_1,1}$}
 \put(5,12){$p_{k_1,1}$}
 \put(11,4){$p_{k_1,2}$}
 \put(70,2){$p_{k_1,3}$}
 \put(5,0){\epsfxsize=80mm
  \epsfbox{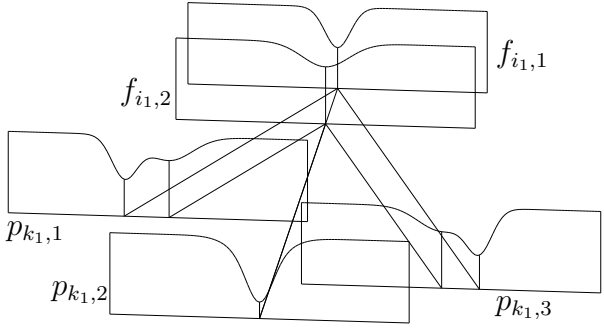}}
\put(90,40){\parbox[t]{65mm}{\caption{
 \label{obr02} Principle of tomographic separation -- projections
 of spectra of individual components (the two spectra in the back)
 into different directions correspond to different Doppler shifts
 in different phases (the three spectra in the foreground).}}}
\end{picture}
\end{figure}

 Bagnuolo and Gies chose the Iterative Least Square Technique
-- cf. Eq.~(\ref{ILST5}).
For a double-lined binary $N_{2}=2$. We suppose $K_{2}>N_{2}$,
$K_{1}=N_{1}$ and approximate each bin of a projection $p_{k_1,k_2}$
as a superposition of both sources $f_{i_1,i_2}|_{i_{2}=1}^{2}$
Doppler shifted for the number of bins corresponding to the radial
velocity $v_{i_2}(t_{k_{2}})$ of the component $i_{2}$ at the exposure
time $t_{k_{2}}$ (i.e. at the proper angle of projection in the view
of tomography)
\begin{equation}
 W_{k_{1},k_{2}}^{i_{1},i_{2}}=\delta_{k_{1}}^{i_{1}+v_{i2}(t_{k2})}\; .
\end{equation}
Following Eq.~(\ref{ILST5}) the correction of an estimate $f_{j_1,j_2}$
reads
\begin{equation}\label{tomsep2}
 \Delta_{j_1,j_2}\simeq\frac{1}{K_{2}}
  \sum_{k_2}\left[p_{j_1+v_{j2}(t_{k2}),k_2}
  -\sum_{i_2}f_{j_1+v_{j2}(t_{k2})-v_{i2}(t_{k2}),i_2}\right]\; .
\end{equation}

 The basic idea of the equivalence of the decomposition with
a tomographic reconstruction of projections of a rigidly rotating
object does not allow a generalization to more components, with
the exception of one, but an important case: it is the case when
the spectrum of a third component does not move in radial velocities.
This is the case of a background star in the view-field of the
spectrograph, or presence of some interstellar lines, in some approximation
also a third companion of the binary (if the amplitude of its radial
velocities and velocities of the mass-centre of the close binary are
small or their variations negligible during the observational run)
or also a presence of telluric lines, if their seasonal variations
of radial velocities may be neglected. Such a case of the third component
has been really studied by Liu et al. (1997). In fact, despite the
assumption of the rigid rotation was an inspiration for a development
of the method, it is no more needed in the final formula (\ref{tomsep2}),
which is in fact equivalent to Eq.~(\ref{Marchen2}). The method is thus
more general than its name suggests and it would be applicable to any
multiple system as well.

 The terminology of this method is misleading also from another point
of view -- it resembles a class of methods designed for the Doppler
tomography, in particular tomographic reconstruction of circumstellar
matter in interacting binaries (cf. e.g. Marsh and Horne 1988).
The difference is that while the method by Bagnuolo and Gies aims at
reconstruction of unknown spectra from their different superpositions
(projections), the Doppler tomography reconstructs the distribution
of emissivity in the velocity-space with an emission line-profile
postulated to be a $\delta$-function in frequency. It would be
desirable to decompose the intrinsic spectra (including their strength)
for a field of sources distributed in the velocity-plane corrotating
with the binary. However, such a problem does not have a unique
solution; obviously, a rotating ring of $\delta$-line emitters results
at a double-peaked profile, which can also be produced by a static source
with the same intrinsic line-profile. Nevertheless, a compromising
problem could be solved if the solution would be constrained to some
reasonably chosen subspaces of functions in both the spectral and
velocity spaces.

 The disadvantage of needing to know the radial velocities first is
common for all mentioned methods of decomposition. However, complementing
this approach with the method of cross-correlation, it is obvious
that both the decomposition of the spectra and the measurement of radial
velocities (and solution of orbital parameters) should be possible
in an iterative process, without the ad hoc choice of the template
spectrum. Such a task is solved by the methods of disentangling of
spectra.

\clearpage
\section{Method of wavelength-domain disentangling of spectra}\label{SSS}
\label{mat}

The first method enabling simultaneous decomposition of the
spectra of binaries and measurement of their radial velocities,
or actually directly the solution of orbital parameters was published by
Simon and Sturm (1994).\footnote{ They also introduced for
 their method the name `disentangling', which is used here to
 denote the problem in this complexity.}

 The decomposition of the spectra is solved by Simon and Sturm
(similarly as in the methods by Ferluga or Bagnuolo and Gies)
directly for $2M$ values of spectral fluxes ${\bf I}_{A}$ and
${\bf I}_{B}$ of components $A$ and $B$ at individual values
of the independent variable $x$ (here $M$ corresponds to the
number of pixels in the typical exposure which is typically
of the order $10^{3}$). A set of $N\times M$ linear equations
\begin{equation}\label{SS}
 \left(\begin{tabular}{cc}
 ${\bf M}_{A1}$& ${\bf M}_{B1}$\\
 \multicolumn{2}{c}{...}\\
 ${\bf M}_{AN}$& ${\bf M}_{BN}$
 \end{tabular}\right) \left(\begin{tabular}{c}
 ${\bf I}_{A}$\\ ${\bf I}_{B}$ \end{tabular}\right)
 = \left(\begin{tabular}{c}
 ${\bf I}(t_{1})$\\...\\ ${\bf I}(t_{N})$ \end{tabular}\right)\; ,
\end{equation}
is obtained for these unknown values, where $N$ is the number
of exposures (it must be $N\ge 2$).
There are $N$ subvectors ${\bf I}(t_{l})$ of dimension $M$ with
spectra observed at times $t_{l}|_{l=1}^{N}$ of exposures on
the right-hand side of this equation.
In the simplest form, the submatrices
\begin{equation}
 {\bf M}_{jl}={\small\begin{tabular}{l}
 $\hspace*{4mm}
 \overbrace{\hphantom{\begin{tabular}{cccc}0& \multicolumn{2}{c}{...}&
 1\end{tabular}}}^{v_{j}(t_{l})}$
 \vspace*{-1mm}\\
 $\left( \begin{tabular}{ccccccc}
 0& \multicolumn{2}{c}{...} &1&\multicolumn{3}{c}{...}\\
  &0& & &1&\multicolumn{2}{c}{} \\
\multicolumn{7}{c}{...}\\
\multicolumn{3}{c}{...}&0&\multicolumn{2}{c}{...} &1\\
\multicolumn{7}{c}{...}\\
 \multicolumn{6}{c}{...}&0
 \end{tabular}\right)$ \vspace*{5mm}\end{tabular}} \vspace*{-5mm}
\end{equation}
of dimension $M\times M$ on the left-hand side have the only
nonzero elements equal to 1 shifted from the main diagonal for
a number of pixels corresponding to the values $v_{j}(t_{l})$ of
Doppler shift of component $j|_{j=A}^{B}$ at time $t_{l}$. The
form of the matrices may be more tricky to include also the
interpolation from input data which would not be sampled in
the same grid points. Another option is to enlarge the matrices
for additional $\max(v_{j}(t_{l}))$ columns on each side and
correspondingly to enlarge the vectors ${\bf I}_{A,B}$ for
wavelengths, which appear in the windows ${\bf I}(t_{j})$
in some exposures only. This enlarging of the window may
help to abandon the edge-errors of the decomposition in the
cases when it is impossible to prevent all lines from escaping
from the window by proper choice of its limits. However, these
regions behind the edges will be less and less overdetermined
and hence the noise may increase therei, like in the method of
the direct subtraction.

 Simon and Sturm solved the set of equations which has a sparse
matrix but of very high dimension using the method of `singular
value decomposition'.
This method has been applied on real data in several studies
(e.g. Sturm and Simon 1994, or Simon et al. 1994).
It could be mentioned that the method was tested also by Hynes
and Maxted (1998) using some simulated data.

 It should be noted that Simon and Sturm introduced their method
in this simple form for two components only. Obviously, it is
straightforward to generalize it for more components by
increasing the number of the columns with the submatrices. The
authors mentioned that they give the simplest unit off-diagonal
form of the submatrices, which requires to rebin all input data
into the same equidistant scale, but they could be constructed
in a more complicated form to interpolate directly in an
arbitrary sampling. Their form of the submatrices implies also
that the Doppler shifts must be rounded to an integer multiple
of the sampling step. As I shall show later, it is also possible
to increase the resolution in radial velocities below the bin
size by an appropriate construction of a three-(off-)diagonal
submatrix.

 Despite of the fact that the decomposition of the spectra is the more
difficult part of the disentangling, including the solution of orbital
parameters into the same procedure represents an essential
qualitative advance in the interpretation of the stellar spectra.
An ideal procedure of any interpretation would be to fit the
observational data by a complete theoretical model and to estimate,
how far the conclusions of the study are determined by the
original observations. If this task is split into subsequent
steps like the determination of radial velocities in individual exposures
first and the solution of radial-velocity curves only after, the information
about the decisive power of the source data is partly lost in the
questionable reliability of the intermediate results. This is
a reason to prefer the direct solution of orbital parameters
instead of disentangling of component spectra together with
their radial velocities\footnote{ Such an option may be preferable if the
 spectra contain a component which need not follow any orbital
 motion, like an absorption in gaseous streams projected on
 photospheres of components in interacting binaries.}
and the same strategy leads us to consider a further generalization
of the method of disentangling to include simultaneously some
other effects, like the line-profile variability.
At the same time, the user must be aware that the choice of
a particular model to fit the data may be biased and even
good results do not exclude that some other model may better
explain the same data or data enriched by some additional
observations.

\clearpage
\section{Fourier disentangling and its generalizations}

Unlike the approach by Simon and Sturm, which is based on the
decomposition of spectra in the wavelength domain, KOREL uses the
least square fit of Fourier transforms of observed spectra (Hadrava
1995, cf. Section~\ref{PFD}), which makes the solution numerically
easier and which thus enables further generalizations (see e.g.
Hadrava 1997, 2004a and Sections~\ref{LSV} and \ref{LPV}).
The mathematical basis of the present method is analogous to the
cross-correlation technique, however, the basic difference from
the standard cross-correlation (e.g. Hill 1993) or its
two-dimensional generalization (Zucker and Mazeh 1994) is that
in KOREL all spectra at different phases of the same variable
star are mutually compared and decomposed instead of performing
cross-correlation of each spectrum of the variable separately
with an ad hoc chosen standard. Numerous applications of this
method to real data have been reviewed by Holmgren (2004).

\subsection{Principle of Fourier disentangling}\label{PFD}
We shall return now to details of the Fourier disentangling outlined
already in Section~\ref{entry}.
Before explaining the procedure of solution of Eqs.~(\ref{Kor1})
or (\ref{FKor1}), let us generalize them for purposes of later
development of the method. In view of the possibility to involve
a broadening of line-profiles similarly as the Doppler shifts by
another convolution with a corresponding broadening profile
(cf. Eq.~(\ref{bf})), we can generalize Eq.~(\ref{Kor1}) to
\begin{equation}\label{Kor2}
 I(x,t)= \sum_{j=1}^{n}I_{j}(x)\ast\Delta_{j}(x,t,p)\; ,
\end{equation}
and its Fourier transform to
\begin{equation}\label{FKor2}
 \tilde{I}(y,t)=
  \sum_{j=1}^{n}\tilde{I}_{j}(y)\;\tilde{\Delta}_{j}(y,t,p)\; ,
\end{equation}
where $\Delta_{j}$ are some general broadening functions, which
may involve now not only the Doppler shifts, but possibly also
some line-profile broadenings at the time $t$; $\tilde{\Delta}_{j}$
are their Fourier transforms. These functions depend on some
parameters $p$ characterizing either the orbital motions of the
components or physical and geometric conditions of formation of
their spectra.

 The principle of disentangling consists in minimization of the
sum of integrated squares of differences between the observed
and model spectra (on the left and right hand sides of
Eq.~(\ref{Kor2}), respectively)
\begin{equation}\label{Kor3}
 0=\delta\sum_{l=1}^{N}\int\left|I(x,t_{l})-\sum_{j=1}^{n}I_{j}(x)
  \ast\Delta_{j}(x,t_{l},p)\right|^{2}dx\; ,
\end{equation}
which is supposed to be due to observational noise in $I(x,t)$.
This expression implicitly assumes, that the noise is the same
for all wavelengths. Only limited spectral regions are available
in practice, which means that outside the corresponding range of
$x$ we take the spectra with zero weight.
The minimization is performed with respect to the component spectra
$I_{j}(x)$ (which gives the decomposition of the spectra) as well
as to the orbital parameters (corresponding to the solution of
radial-velocity curves with implicitly involved radial velocity
measurements) or other free
parameters $p$.

 According to the Parseval theorem (\ref{Parseval}), the condition
(\ref{Kor3}) can be equivalently rewritten as minimization (i.e.
zero variation) of sum of integrals of the Fourier transforms
\begin{equation}\label{FKor3a}
 0=\delta\sum_{l=1}^{N}\int\left|\tilde{I}(y,t_{l})
  -\sum_{j=1}^{n}\tilde{I}_{j}(y)
  \tilde{\Delta}_{j}(y,t_{l},p)\right|^{2}dy\; .
\end{equation}
This form of the condition assumes implicitly, that the noise is
white, which need not be always the case, as we shall discuss
later. It can thus be better to involve some weights $w_{l}(y)$
of individual Fourier modes and to write the condition in the
form
\begin{equation}\label{FKor3b}
 0=\delta S\; ,
\end{equation}
where
\begin{equation}\label{FKor3}
 S=\sum_{l=1}^{N}\int\left|\tilde{I}(y,t_{l})-\sum_{j=1}^{n}\tilde{I}_{j}(y)
  \tilde{\Delta}_{j}(y,t_{l},p)\right|^{2}w_{l}(y)dy\; .
\end{equation}

 The equation for decomposition of spectra can be obtained by
varying $S$ with respect to individual Fourier modes,\footnote{
 Because $\tilde{I}$ are complex values, the partial derivatives
 of $S$ must be calculated either independently with respect to
 their real and imaginary parts, or with respect to $\tilde{I}$
 and its complex conjugate as independent variables.}
\begin{eqnarray}\nonumber
 0&=&\frac{\partial S}{\partial\tilde{I}^{\ast}_{m}(y)}=
 -\sum_{l=1}^{N}\left[\tilde{I}(y,t_{l})
 -\sum_{j=1}^{n}\tilde{I}_{j}(y)\tilde{\Delta}_{j}(y,t_{l},p)\right]
 \tilde{\Delta}_{m}^{\ast}(y,t_{l},p)w_{l}(y)=\\
 &=&
 \sum_{j=1}^{n}\left[\sum_{l=1}^{N}\tilde{\Delta}_{j}(y,t_{l},p)
  \tilde{\Delta}_{m}^{\ast}(y,t_{l},p)w_{l}(y)\right]\tilde{I}_{j}(y)
 -\sum_{l=1}^{N}\tilde{I}(y,t_{l})
  \tilde{\Delta}_{m}^{\ast}(y,t_{l},p)w_{l}(y)\; , \label{SI}
\end{eqnarray}
which is obviously a set of $n$ linear equations for each Fourier
mode separately.

The solution is similar to quantum mechanics, where any linear
operator can be equivalently written in different representations,
however, it is easier to calculate its functions and in particular
its inversion in a reducible representation.
This is the basic trick which makes the Fourier disentangling
more versatile compared to the wavelength domain disentangling,
despite in principle both methods are equivalent and their general
features, which are easier to understand in one representation
are in fact valid for the other as well.\footnote{ It is obvious
 that for the special case $n=2$, Eqs.~(\ref{Kor1}) or
 (\ref{FKor1}) are equivalent with Eq.~(\ref{SS}). Consequently,
 it is incorrect, what Simon and Sturm (1994, p. 291) claim, that
 unlike the problem of tomography, which is analytical by its
 nature, the decomposition or disentangling is the algebraic one.
 Both these problems, as well as others related with them, can
 be formulated analytically and solved algebraically
 in a chosen numerical representation. The generalizations of
 disentangling given below could be, in principle, done also in
 the wavelength domain as it was done in the pioneering work
 by Simon and Sturm. However, the numerical solution would be
 then more difficult.}

Regarding that $I_{j}(x)$ are real values, the
Fourier transform $\tilde{I}_{j}$ must satisfy
\begin{equation}
 \tilde{I}_{j}(-y)=\tilde{I}_{j}^{\ast}(y)\; ,
\end{equation}
i.e. its real part must be symmetric and the imaginary part
antisymmetric. It is thus sufficient to solve $n/2$ independent
complex linear equations. Similarly, the matrix $a_{mj}\equiv
\sum_{l}\tilde{\Delta}_{j}(y,t_{l},p)\tilde{\Delta}_{m}^{\ast}(y,t_{l},p)
w_{l}(y)$ to be inverted for the solution is Hermitean, so that
only one triangular part of it must be computed according to the
definition and the other part is given by the symmetry.

 To optimize the solution with respect to the parameters $p$ we
can use the conditions
\begin{equation}\label{Sp}
 0=\frac{\partial S}{\partial p}=
 -2\Re\sum_{l=1}^{N}\int\left[\tilde{I}(y,t_{l})
 -\sum_{j=1}^{n}\tilde{I}_{j}(y)\tilde{\Delta}_{j}(y,t_{l},p)\right]
 \sum_{m=1}^{n}\tilde{I}_{m}^{\ast}(y)
 \frac{\partial\tilde{\Delta}_{m}^{\ast}(y,t_{l},p)}{\partial p}
 w_{l}(y) dy \; .\nonumber
\end{equation}
However, the dependence of $\tilde{\Delta}_{j}$ on $p$ can be
generally quite complicated and consequently also the solution
of these equations may be difficult. It can thus be easier to
minimize directly the expression for $S$ in the form (\ref{FKor3})
using some numerical method of optimization, like the simplex method.

 Let us note that despite the determination of radial velocities
and subsequent solution of radial-velocity curve is an obsolete
procedure compared to the disentangling of spectra, it may be still
useful to determine radial velocities for individual exposures
first to enable their combination with older data from literature
or with solution of light-curves. Having already disentangled
a set of spectra, it is possible to measure radial velocities
of the component stars in each of them by fitting
the spectrum as a superposition of the disentangled spectra
Doppler-shifted for velocities independent of the found orbital
parameters.

 As noted by D. Holmgren, it is interesting to mention that the
Fourier method of spectra decomposition was in a way anticipated
by the analysis of line-profile variability in the form of traveling
features in the spectra of $\zeta$ Oph by Walker et al. (1979).

\subsection{Simple Fourier disentangling}
Let us illustrate the Fourier disentangling first in the simple
case, which is equivalent to the above explained method by
Simon and Sturm (cf. Section \ref{SSS}), with the only
generalization that the observed system need not be a binary
only, but it may consist of $n$ stars. Then $\Delta$-functions
corresponding to the pure Doppler shifts are given by
\begin{equation}\label{Dopdel}
 \Delta_{j}(x,t,p)=\delta (x-v_{j}(t,p))\; ,
\end{equation}
their Fourier transforms are
\begin{equation}\label{FDopdel}
 \tilde{\Delta}_{j}(y,t,p)=\exp(iyv_{j}(t,p))\; ,
\end{equation}
and consequently (if we skip the weights $w_{l}(y)$ for simplicity)
Eq.~(\ref{SI}) reads
\begin{equation}\label{FKor4}
 \sum_{j=1}^{n} \left[\sum_{l=1}^{N}
  \exp(iy(v_{j}(t_{l},p)-v_{m}(t_{l},p)))\right]\tilde{I}_{j}(y)
  =\sum_{l=1}^{N}\exp(-iyv_{m}(t_{l},p))\tilde{I}(y,t_{l})\; .
\end{equation}
This set of equations can be solved with respect to
$\tilde{I}_{j}(y)$ whenever the matrix (with indices $j,m$) on
the left-hand side is non-singular.

 It is obvious that the singularity occurs always for $y=0$
(which can be seen also in Eq.~(\ref{FKor1})), when this equation
reduces for all $m$ to a single condition\label{cont}
\begin{equation}\label{FKor5}
 \sum_{j=1}^{n}\tilde{I}_{j}(0)=\frac{1}{N}\sum_{l=1}^{N}\tilde{I}(0,t_{l})
\end{equation}
for the sum of mean values of the component spectra to give
the mean value of the observed spectra. It means, that the
continua cannot be directly decomposed, because they are not
influenced by the Doppler shifts. An indirect method of
distinguishing the contributions to the continuum is described
in Section~\ref{LF}. This problem will be discussed in detail
in Section~\ref{norm}. It can happen also for non-zero
Fourier modes that the matrix is nearly singular and the solution
is thus unstable. This danger is higher for low modes, for which
the Doppler effect is smaller. These modes can be more influenced
by errors in rectification of spectra. This is why it may be
better to filter them out using some weights $w_{l}(y)$ as
additional multiplicators in the integrals in expression
(\ref{FKor3}). These multiplicators will not influence the
decomposed spectra directly by the amplitudes and phases of
individual Fourier modes but through the optimal values of the
parameters $p$.
It is possible in KOREL to cut out a chosen number of lowest
modes.\footnote{ Iliji\'{c} et al. (2000) advocate a filtering
 of high frequency noise from spectra before the disentangling.
 It would be possible to involve the filtering directly into
 the procedure of disentangling, if such a need will be
 confirmed. The present version of KOREL does not allow it
 not to complicate much its use. Some filtering of high frequencies
 is performed already by the interpolation of the original data
 into the chosen scale of $x$. \label{filtr}}

 As mentioned in the previous Section, limited ranges with a
finite sampling are used in practice both in the wavelength
domain as well as in the domain of the Fourier transforms of
spectra. The highest efficiency of numerical calculation is
achieved if the numbers of bins in both representations are
comparable. The Fourier transform between these limited sets
of discretized data points assumes a periodic repeating.
It means that a spectral line which disappears at some phase
behind one edge of the spectral region to be decomposed is
expected to appear at the other. Because this will not generally
happen (unless by chance a similar line exists at the other
edge), solution of decomposition cannot fit at the edges all
observed spectra with and without the line. This leads to errors
propagating from edges toward the middle of the decomposed range,
as described by Iliji\'{c} et al. (2000). To prevent these errors
the spectral regions chosen for the disentangling should have
continua\footnote{ Naturally the continua on both edges of each
 spectral region should have the same level to prevent also
 a discontinuity in jump between the edges of decomposed spectra.
 This can be  ensured by the rectification of input spectra to
 their continua.}
without spectral lines at both edges of the decomposed spectral
range, best of all wider than the expected amplitude of radial
velocities of the components. To facilitate the choice of convenient
spectral range code PREKOR has been written, which interpolates
from the data-files with individual exposures into chosen regions
with proper discretization and displays the result to enable
a visual check of the input data for KOREL. It may be difficult
to satisfy the demand for pure continua at edges of the region
when disentangling is applied to late-type stars. In such a case
the edge defects can be partly suppressed by the so called
tapering\label{taper} of the signal towards the edges
(or the hemming of signal window), 
i.e. a smooth suppressing the signal of source data in narrow
strips on both edges towards the continuum, as it is recommended
also in the technique of cross-correlation.
This trick is in fact similar to the choice of slightly higher
dimension of decomposed vectors ${\bf I}_{A,B}$ on the left-hand
side of Eq.~(\ref{SS}) of the method by Simon and Sturm than is
the dimension of the source data ${\bf I}(t_{l})$ on
the right-hand side.

\subsection{Line-strength variations and removal of telluric
 lines}\label{LSV}
The simplest generalization of the disentangling is to abandon
the assumption of constant component spectra and to admit a change
of strength of lines of some component. The original motivation
for this step was the experience that in some binaries errors
of radial velocities increased significantly close to conjunctions
where an eclipse could be expected. If the contribution of one eclipsed
star is missing in the spectrum of the whole system, the spectra
of the remaining are more pronounced and their sum cannot be
fitted so well as a superposition of lines with the same depths
as out of the eclipse.

 The effect of eclipse may be even more complex e.g. due to
the Schlesinger -- Rossiter -- Mc Laughling rotation effect,
however variations of line strengths are observed also as
the so called Struve -- Sahade effect (cf. Stickland 1997),
they are characteristic for circumstellar lines or telluric
lines etc. The calculation of the line-strength factors may
also partly compensate some instrumental and data-processing
imperfections like incorrect flatfielding or rectification.

 Let us thus generalize Eqs.~(\ref{Dopdel}) and (\ref{FDopdel})
by involving multiplicative line-strength factors $s_{j}(t)$,
i.e.
\begin{equation}\label{sDopdel}
 \Delta_{j}(x,t,p)=s_{j}(t)\delta (x-v_{j}(t,p))\; ,
\end{equation}
the Fourier transforms of which read
\begin{equation}\label{FsDopdel}
 \tilde{\Delta}_{j}(y,t,p)=s_{j}(t)\exp(iyv_{j}(t,p))\; .
\end{equation}
In this case Eq.~(\ref{SI}) gets slightly more complicated form
compared to (\ref{FKor4}), namely
\begin{eqnarray}\label{decomp}
 \sum_{j=1}^{n} \left[\sum_{l=1}^{N} w_{l,X}(y)s_{jl}s_{ml}
  \exp(iy(v_{j}(t_{l},p)-v_{m}(t_{l},p)))\right]\tilde{I}_{X,j}(y)
 &=&\nonumber\\
 =\;\sum_{l=1}^{N} w_{l,X}(y)s_{ml}\exp(-iyv_{m}(t_{l},p))
  \tilde{I}_{X}(y,t_{l}) \vspace*{-50mm}\; . &&
\end{eqnarray}
Here $s_{jl}\equiv s_{j}(t_{l})$ and the subscript $X$ refers
to different regions of the observed spectra, each one being
characterized by its initial wavelength $x$ and its dispersion
(in the value of radial velocity per one bin of the sampling in $x$).
The weights $w_{l,X}(y)$ could be, in principle, different for each
Fourier mode $y$ in each spectral region $X$ of the exposure $l$.
However, in the present version of KOREL, we choose\footnote{
 The weights $w_{l,X}$ are part of input data for KOREL. They
 can be chosen before running PREKOR and altered for selection of
 different regions $X$ if these are merged from different runs
 of PREKOR. The filter $w(y)$ is taken as the same function of
 the order of Fourier harmonic for all $l,X$ which means that its
 scale in wavelengths of the original spectra is dependent on
 sampling in $X$.}
\begin{equation}\label{w}
 w_{l,X}(y)=w_{l,X}w(y)\; .
\end{equation}
The weight $w_{l,X}$ of each exposure can be chosen, e.g., in
dependence on the number of photon counts. The spectral filter
$w(y)$ enables to cut out the lowest Fourier modes, as mentioned
at the note $^{\ref{filtr}}$ on page~\pageref{filtr}.

 Following (\ref{FsDopdel}), $S$ given by (\ref{FKor3})
is bilinear also in coefficients $s_{jl}$. Hence, varying with
respect to $s_{ml}$, we get from Eq.~(\ref{Sp}) for each chosen
exposure $l$ the following set of linear equations
\begin{eqnarray}\label{sjl}
 \sum_{j=1}^{n}\Re \left[\sum_{X}\int w_{l,X}(y)
  \tilde{I}_{X,j}(y) \tilde{I}_{X,m}^{\ast}(y)
  \exp(iy(v_{j}(t_{l},p)-v_{m}(t_{l},p)))dy\right]s_{jl}
 &=&\nonumber\\
 =\;\Re\sum_{X}\int w_{l,X}(y)\tilde{I}_{X}(y,t_{l})
  \tilde{I}_{X,m}^{\ast}(y) \exp(-iyv_{m}(t_{l},p))dy
\vspace*{-50mm} &&
\end{eqnarray}
for these coefficients corresponding to different components.
It is obvious from Eqs. (\ref{FKor2}) and (\ref{FsDopdel})
that for each component its spectrum $\tilde{I}_{j}(y)$ and
strengths $s_{jl}$ are defined by the observations up to
a reciprocal multiplicator. This must be fixed by a normalization
condition.

 If strengths of some components are fixed, their terms must be
transferred from the left- to the right-hand side of this
equation.
Because the coefficients $s_{jl}$ are generally still quite
numerous (but less in number than the Fourier modes of the component
spectra), it is advantageous to solve for them directly from
equations (\ref{sjl}) before optimizing $S$ with respect
to either $v_{j}(t_{l})$ or $p$, in which it is non-linear.

 It is important to hold in mind that the solutions of orbital
elements (or individual independent radial velocities),
the decomposition of the spectrum and the solution of component
strengths are inter-related and their self-consistent solution
should be found. To find this solution, an iterative procedure
is used if all these kinds of unknowns are allowed to converge.
However, there is no guaranty that this scheme will converge from
every arbitrarily chosen initial condition. Instead, it can
achieve some false local minimum by suppressing lines in exposures
for which the true radial velocites differ from the instantaneous
approximation or orbital parameters. It is thus often more efficient
to approximate the solution of spectra and orbital parameters with
fixed strengths (either found in some other spectral region,
where lines of given components are better pronounced, or simply
let them equal to one) first and to allow them to converge for
a final tuning of the solution only.

 The option of line strength solution enables a decomposition of
the telluric lines (or, in principle, also some interstellar
lines) from the observed spectra.
Exactly speaking, the telluric component of the spectrum is not
additive, but multiplicative, because the observed spectrum
\begin{equation}\label{tellur1}
 I_{obs}(x,t)=\exp(-\tau(x,t))I_{0}(x,t)
\end{equation}
is proportional to the true composed spectrum $I_{0}$ of the
studied stellar system as seen outside the Earth's atmosphere.
However, this formula can be approximated as
\begin{equation}\label{tellur2}
 I_{obs}(x,t)=I_{0}(x,t)+I_{tell}(x,t)\; ,
\end{equation}
where telluric spectrum
\begin{equation}\label{tellur3}
 I_{tell}(x,t)=({\rm e}^{-\tau(x,t)}-1)I_{0}(x,t)
 \simeq-\tau(x,t)I_{0}(x,t)
\end{equation}
is a negative contribution in lines with no continuum.\footnote{
 The small telluric absorption in continuum is eliminated by
 the rectification of the observed spectra.}
The optical depth $\tau$ and hence also strength of telluric
lines is very sensitive to the air-mass and humidity at each
exposure. Because usually we are not interested in the true
telluric spectrum $\tau(x,t)$ but only in how to eliminate its
influence, we can decompose its lines traveling with the annual
motion of the Earth as they are imprinted on the mean $I_{0}(x)$.
A small discrepancy may arise only with telluric lines falling
on steep slopes of line profiles in $I_{0}(x)$, the strength
of which in ratio to the strengths of lines falling to continuum
of $I_{0}(x)$ can differ in each exposure depending on the
instantaneous radial velocities.

\subsection{Line photometry}\label{LF}
Taking into account the connection between formation of
continuum and spectral lines in stellar atmospheres, the
above described method for calculation of line-strength
variations yields a possibility to find differential magnitude
changes between the components and also to determine the ratio
of component continua in the case that the intensity variations
are caused by some overall darkening of a component e.g. by an
eclipse. Let in the `normal' state of a binary (i.e. out of
eclipse) the intensities $I_{1,2}$ of components continua be
normalized
\begin{equation}\label{I}
 I_{1}+I_{2}=1\; .
\end{equation}

\setlength{\unitlength}{1mm}
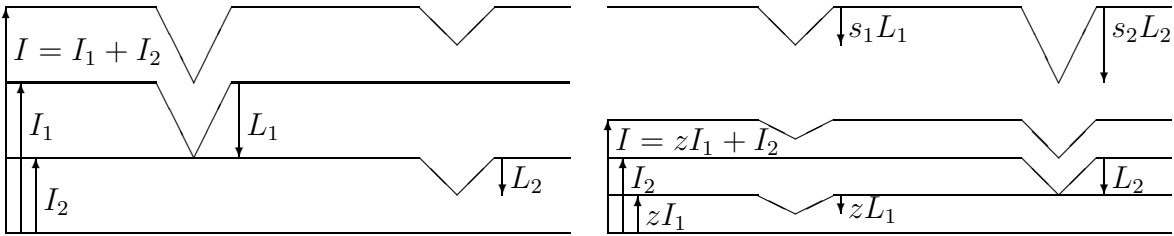
\begin{figure}[hbt]
\unitlength=1mm
\special{em:linewidth 0.4pt}
\linethickness{0.4pt}
\begin{picture}(160.00,30.00)
\put(0.00,0.00){\line(1,0){75.00}}
\put(0.00,30.00){\line(1,0){20.00}}
\put(20.00,30.00){\line(1,-2){5.00}}
\put(25.00,20.00){\line(1,2){5.00}}
\put(30.00,30.00){\line(1,0){25.00}}
\put(55.00,30.00){\line(1,-1){5.00}}
\put(60.00,25.00){\line(1,1){5.00}}
\put(65.00,30.00){\line(1,0){10.00}}
\put(0.00,0.00){\vector(0,1){30.00}}
\put(1.00,23.00){\makebox(0,0)[lb]{$I=I_{1}+I_{2}$}}
\put(0.00,20.00){\line(1,0){20.00}}
\put(20.00,20.00){\line(1,-2){5.00}}
\put(25.00,10.00){\line(1,2){5.00}}
\put(30.00,20.00){\line(1,0){45.00}}
\put(2.00,0.00){\vector(0,1){20.00}}
\put(3.00,13.00){\makebox(0,0)[lb]{$I_{1}$}}
\put(31.00,20.00){\vector(0,-1){10.00}}
\put(32.00,13.00){\makebox(0,0)[lb]{$L_{1}$}}
\put(0.00,10.00){\line(1,0){55.00}}
\put(55.00,10.00){\line(1,-1){5.00}}
\put(60.00,5.00){\line(1,1){5.00}}
\put(65.00,10.00){\line(1,0){10.00}}
\put(4.00,0.00){\vector(0,1){10.00}}
\put(5.00,3.00){\makebox(0,0)[lb]{$I_{2}$}}
\put(66.00,10.00){\vector(0,-1){5.00}}
\put(67.00,6.00){\makebox(0,0)[lb]{$L_{2}$}}
\put(80.00,0.00){\line(1,0){75.00}}
\put(80.00,30.00){\line(1,0){20.00}}
\put(100.00,30.00){\line(1,-1){5.00}}
\put(105.00,25.00){\line(1,1){5.00}}
\put(110.00,30.00){\line(1,0){25.00}}
\put(135.00,30.00){\line(1,-2){5.00}}
\put(140.00,20.00){\line(1,2){5.00}}
\put(145.00,30.00){\line(1,0){10.00}}
\put(111.00,30.00){\vector(0,-1){5.00}}
\put(112.00,26.00){\makebox(0,0)[lb]{$s_{1}L_{1}$}}
\put(146.00,30.00){\vector(0,-1){10.00}}
\put(147.00,26.00){\makebox(0,0)[lb]{$s_{2}L_{2}$}}
\put(80.00,15.00){\line(1,0){20.00}}
\put(100.00,15.00){\line(2,-1){5.00}}
\put(105.00,12.50){\line(2,1){5.00}}
\put(110.00,15.00){\line(1,0){25.00}}
\put(135.00,15.00){\line(1,-1){5.00}}
\put(140.00,10.00){\line(1,1){5.00}}
\put(145.00,15.00){\line(1,0){10.00}}
\put(80.00,0.00){\vector(0,1){15.00}}
\put(81.00,11.00){\makebox(0,0)[lb]{$I=zI_{1}+I_{2}$}}
\put(80.00,5.00){\line(1,0){20.00}}
\put(100.00,5.00){\line(2,-1){5.00}}
\put(105.00,2.50){\line(2,1){5.00}}
\put(110.00,5.00){\line(1,0){45.00}}
\put(84.00,0.00){\vector(0,1){5.00}}
\put(85.00,1.00){\makebox(0,0)[lb]{$zI_{1}$}}
\put(111.00,5.00){\vector(0,-1){2.50}}
\put(112.00,2.00){\makebox(0,0)[lb]{$zL_{1}$}}
\put(80.00,10.00){\line(1,0){55.00}}
\put(135.00,10.00){\line(1,-1){5.00}}
\put(140.00,5.00){\line(1,1){5.00}}
\put(145.00,10.00){\line(1,0){10.00}}
\put(82.00,0.00){\vector(0,1){10.00}}
\put(83.00,6.00){\makebox(0,0)[lb]{$I_{2}$}}
\put(146.00,10.00){\vector(0,-1){5.00}}
\put(147.00,6.00){\makebox(0,0)[lb]{$L_{2}$}}
\end{picture}\\
\caption{ 
 Continuum and line strengths of uneclipsed components (left)
 and in the primary eclipse (right).\label{obr2}}
\end{figure}

\noindent{}The line depths $L_{1,2}$ of the components found by
solution of Eq.~(\ref{decomp}) are expressed in units of this
common continuum with respect to which the input spectra were
rectified. If in another exposure the spectrum of component `1'
is decreased by factor $z\times$ (see Fig.~\ref{obr2}),
then the decomposed line depths of both components referred to
the instantaneous common continuum will be changed by factors
$s_{1,2}$ to values
\begin{eqnarray}\label{s1}
 s_{1}L_{1}&=&\frac{z_{lin}L_{1}}{z_{cont}I_{1}+I_{2}} \\
 s_{2}L_{2}&=&\frac{L_{2}}{z_{cont}I_{1}+I_{2}} \; .\label{s2}
\end{eqnarray}
The factor $z_{lin}$ can thus be simply expressed as
\begin{equation}\label{z}
 z_{lin}=\frac{s_{1}}{s_{2}}\; ,
\end{equation}
and assuming that the factor $z$ in continuum is the same,
$z_{cont}\equiv z_{lin}$, the ratio of continua intensities
can be also found as
\begin{equation}\label{I12}
 \frac{I_{1}}{I_{2}}=\frac{1-s_{2}}{s_{1}-1}\; .
\end{equation}
Obviously, if $z<1$, then $s_{1}<1$ and $s_{2}>1$. This behavior
can help to distinguish the variations caused by the
`geometrical' reasons (or their equivalent) from intrinsic
variations of line intensities of a component or from the
observational noise.

 In the usual case of $N$ exposures, the factors $z_{l}|_{l=1}^{N}$
of the darkenings of the component `1' can be calculated independently
for each exposure from $s_{jl}$ according to Eq.~(\ref{z}).
The ratio of continua intensities can be then obtained by least
square fit of Eqs.~(\ref{s1}) and (\ref{s2}) e.g. in logarithmic
(i.e. magnitude) scale, i.e. by solving the condition
\begin{equation}\label{I1}
 0=\delta\sum_{l}\left[\ln s_{2l}+\ln(1+I_{1}(z_{l}-1))\right]^{2}
\end{equation}
for the variation $\delta I_{1}$.

 The applicability of this simple estimate of ratio of continua
from line photometry is limited by the above mentioned assumption
that the change of intensity is the same for the line and the
continuum and certainly also by assumption that only two
components are present as well as that there are no intrinsic
changes of line profile shapes. Practical experience indicated
that this is not exactly true even for eclipsing binaries, not to
speak about stars where intrinsic line-profile variability
can be expected due to ellipticity, radial or non-radial
pulsations, spots or reflection. The reason can be simply
understood in terms of limb-darkening variations within the
line-profile. If the limb darkening is different in line and
in continuum, the portions $z$ of eclipsed light in numerator
and denominator in Eq.~(\ref{s1}) are different and cannot be
solved together with $I_{1,2}$ from $s_{1,2}$, unless their
relation is known from theory. On the other hand, fitting of
$s_{1,2}$ by a more detailed model of eclipse light-curves in
each $x$ can reveal the limb-darkening variations within the
line and thus yield information about the structure
of atmosphere of the eclipsed component.

 As shown in a preliminary study of this problem (Hadrava and
Kub\'{a}t, 2003), the variations of line profile across the stellar
disk are generally very complex, so that even the often used
expression of rotational broadening as a convolution with some
rotational profile is in fact wrong. However, some relatively
good approximations can be developed from the theory of stellar
atmospheres, which will simplify the task to a solvable and
quite powerful method.

 If source-function in a plane-parallel atmosphere can be
expanded into a Taylor-series in monochromatic optical depth
\begin{equation}\label{S}
 S(x,\tau)=\sum_{k}\frac{1}{k!}S_{k}(x)\tau^{k}\; ,
\end{equation}
the surface intensity is polynomial in directional cosine $\mu$
with coefficients $S_{k}$,
\begin{equation}\label{Inten}
 I(x,\mu)|_{\tau=0}=\sum_{k}S_{k}(x)\mu^{k}\; .
\end{equation}
In the Milne-Eddington approximation these sums reduce to the
first two terms, so that the distribution of intensity over the
visible stellar disk
\begin{equation}\label{Inten1}
 I(x,\mu)|_{\tau=0}=S_{0}(x)+S_{1}\mu
    = I(x,1)|_{\tau=0}(1 - u + u \mu)
\end{equation}
corresponds to linear limb darkening
\begin{equation}\label{limbd1}
 u=\frac{S_{1}}{S_{0}+S_{1}}\; ,
\end{equation}
which is according the theory $u=\frac{3}{5}$ for the light
integrated in frequencies and according to solutions of observed
light-curves around $u\simeq 0.3$ for the visible light in
wide frequency bands, i.e. in continuum.

\setlength{\unitlength}{1mm}
\begin{figure}[hbt]
\noindent\begin{picture}(160,65)
 \put(5,0){\epsfxsize=80mm
  \epsfbox{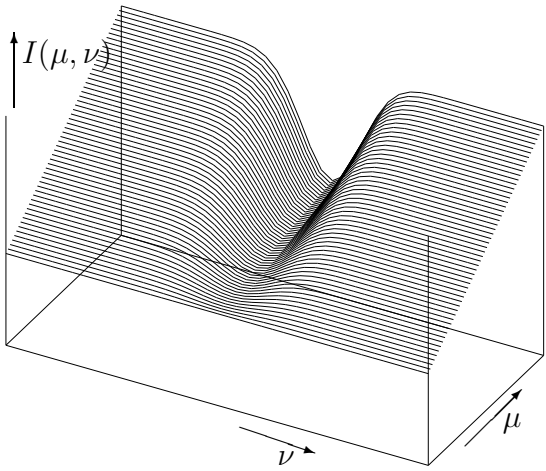}}
 \put(11.,56.){$I(\mu,\nu)$}
 \put(10.00,50.00){\vector(0,1){10.00}}
 \put(45.,2.5){$\nu$}
 \put(40.00,7.50){\vector(3,-1){10.00}}
 \put(75.,7.5){$\mu$}
 \put(70.00,5.00){\vector(1,1){7.50}}
\put(90,65){\parbox[t]{65mm}{\caption{
 Schematic dependence of limb darkening across line-profile of
 an absorption line. Intensity is low on edge of stellar disk
 ($\mu=0$) both in continuum and the line, but it is more
 brightened toward the disk centre ($\mu=1$) in continuum than
 in the line, because the radially escaping photons in continuum
 originate in deeper and hotter layers of the atmosphere.
 The central parts thus contribute to the formation of spectral
 lines with increasing weight.}}}
\end{picture}
\end{figure}

 Suppose that in all geometric depths (radii of the star $r$)
the monochromatic opacity across a line-profile is proportional
to the opacity in continuum with the same line-profile function
$\phi$ dependent on the frequency $x$ only.\footnote{ This is not
 true generally. However, we can take the assumption of separability
 in variables $r$ and $x$ as the first approximation at least in
 a small region where the core of the line is formed.}
Then also the monochromatic optical depth $\tau$ is proportional
to the optical depth in continuum,
\begin{equation}\label{taux}
 \tau(x,r)=[1+\phi(x)]\tau_{cont}(r)\; .
\end{equation}
If conditions of LTE are satisfied for studied lines, $S$ is
a smooth (Planckian) function of $x$, so that its change within
the line profile can be neglected and its slope in monochromatic
optical depth varies only due to the opacity profile,
$S_{1}(x)[1+\phi(x)]\simeq S_{1,cont}$.
The linear limb-darkening is thus decreasing toward the center
of line
\begin{equation}\label{Inten2}
 I(x,\mu)|_{\tau=0}=S_{0,cont}+S_{1,cont}[1+\phi(x)]^{-1}\mu\; ,
\end{equation}
and the line-contribution to the spectrum
\begin{equation}\label{Intlin}
 \left[I(x,\mu)-I_{cont}(\mu)\right]_{\tau=0}
  =-S_{1,cont}\frac{\phi(x)}{1+\phi(x)}\mu\; ,
\end{equation}
which is negative, has distribution over the stellar disk
corresponding to limb-darkening
\begin{equation}\label{limbd2}
 u_{lin}=1\; .
\end{equation}
It means that at initial phases of an eclipse when only a part
of disk edge is hidden, the light missing in line represents
larger portion of the overall flux in that frequency than the
light missing in the continuum ($z_{lin}<z_{cont}$), so that
in some cases line-strengths of both components can be enhanced.
The $z$-factors from Eqs.~(\ref{s1}) and (\ref{s2}) for an
eclipse must be thus modelled simultaneously for the continuum
and lines with their corresponding limb-darkenings. The geometric
parameters of the eclipse can be converged to fit the observed
line-strength variations like in the standard procedure of
light-curve solution. However, despite such a procedure is
a significant improvement compared to the standard methods,
it is still limited by several assumptions, which are never
exactly satisfied.

\begin{figure}[ht]
\setlength{\unitlength}{1mm}
\hfil\begin{picture}(160,60)
\put(10,0){\epsfxsize=60mm \epsfbox{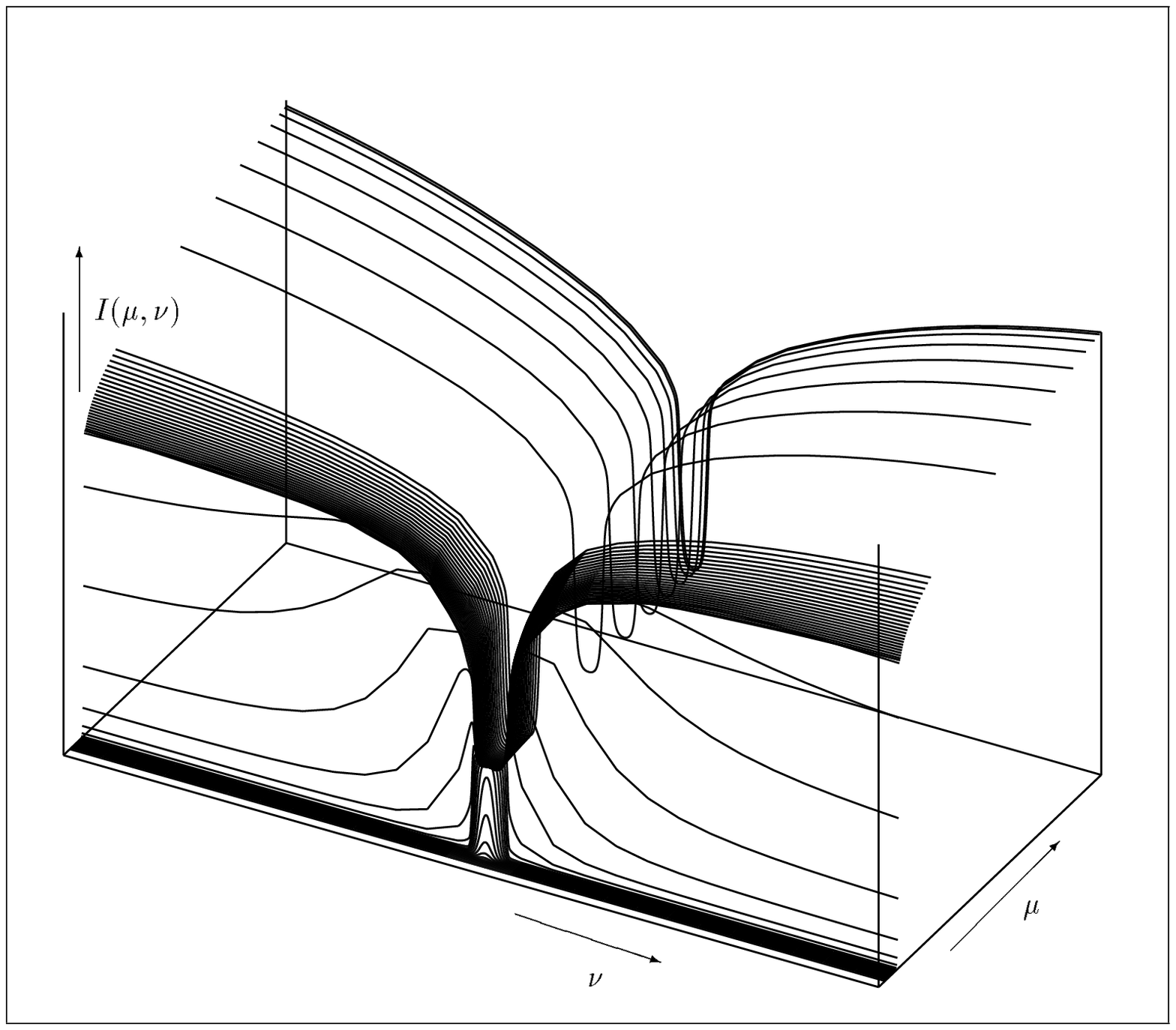}}
\put(90,0){\epsfxsize=60mm \epsfbox{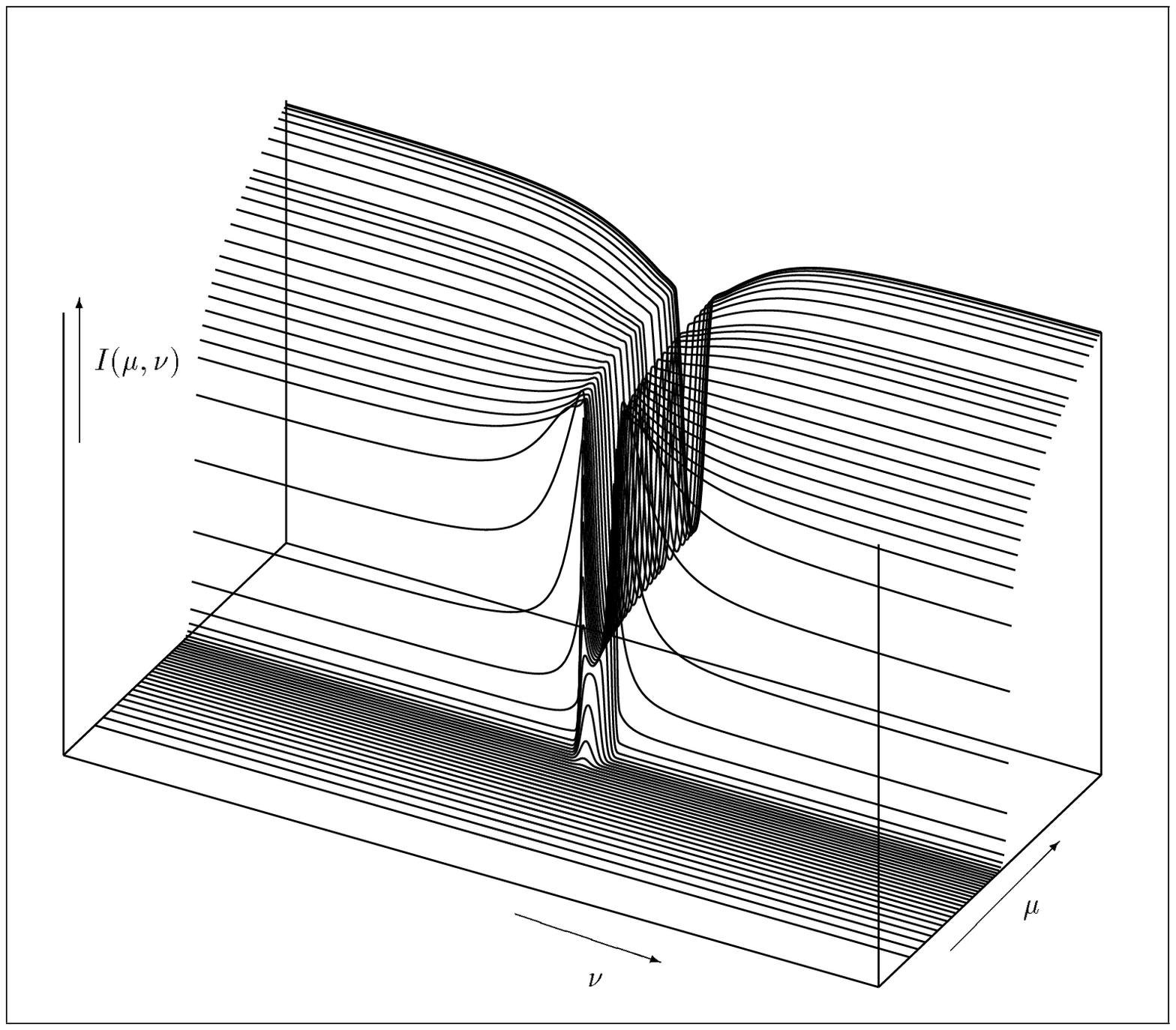}}
\end{picture}\hfil
\caption{ Limb darkening across the H$\alpha$
 line-profile of hydrogen-helium non-LTE spherical model atmospheres with
 (left:) $T_{\rm eff}=15000\,{\rm K}$, $\log g = 4.5$
 ($L=1.7\cdot 10^2 L_\odot$, $M=4.3 M_\odot$, $R=1.9
 R_\odot$) and
 (right:)  $T_{\rm eff}=15000\,{\rm K}$, $\log g = 2.0$
 ($L=5\cdot 10^5 L_\odot$, $M=4 M_\odot$, $R=33.1
 R_\odot$).
\newline
}
\label{obr3a}
\end{figure}

\subsection{Disentangling with intrinsic line-profile
 variations}\label{LPV}
In view of the fact that components of binaries are often
subjected to different kinds of variations and asymmetries
(already mentioned ellipticity, reflection, spots, radial as
well as nonradial pulsations), which are manifested as
line-profile variations, the method of disentangling needs a
further generalization. If some of these effects becomes
non-negligible, the basic equation (\ref{Kor1}) must replaced
by a more detailed expression
\begin{equation}\label{comp2}
 I(x,t)= \sum_{j=1}^{n}\int_{s}\mu I_{j}(x,s,\mu,t)\ast
 \delta(x-v_{j}(s,t)) d^{2}s\; ,
\end{equation}
for the spectral flux as an integral of the local monochromatic intensities
over the visible parts of surfaces $s$ of individual components
$j$, each one Doppler shifted according to the local radial
velocity, which can reflect now not only the orbital motion, but
also rotation and pulsation of the stellar atmosphere. It should
be noted that this expression is already simplified substantially,
because it neglects the changes of radial velocity along the line
of sight. This model of the atmosphere moving as a whole is commonly
used in studies of non-radial stellar pulsations (cf., e.g., Townsend
1997), however, observations of different Doppler shifts of lines
with different excitation potentials in Cepheids (e.g. Butler 1993)
reveal, that it is not exactly the truth. A similar problem may arise
in the case of stellar winds or disk-like envelopes, etc.

 The unknown functions $I_{j}$ depend now on a larger number of
variables than the observed spectra on the left-hand side,
equally as the velocities $v_{j}$. Consequently these functions
cannot be reconstructed from the observed spectra without
some additional conditions like it was in the previous case of
surface homogeneity. Such a condition can be based either on
some geometric or physical assumption. For a suitable choice of
some sets of functions for $I_{j}$ and $v_{j}$, their free
parameters can be adjusted to fit the observed spectra. However,
it must not be forgotten that the solution is model dependent and
that in principle it cannot be excluded that some other model may
fit the data equally well or even better.

 Quite generally (c.f. Hadrava 1997) the local intensity can be
expressed as a linear combination of a few spectral functions $I_{j}^{k}$
\begin{equation}\label{dark}
 I_{j}(x,s,\mu,t)= \sum_{k} f_{j}^{k}(s,\mu,t)I_{j}^{k}(x)
\end{equation}
with coefficients $f_{j}^{k}$ whose dependence on the position
on the star-surface $s$, directional cosine $\mu$ and time $t$
can be modelled up to a few free parameters. Substituting this
into Eq.~(\ref{comp2}) we arrive at equation
\begin{equation}\label{dark2}
 I(x,t)= \sum_{j,k}I_{j}^{k}(x)\ast\Delta_{j}^{k}(x,t,p)\; ,
\end{equation}
which is formally identical with (\ref{Kor2}), apart of the fact
that each component $j$ can be now characterized by several
spectra $I_{j}^{k}(x)$ (e.g. corresponding to different terms in
the expression (\ref{Inten}) for the limb darkening, or to
spectra inside and outside a spot etc.) with different spectral
broadenings
\begin{equation}\label{broad}
 \Delta_{j}^{k}(x,t,p)= \int_{s}\mu f_{j}^{k}(s,\mu,t)
   \delta(x-v_{j}(s,t)) d^{2}s\; .
\end{equation}
Differences in these broadenings can ensure that the corresponding
spectral functions can be decomposed from the observations using
the general procedure described in Section \ref{PFD}.

\subsection{Broadening by pulsations}\label{puls}
One of the simplest generalizations of disentangling for a case
of intrinsic line-profile variations, which was outlined
in the Section~\ref{LPV}, is the problem of pulsating stars.
Suppose first, that the atmosphere of a spherical star moves as
a whole radially (with respect to its center) with instantaneous
velocity $v_{p}(t)$. In agreement with conclusion~(\ref{limbd2})
let us simplify the sum (\ref{dark}) to a single term linear
in the directional cosine $\mu=\cos\vartheta$, which is a function
of the projected distance $r=R\sin\vartheta$ from the centre of
stellar disk
\begin{equation}\label{darkpul}
 I_{j}(x,s,\mu,t)= s_{j}\mu \;I_{j}(x)\; .
\end{equation}
The total velocity of a projected surface element $\mu d^{2}s=
rdrd\varphi$ is the velocity $v_{j}(t)$ of the star as before,
plus projection $\mu v_{p}$ of the pulsational motion into the
line of sight. The broadening function given by (\ref{broad})
thus reads\footnote{ Here we use the relation $rdr=-R^{2}\mu
 d\mu$. The bracket [\hphantom{x}] in (\ref{broadpul}) means
 that the quadratic function inside is multiplied by the
 characteristic function of the interval in subscript (or
 interval $(v_{j}+v_{p},v_{j})$ if $v_{p}<0$), i.e. by 1 inside
 and 0 outside the interval. Note that if the intensity in the lines
 is taken to be homogeneous instead of (\ref{limbd2}), the pulsational
 broadening is linear instead of quadratic and it thus leads to
 a smaller amplitude of observed pulsational Doppler shifts.}
\begin{equation}\label{broadpul}
 \Delta_{j}(x,t,p)= 2\pi s_{j}(t)\int \mu
 \delta(x-v_{j}(t)- \mu v_{p}(t)) r dr
 =\frac{ 2\pi R^{2} s_{j}(t)}{v_{p}^{3}(t)}
  \left[(x-v_{j}(t))^{2}\right]_{x\in(v_{j},v_{j}+v_{p})}\; .
\end{equation}
Its Fourier transform reads
\begin{equation}\label{fbroadpul}
 \tilde{\Delta}_{j}(x,t,p)=\frac{2\pi i R^{2}}{v_{p}^{3}(t)}
  s_{j}(t)\exp(iyv_{j}(t,p))y^{-3}
  \left[\exp(iyv_{p}(t))(2-2iyv_{p}-y^{2}v_{p}^{2})-2\right]\; ,
\end{equation}
it means, that in comparison with Eqs. (\ref{FDopdel}) and
(\ref{FsDopdel}) it contains now also additional broadening terms
corresponding to the line-profile variations. A more detailed description
of the disentangling with LPVs due to pulsations is given in recent
publication (Hadrava, \v{S}lechta \& \v{S}koda 2009).

 Similar generalization can be done also for rotational broadening
which causes LPVs in elliptic variables and can thus be found in
the data. Non-zero limb-darkening leads in this case to asymmetry
of lines and it complicates substantially the expression for
$\Delta$ and its Fourier transform, equally as in the case of
non-radial pulsations. However, $\Delta$ can be modelled and
$\tilde{\Delta}$ calculated numerically in these cases.


\subsection{Rotational (Schlesinger -- Rossiter -- McLaughlin)
 effect}\label{rot}
Another important case of line-profile (and consequent radial-velocity)
variations is the rotational effect (Schlesinger 1909, Rossiter 1924,
McLaughlin 1924) which may occur during eclipses.

 Let us assume that the observed spectral flux is given by
Eq.~(\ref{dark2}) with the broadening functions (\ref{broad}).
Note, that it is customary (cf., e.g., Gray 2005, p.~436, 464 etc.
as a classical reference)
to simplify Eq.~(\ref{dark2}) to a single
term ($k=1$) for each component $j$, i.e. to assume the local intensity
$I_{j}(x)$ to be the same function of $x$ at each $s$ and $\mu$.
This may be an acceptable approximation for some lines, however, it
is incorrect to fix the limb darkening in lines to the values
obtained for continua either from light-curve solution or from model
atmospheres (cf. Hadrava 1997). Gray (2005, p. 436) argues that
the limb-darkening coefficient ``varies slowly with the wavelength
and can be taken to be constant over the span of a spectral line".
However, this argument based on the behaviour of the limb darkening
in continuum is incorrect, similarly as the slow (Planckian) variation of
specific intensity in the continuum does not exclude its fast variation
within the spectral line.
It follows from calculations of $\mu$-dependent synthetic spectra, that
the limb-darkening varies across the line-profile and hence at
least two terms (namely those proportional to $\mu^{0}$ and
$\mu^{1}$) can fit strong lines like H$\alpha$ much better
(cf. Hadrava and Kub\'{a}t 2003).

 On the other hand, we neglect here the coefficient
$(1+v/c)^{3}$, which results from relation between the specific
intensity and the Lorentz-invariant distribution function of
photons (cf. Ohta et al. 2005), which, however is well negligible
for non-relativistic velocities $v$.

\begin{figure}
\setlength{\unitlength}{1mm}
\hfil\begin{picture}(160,40)
\put(10,0){\epsfxsize=60mm \epsfbox{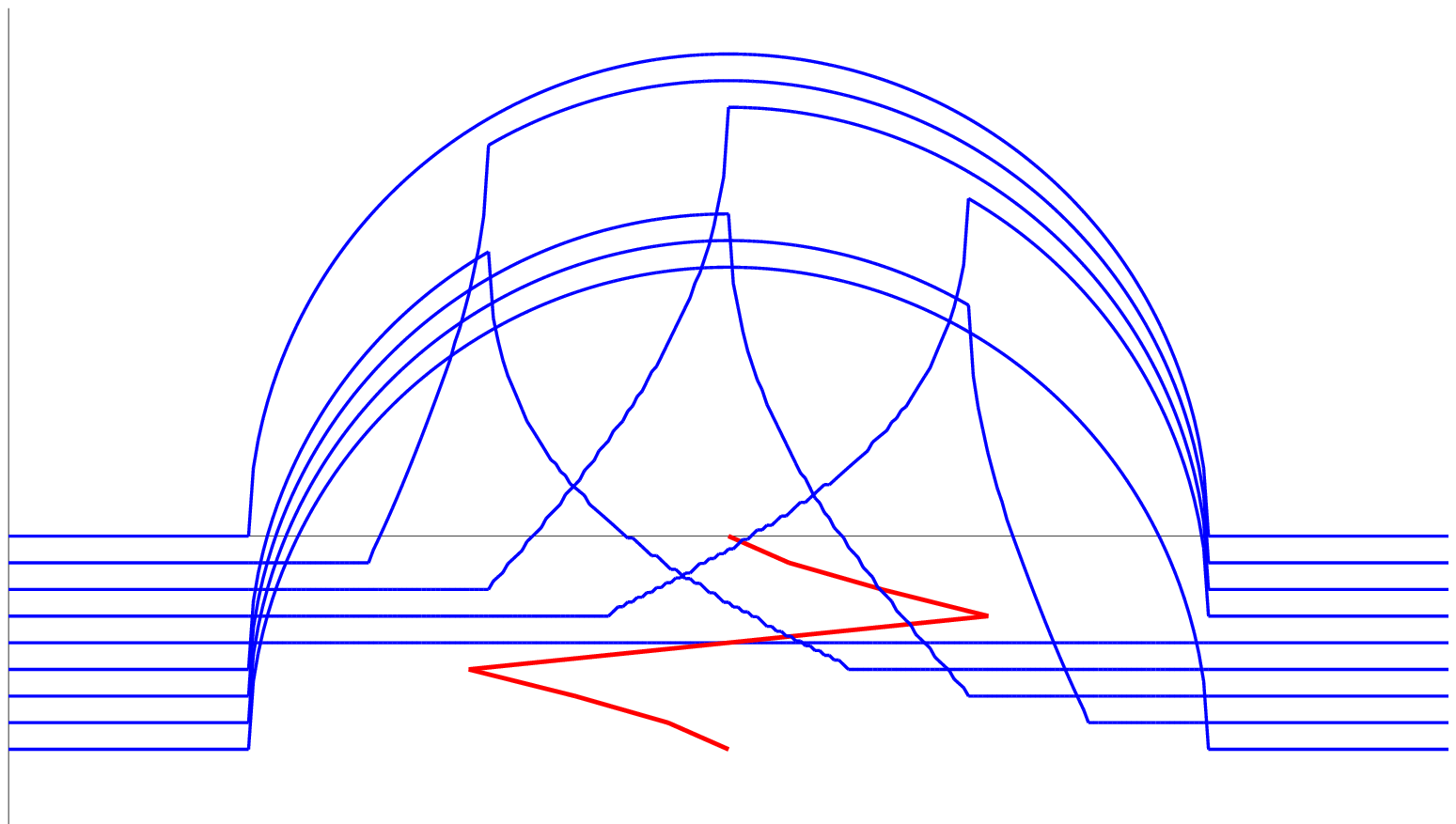}}
\put(90,0){\epsfxsize=60mm \epsfbox{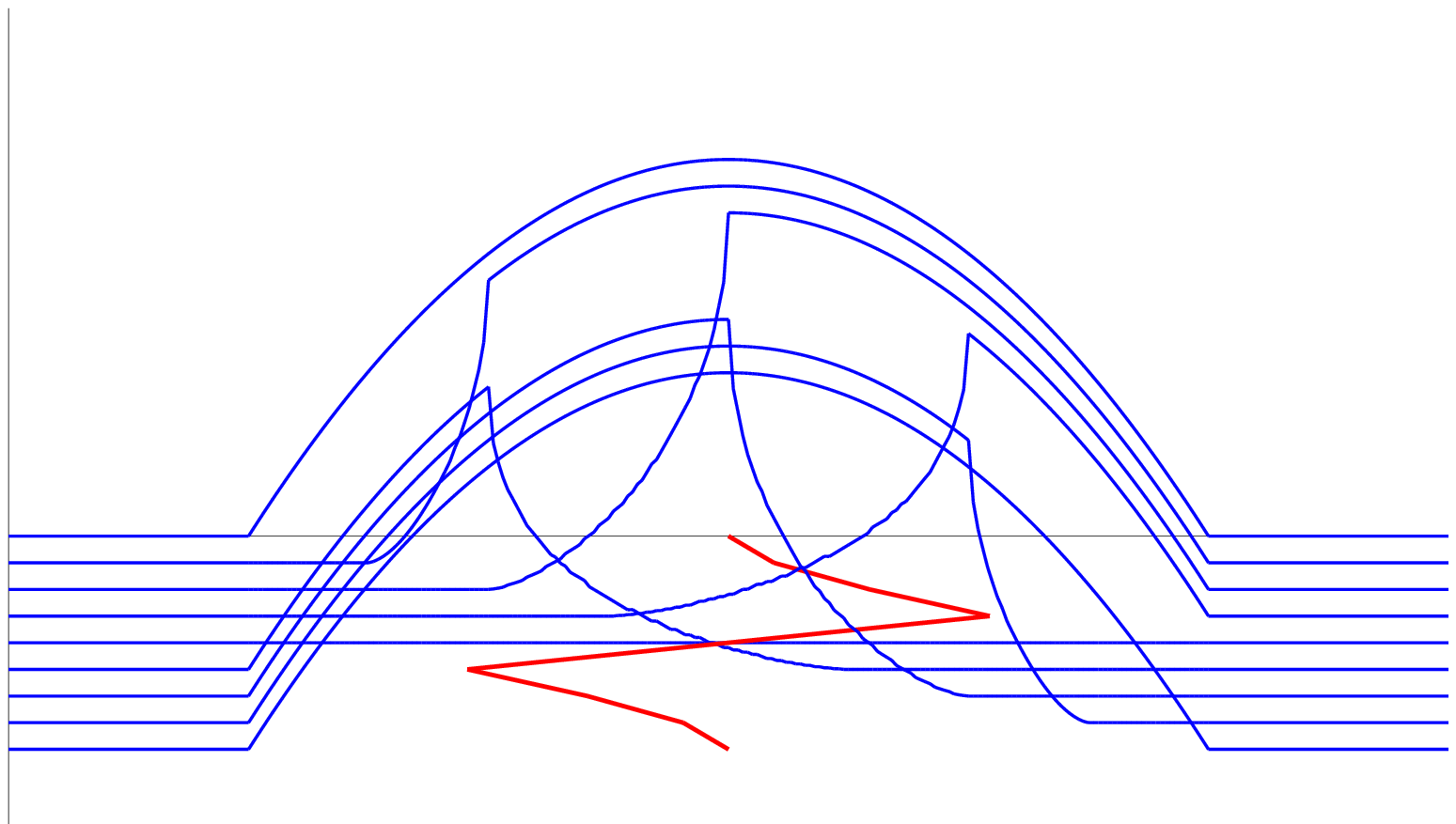}}
\end{picture}\hfil
\caption{Broadening profiles $\Delta_{j}^{0}$ (left) and
 $\Delta_{j}^{1}$ (right) for a series of phases of total
 eclipse ($r_{i}=r_{j}$) with rotational axis perpendicular
 to the path of the eclipsing (i.e. the front) star. The
 broken line joining the profile centers indicates the observable
 RV- perturbation due to the rotational effect. }\label{Rossit}
\end{figure}

 We shall assume now, that the stars rotate with angular
velocities $\omega_{j}$ as rigid bodies, their rotations do
not oblate their spherical shapes and the gravity darkening is
negligible, so that
\begin{equation}\label{darksf}
 I_{j}(x,s,\mu,t)= \sum_{k} \mu^{k}I_{j}^{k}(x)\; ,
\end{equation}
\begin{equation}\label{vj}
 v_{j}(s,t)= \omega_{j}r_{j} v\; ,
\end{equation}
and
\begin{equation}\label{mu}
 \mu= (1-u^{2}-v^{2})^{1/2}\; ,
\end{equation}
where $u$ and $v$ are Cartesian coordinates on the projected visible
stellar disc ($\mu d^{2}s_{j}=r_{j}^{2}dudv$). The coordinate $u$
is chosen in the direction of projected rotational axis. If at time $t$
the star $j$ is partly eclipsed by star $i$ with centre projected to
coordinates $u_{i}(t), v_{i}(t)$, then the visible part of stellar
disc $s_{j}$ is given by conditions
\begin{equation}\label{d2s}
 u^{2}+v^{2}< 1\; , \hspace*{5mm}{\rm and}\hspace*{5mm}
 (u-u_{i})^{2}+(v-v_{i})^{2}< (r_{i}/r_{j})^{2}\; .
\end{equation}
The broadening profile thus reads
\begin{equation}\label{broadsf}
 \Delta_{j}^{k}(x,t,p)=\frac{r_{j}}{\omega_{j}}\int_{u\in s_{j}}
   (1-v^{2}-u^{2})^{k/2}|_{v=x/r_{j}\omega_{j}} du\; .
\end{equation}
The first two of these broadenings have the form
\begin{eqnarray}\label{broadsf0}
 \Delta_{j}^{0}(x,t,p)&=&\frac{r_{j}}{\omega_{j}}
 \left[ u\right]_{v=x/r_{j}\omega_{j}, u\in s_{j}}\\
 \Delta_{j}^{1}(x,t,p)&=&\frac{r_{j}}{2\omega_{j}}
 \left[u(a^{2}-u^{2})^{1/2}+a^{2}\arcsin\frac{u}{a}
 \right]_{v=x/r_{j}\omega_{j}, u\in s_{j}} du\; , \label{broadsf1}
\end{eqnarray}
where $a^{2}=1-v^{2}$. Examples of these broadening profiles are
shown in Fig.~\ref{Rossit}.

 The standard treatment of the rotational effect does not deal
with the shape of line-profile distortions during the eclipse,
but with the resulting RV-shifts only. To assign an observed
wavelength and hence also the RV to a spectral line with asymmetric
profile is an ambiguous task.
Different definitions are used, which imitate different methods
of measurement -- e.g. the wavelength of the extremal intensity
(i.e. bottom of an absorption line) or centre of profile at half
depth (or another fraction of the line depth) are taken. Here we
use wavelength defined by Eq.~(\ref{mom4}) using the moments (calculated
in the logarithmic scale) of line profile $\phi(x)$.
An advantage of this definition is that the radial-velocity shift
is additive with respect to subsequent broadenings of the profile.
Let the broadened profile $\phi'=B\ast\phi$ arises from original
profile $\phi$ by broadening $B$. Then according to Eq.~(\ref{mom5})
really the centre of the profile $\phi'$ is
\begin{equation}
 x_{\phi'}=x_{B}+x_{\phi}\; ,
\end{equation}
where $x_{B}$ is centre of the broadening profile $B$ defined also
by Eq.~(\ref{mom4}). It means that spectrographs with different
instrumental profiles should give the same wavelengths (if their
possible instrumental shifts $x_{B}$ are subtracted), while the
wavelengths given by alternative definitions depend on resolution.
It is important to note, that centers of different broadening
profiles $\Delta_{j}^{k}$ are generally different. Because different
lines in spectrum of the same star are different superpositions of
all of them, then the rotational effect may have different values
for different lines.

\subsection{Disentangling with constraints}\label{constrain}
It is obvious that from the view-point of generality of a model
used for data interpretation as discussed in Section~\ref{mysl},
the disentangling is less restrictive on the component spectra
than the methods based on template spectra. On the contrary, in
its standard form it is more restrictive regarding the broadening
function compared to the Rucinski's method and, in some sense,
even compared to any RV-measurement if the assumption of Keplerian
motion is imposed on the disentangling (KOREL enables also arbitrary
RVs for chosen components). Because each degree of freedom of the
model can be useful or harmful for the solution depending on the
particular case in question, it is desirable to have a tool for
the data interpretation, in which the level of the model freedom
could be tuned in details.

  The above described generalization for line-profile variability
enlarges the freedom of broadening function in the disentangling.
It is thus needed to enable also some restrictions for component spectra.
One straightforward possibility is to put a constraint on one
or of the components in the form of template spectrum $J_{j}$
(either from another star, from model atmospheres or even from
other solutions of the same system) and to disentangle the observed
spectra with respect to a subset of $m$ component spectra only (and
the parameters $p$ of any broadening $\Delta_{j}$) modifying
Eq.~(\ref{Kor2}) to the form
\begin{equation}
 \sum_{j=1}^{m}I_{j}(x)\ast\Delta_{j}(x,t,p)=
  I(x,t)-\sum_{j=m+1}^{n}J_{j}(x)\ast\Delta_{j}(x,t,p)\; .
\end{equation}
This may be useful e.g. for disentangling of telluric spectrum,
which is basicaly known, but quite often its continuum is distorted
due to an imperfect disentangling of lower Fourier modes of wide
stellar lines.

  Another type of constraints can be imposed on the orbital
and other parameters of $\Delta$-functions. Like in light- or
RV-curve solution or any other data fitting, different parameters
$p$ can be either converged or fixed, if they are better
determined by an independt way. However, a constraint on the
parameters may also have a form of mutual dependence, which
can be described by one or more conditions $F_{k}(p)=0$ in the
space of $p$. In such a case, the parameters can be substituted
in a form $p=p(q)$ and the observations fitted with respect to
the less dimensional space of $q$, or, invoking the method of
Lagrange multiplicators, the fit can be performed by numerical
minimization of expression
\begin{equation}\label{disent4}
 0=\delta\left\{\sum_{t}\int|I-\sum_{j}I_{j}\ast\Delta_{j}|^{2}dx+
 \sum_{k} \lambda_{k} F_{k}^{2}(p)\right\}\; ,
\end{equation}
where sufficiently high values of coefficients $\lambda$ are
chosen. The later aproach is predominant, if the parameter
constraints are also of an observational nature with some
uncertainty. In such a case we can put $F^{2}\equiv (O-C)^{2}$
and Eq.~(\ref{disent4}) is then an equation for simultaneous
disentangling and solution of other type of data. A merging
of disentagling (e.g. using KOREL) with solution of light- and RV-
curves, astrometry etc. (e.g. using FOTEL) is thus a direction
to be followed in a near future (cf. Wilson 1979, Holmgren 2004,
Hadrava 2004b, 2005).

\subsection{Discretization of the input spectra}
Both the contemporary observational technique of spectroscopy as
well as the numerical data-processing on digital computers force
us in practice to treat the stellar spectra not as smooth functions
in the whole region of wavelengths but represented as a finite set
of discrete numbers sampling the spectrum with limited spectral
resolution within a limited spectral window. The detectors nowadays
in use are mostly some CCD- or other electronic detectors, which
integrate the light during the exposure in many pixels with individual
characteristics of spectral sensitivity (hopefully linear in the intensity
and very narrow-band, but never point-like in the wavelength). Depending
on the spectrograph configuration, the individual pixels of detectors
are usually arranged in the spectrograph focal plane both along the
dispersion as well as in the transverse direction. The original data from
the pixels are then (after a necessary flat-fielding etc.) interpolated
to the desirable one-dimensional representation, usually calibrated
in wavelength according to an exposure of properly chosen comparison
spectrum. Similarly, the spectral resolution of previously used
photographic spectrograms was limited by the finite size of emulsion
grains in addition to the spectrograph limitations due to the slit-width
or size and quality of the dispersion element, so that a scanning of
the spectrogram yields again qualitatively comparable representation
of the spectrum. Regarding the extension of the spectral window, it
is usually limited by the spectrograph construction in combination with
the size of detector (in addition to the limited spectral sensitivity of
the detector and the transparency of the atmosphere and the instrument).

 The discretized spectrum $D[I]$, which we use as an input for
the disentangling method or for any other spectroscopic analysis
is thus related to the original observed spectrum $I(x)$ by a relation,
which can be expressed by means of a linear operator
\begin{equation}\label{diskr1}
 D[I](x)=\sum_{k=1}^{N}\delta(x-x_{k})D_{k}[I]
 \equiv\sum_{k=1}^{N}\delta(x-x_{k})\int D_{k}(x')I(x')dx'\; ,
\end{equation}
where $\{x_{k}|_{k=1}^{N}\}$ is the set of sampling frequencies
(we assume $x_{k}<x_{k+1}$) and $D_{k}(x)$ is the overall spectral
characteristic of sensitivity of the bin number $k$. For a simple
approximation we can choose
\begin{equation}\label{diskr2}
 D_{k}(x)\simeq\frac{x_{k+1}-x_{k-1}}{2}\delta(x-x_{k})\; ,
\end{equation}
so that
\begin{equation}\label{diskr3}
 D[I](x)=\sum_{k=1}^{N}\frac{x_{k+1}-x_{k-1}}{2}I(x_{k})\delta(x-x_{k})\; .
\end{equation}
For an equidistant sampling ($x_{k+1}=x_{k}+\Delta_{x}$) we thus arrive
at expression used for sampling e.g. by Gray (2005, p. 35), with the
difference that the above expression includes also the limitation
due to the finite spectral region (instead of $k\in(-\infty,\infty)$)
and that it is multiplied by $\Delta_{x}$ to get $\int I(x)dx\simeq \int
D[I](x)dx$ (and to preserve the same physical dimension of $I$ and $D[I]$
also if $x$ is not dimensionless). Another possible choice of the
discretization, which better preserves the flux integrated over the
frequencies is the box-function for $D_{k}$,
\begin{equation}\label{diskr4}
 D_{k}[I]=\int_{x_{k-1}/2+x_{k}/2}^{x_{k}/2+x_{k+1}/2}I(x)dx\; .
\end{equation}
A more detailed model of discretization by a real exposure on
a one-dimensional electronic detector (e.g. of the Reticon type)
could be reproduced by a box-function narrowed for the width of
gaps between the detector pixels. On the other hand, a misalignment
of the detector with the dispersion or the instrumental profile of
the spectrograph may cause a tail of $D_{k}$ reaching even over
several neighbouring pixels. These problems are particularly difficult
in echelle spectrographs, where the tracks of individual orders are
curved and thus necessarily also skewed with respect to the rows and
lines of the matrix of detector pixels. When a one-dimensional
representation of the echelle spectrum is obtained, value in each bin
is a result of an interpolation between several detector pixels
properly weighted to account for the blaze-function and the intensity
profile in the direction of cross-dispersion of the orders (cf. e.g.
Piskunov and Valenti 2002, \v{S}koda et al. 2008 and references therein).
Errors in this weighting, e.g. due to bad columns in the detector
and a deformation between the exposure of the target star and
the flat-field may induce a wavy patterns in the spectrum (cf. De Cuyper
and Hensberge 2002, Hensberge 2004), which may then hamper the disentangling.

 Despite the basic equations of the Fourier disentangling are valid
for the spectra as functions of the logarithmic wavelength $x$ regardless
of their numerical representation, for practical reasons of computational
performance the code KOREL uses the Fast Fourier Transform, which
requires the data to be represented by values in equidistant spacing
of $x$, the number of data points $N=2^m$ (i.e. being given as an
integer power of 2).
Not to lose resolution by subsequent interpolations, it would thus
be desirable to reduce the original read-out of the detector chip
directly into the spacing required for the KOREL input.
On the other hand, in view of all the above mentioned effects decreasing
the spectral resolution, such an additional loss by interpolation is not
substantial if the spacing of the KOREL input data is comparable or
even better somewhat smaller than the resolution of the spectrograph.
If we are forced to use a sparse data spacing significantly worse than
that of the original data (e.g. in order to disentangle very wide spectral
regions), the proper choice of the interpolation method may be important.
The formula (\ref{diskr4}) preserving the integral flux is then
advantageous, because the simple sampling (\ref{diskr2}) may skip narrow
lines.

 Hensberge et al. (2008) present the use of common sampling grid of
all input data as a limitation of the Fourier method of decomposition
``at the expense" of its efficiency. This is misleading, because
a rebinning is an inevitable feature of any method of the disentangling;
once the input spectra are sampled in different grids and, moreover,
for different components in the same exposure these grids are shifted
for different radial velocities, resulting value in each bin of the
disentangled spectrum must be a weighted mean of input values at
wavelengths, which are generally interpolated between the grid points of
input exposures. Fig.~1 by Hensberge et al. (2008) illustrates well
that a common grid for all input exposures is taken in the standard
wavelength-domain disentangling as well (cf. the original work by
Simon and Sturm 1994). The numerical implementation of the method could
be generalized in this respect, but, effectively, it would be equivalent
to the resampling of the input. The problem is thus not which method
to choose in order to minimize the loss by rebining, but how to rebin
the data optimally and the answer can checked from the point of view
of both these methods, which are equivalent in this respect.
Regarding to the fact that the Fourier method solves the decomposition
of spectra in terms of Fourier modes, just the Fourier modes of the
input spectra are needed and these could be computed directly from
the original sampling of the exposures. An interpolation scheme
reproducing after FFT the same values of Fourier modes could thus be
elaborated. A problem remains that the discretization decreases the
amount of information in any case. It means, for instance, that some
wavy patterns in the input spectra with frequencies in between the
grid points could be filtered out, unless the representation will
smear-out the input in terms of its power-spectrum.

 A consequence of discretization of the spectra is the limitation
of accuracy with which the radial velocities (and hence also the
orbital elements) are determined. In the method by Simon and Sturm,
the expected Doppler shift is rounded to an integer multiple of
the radial-velocity step to determine the proper displacement of
the unit off-diagonal. In the Fourier transform, the shift of the
spectrum for value $v$ is given by multiplication with function
$\exp(iyv)$ (cf. Eq.~(\ref{FKor1})), which could be evaluated
precisely at each frequency $y$, however, due to the limited number
$N$ of the modes taken into account, its inverse Fourier transform
will generally produce a wider peak with some ghosts aside resembling
interference fringes. Only in the special case of $v$ being an integer
multiple of the grid step, $\exp(iyv)$ is in resonance with the interval
length in $y$-representation and a sharp shifted $\delta$-function
coinciding with a point of the $x$-representation can be reproduced.
This is why the radial velocity is rounded to the nearest grid point
in the Fourier disentangling also and why the radial velocities or
their residuals calculated in the original KOREL-code are quantized
depending on the radial-velocity step.

\setlength{\unitlength}{1mm}
\begin{figure}[hbt]
\noindent\begin{picture}(140,80)
 \put(5,0){\epsfxsize=80mm
  \epsfbox{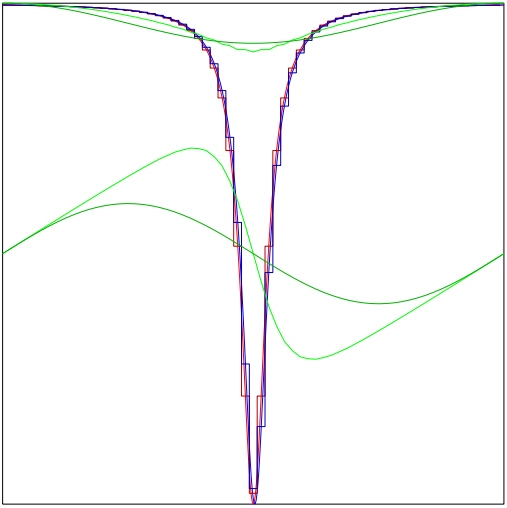}}
\put(90,75){\parbox[t]{65mm}{\caption{
 \label{obr0x} Discretization of a Lorentzian profile (the smooth thick
 line) centered with pixel position should yield a symmetric distribution
 of counts in neighbouring bins (the thick step function). A slightly
 shifted profile (for 0.2 pixel-width in this figure -- see the thin lines)
 results in an asymmetry of the counts, which in turn enables to determine
 the position of the line with precision below the pixel width.
}}}
\end{picture}
\end{figure}

 However, owing to the resolution in the digitalized values of intensity
read from individual detector pixels, the position of spectral lines wider
than the sampling step can be deduced with an accuracy exceeding the step
width (cf. Fig~\ref{obr0x}). As described in the work Hadrava (2009), it
is thus possible to reach a sub-pixel resolution in the disentangling. This
method is implemented in the version KOREL08 of the KOREL code.
Even higher precision could be reached in disentangling constrained
by templates and the method could be applied in data-processing
not only in spectroscopy.

\subsection{Normalization of disentangled spectra}\label{norm}
Let us investigate now in detail the obstacle of the
decomposition, which was already mentioned in Section~\ref{cont}
on page~\pageref{cont} and which is immediately obvious from
Eq.~(\ref{FKor1}).
If we choose $y=0$, the equation (\ref{FKor1}) received for the
integral mean values $\int I_{j}dx$ reduces to a singular system
of linear equations
\begin{equation}\label{FKor0}
 \tilde{I}(0,t)=\sum_{j=1}^{n}\tilde{I}_{j}(0) \; .
\end{equation}
According to the assumption of invariability of $I_{j}$ in the
simple disentangling, the mean intensity on the left-hand side
of this equation should not depend on the time $t$. Assuming such
a dependence to be caused only by a noise in individual
exposures, the left-hand side can be replaced by its mean value,
as it is done in Eq.~(\ref{FKor5}). However, instead of a set
of equations determining uniquely the other Fourier modes of
component spectra, we have then this single condition restricting
solutions for $n$ values $\tilde{I}_{j}(0)|_{j=1}^{n}$ only to
an infinite $(n-1)$-parametric set. In words it means that
constant parts (continua) of the spectra cannot be decomposed,
because they are invariant with respect to the Doppler shift.
The continuum is never constant in the whole range of frequencies,
so that in principle it should be possible to decompose the
spectrum completely if it were to be available in the whole range
of $x$ with an unlimited precision. From the mathematical point
of view it would require the use of additional conditions for limits,
$\lim_{x\rightarrow\pm\infty}I(x,t)=0$.
In practice the whole spectrum is never available and its
precision is insufficient to determine the Doppler shifts
of continua.\footnote{ In fact just such an overestimation of the
 effect was a shortcoming of the historical paper by Christian
 Doppler (1842).}
Usually we decompose limited parts $x\in(x_{1},x_{2})$ of the
whole spectrum which are rectified with respect to the local
continuum. For these intervals the mean intensities
\begin{equation}\label{E1}
 \tilde{I}_{j}(0)\equiv\langle I_{j}\rangle\equiv
 \int_{x_{1}}^{x_{2}}I_{j}(x)\frac{dx}{x_{2}-x_{1}}
\end{equation}
differ from the individual continua $C_{j}$. These differences
can be found by a new rectification of the decomposed spectra,
however the ratios of values $C_{j}$ must be estimated
from some additional information.\footnote{ As it has been
 mentioned already in Section~\ref{cont}, the sets of linear
 equations (\ref{FKor1}) may be nearly singular also for some
 other low harmonics and due to the limited precision of the
 source data their solution may be unstable, especially if the
 radial velocities are not yet well converged. This results in some long-period
 waves on the decomposed spectra, which are frequently met during
 the procedure of convergence of the disentangling. Usually these
 waves disappear when a better solution is reached, however, a
 better rectification of the input spectra can also help.}
The uncertainty of $C_{j}$ is also why KOREL gives the
disentangled component spectra $I_{j}$ on its output only
relatively, i.e. in the same units used at input and with an
unknown shift of the zero level (and of the continuum). Because
the input spectra for disentangling are supposed to be rectified
with respect to the total continuum (or a pseudocontinuum)
\begin{equation}\label{E4}
 C\equiv\sum_{j=1}^{n} C_{j}=1\; ,
\end{equation}
the output
\begin{equation}\label{E0}
 I'_{j}(x)=I_{j}(x)-\langle I_{j}\rangle+1
\end{equation}
is shifted for this value $C=1$ to prevent negative values
in absorption lines and to give an approximate information about
the depths of the lines of individual components.
However, to enable a comparison of the disentangled
spectra with theoretical spectra from model atmospheres it would
be desirable to calculate the component spectra rectified with
respect to their (unknown) individual continua $C_{j}$, i.e. to
find the functions $I_{j}(x)/C_{j}$.

The only value we can calculate from the input spectra is the total
mean intensity
\begin{equation}\label{E2}
 \langle I\rangle=\sum_{j=1}^{n}\langle I_{j}\rangle\; .
\end{equation}
This value (which is decreased below the level of total continuum
for the ratio of sum of equivalent widths of lines in the spectral
region and the length of the region) is given at the output of KOREL for
each spectral region.
Even if we know, e.g. from a broad-band photometry, the ratio of colour
luminosities
\begin{equation}\label{E3}
 L_{j}=\int I_{j}(x)\phi(x)dx
\end{equation}
of the components, it does not tell us directly the ratio of
$\langle I_{j}\rangle$, because $L_{j}$ are integrated over a broader
region with wavelength-dependent sensitivity $\phi$ of the photometric
channels (given by filters/detectors). Both integrals (\ref{E1}) and
(\ref{E3}) are decreased with respect to the levels of continua $C_{j}$
by the sum of equivalent widths of absorption spectral lines
contained in the corresponding region. However, for disentangling
we usually choose narrow neighborhoods of strong (mostly
absorption) lines, so that the values of $\langle I_{j}\rangle$ can
be expected to be sensibly smaller in comparison with levels of continua
$C_{j}$, while the photometric luminosities can nearly reach the values
of integrals of the continua\footnote{ The correction could be obtained
 from model atmospheres. In the next we will thus suppose that we know
 the ratio of the continua $C_{j}$ from the photometry.}
\begin{equation}\label{E6}
 \langle I_{j}\rangle < C_{j}\; ,\hspace*{20mm}
 L_{j} \leq C_{j}\; .
\end{equation}
Fortunately, the shifts of components' mean intensities with respect
to their continua can be determined from the output disentangled
spectra
\begin{equation}\label{E7}
 \Delta_{j}\equiv C_{j}-\langle I_{j}\rangle = [I'_{j}(x)-1]_{x\in cont.}
\end{equation}
(cf. Eq.~(\ref{E0})) simply by fitting the level of continuum in
$I'_{j}$. These $n$ values should satisfy the bounding condition
which follows from Eqs.~(\ref{E4}) and (\ref{E2}),
\begin{equation}\label{E8}
 Q\equiv 1-\langle I\rangle -\sum_{j=1}^{n} \Delta_{j}=0\; ,
\end{equation}
the right-hand side of which is given at the output of KOREL
as the integral of the input spectra.
Neglecting this condition, the continuum shift of each disentangled
component spectrum could be calculated independently by a new
rectification of the output as its mean value in the continuum.
Such a result could be then substituted into Eq.~(\ref{E8}) to
check it precision. An alternative approach is to solve for all
$\Delta_{j}$ simultaneously by minimizing the sum
\begin{equation}\label{E10}
 S\equiv \sum_{j}w_{j} \int[I'_{j}(x)-1-\Delta_{j}]^{2}
\end{equation}
with the condition (\ref{E8}). Here the weight $w_{j}$ of each
component spectrum can be chosen inversely proportionally to the
square of its noise. The minimization can be done using the standard
Lagrange multipliers method, i.e. solving the set of linear equations
\begin{equation}\label{E11}
 0=\frac{\partial}{\partial \Delta_{j}}[S-\lambda Q]=
  2w_{j} \int[\Delta_{j}+1-I'_{j}(x)]-\lambda\; .
\end{equation}
The solution reads
\begin{equation}\label{E12}
 \Delta_{j}=\left(\int[I'_{j}(x)-1]+\frac{\lambda}{2w_{j}}\right)
            /\int 1\; ,
\end{equation}
where
\begin{equation}\label{E13}
 \lambda\sum_{j}\frac{1}{2w_{j}}=
  [1-\langle I\rangle]\int 1 - \sum_{j}\int[I'_{j}(x)-1]\; .
\end{equation}

Once the shifts $\Delta_{j}$ are determined (and continua $C_{j}$
chosen), the rectified component spectra can be calculated according to
\begin{equation}\label{E9}
 I_{j}(x)/C_{j}=1+(I'_{j}(x)-1-\Delta_{j})/C_{j}\; .
\end{equation}
Note, that a telluric component contributes also by its $\Delta_{j}$,
despite it has $C_{j}\equiv 0$ (but its $\langle I_{j}\rangle<0$).
It is natural to let the telluric lines be normalized with respect
to the total continuum $C$.

\subsection{Numerical method}\label{kor}
The Fourier transform is calculated in KOREL using the procedure
Fast Fourier Transform (FFT). Consequently, the input spectra
must be discretized into $2^n$ points equidistant in logarithmic
wavelength.\footnote{ The number of data points
 in each spectral region is 256 in the PC-version or its multiple
 by a power of 2 in the workstation version of the code KOREL.}
The shift ($v_{j}(t)$) of $\delta$-function in Eq.~(\ref{Kor1})
or its generalizations must be discretized with
the sampling frequency (to give the function $\exp(iyv_{j}(t))$
in Eq.~(\ref{FKor1}) periodic with the period of the data
interval). Consequently, both $v_{j}(t)$ as well as the minimized
sum $S$ given by Eq.~(\ref{FKor3}) are step functions
(in $t$ and $p$, resp.) in the numerical representation.
To achieve a good resolution in velocity, it can thus be
advantageous to interpolate the input spectra into higher
sampling density than the original one read from the detector.
The Fourier transforms of these spectra must be stored in the
computer memory in the course of the solution. To enable the
use of large number of spectra with high resolution, the spectra
can be represented by several spectral regions only (containing
spectral lines chosen for the solution), each one characterized
by the initial wavelength and step in radial velocity per one
bin. For each spectral region the functions $S$ are calculated
according to Eq.~(\ref{FKor3}) and the corresponding sets of
Eqs.~(\ref{decomp}) are independent. The total $S$ summed over
spectral regions is calculated only for the purpose of
convergence of parameters $p$.

 The spectra are supposed to originate from a multiple
stellar system with a hierarchical structure shown in
Fig.~\ref{obr1}, where the numbers in circles are indexes of
the star position, numbers in parenthesis give indexes of
the corresponding orbit. The occupation of each position by a visible
component is to be given by special key on input. The orbits
Nos.~1, 2 and 3 can be suppressed by the choice of the corresponding
period equal to 0.

\setlength{\unitlength}{1mm}
\begin{figure}[hbt]
\setlength{\unitlength}{1mm}
\begin{picture}(160,50)
\put(5,0){\framebox(80,50){}}
\put(12,37){\makebox(6,6)[cc]{1}}
\put(15,40){\circle{6}}
\put(32,37){\makebox(6,6)[cc]{2}}
\put(35,40){\circle{6}}
\put(18,40){\line(1,0){14}}
\put(23,42){(0)}
\put(22,7){\makebox(6,6)[cc]{3}}
\put(25,10){\circle{6}}
\put(42,7){\makebox(6,6)[cc]{4}}
\put(45,10){\circle{6}}
\put(28,10){\line(1,0){14}}
\put(33,6){(1)}
\put(35,10){\line(-1,3){10}}
\put(24,24){(2)}
\put(72,22){\makebox(6,6)[cc]{5}}
\put(75,25){\circle{6}}
\put(30,25){\line(1,0){42}}
\put(45,27){(3)}
\put(90,50){\parbox[t]{65mm}{
\caption{The hierarchical structure of the stellar system.
The numbering of component stars (in circles) and their
orbits (in parenthesis) as used in KOREL is shown.}\label{obr1}
}}
\end{picture}
\end{figure}
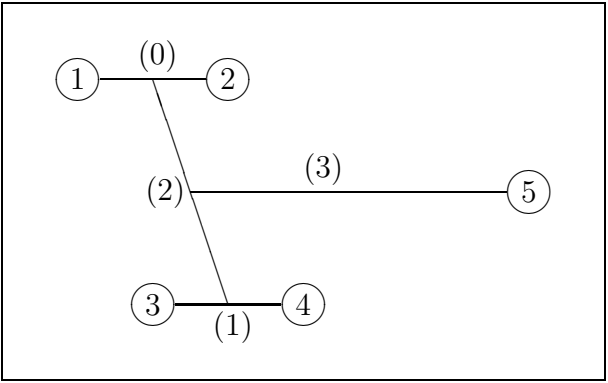

 The radial velocity of a component is thus given by
\begin{equation}\label{RV}
 v_{j}(t;p)=\sum_{o} K(\cos(\omega+\upsilon)+e\cos\omega)\; ,
\end{equation}
where the summation is performed over the orbits influencing
the motion of the star (e.g. orbits Nos.~0, 2 and 3 for the star
No.~1; note that the $\gamma$-velocity of the system does not
appear in this formula -- it can be specified only after the
identification of lines in the component spectra).
The true anomaly $\upsilon$ is calculated according to
\begin{equation}\label{anom}
 \upsilon = 2 {\rm arctg}\left( \sqrt{\frac{1+e}{1-e}}
  {\rm tg}\frac{E}{2}\right)
\end{equation}
from the solution of Kepler's equation
\begin{equation}\label{Kepler}
 2\pi\frac{t-t_{0}-\Delta t}{P}
 \left(1-\frac{\dot{P}}{2}\frac{t-t_{0}-\Delta t}{P}\right)=M=E-e\sin E
\end{equation}
for the time $t$ corrected by
\begin{equation}\label{lte}
 \Delta t = \sum_{o}\frac{PK}{2\pi c}(1-e^{2})^{3/2}
  \frac{\sin (\omega+\upsilon)}{1+e \cos (\upsilon)} \; ,
\end{equation}
for the light-time effect due to the higher orbits in the hierarchical
system (e.g. orbits Nos.~2 and 3 for stars Nos.~1 and 2).
The pericenter longitude is given by
\begin{equation}\label{omg}
 \omega=\omega_{0}+\dot{\omega}(t-t_{0}-\Delta t)\; ,
\end{equation}
i.e., the secular periastron advance (linear in time) can be
taken into account. Similarly
\begin{eqnarray}\label{ecc}
 e&=&e_{0}+\dot{e}(t-t_{0}-\Delta t)\; ,\\
 K&=&K_{0}+\dot{K}(t-t_{0}-\Delta t)\; ,\\
 q&=&q_{0}+\dot{q}(t-t_{0}-\Delta t)\; ,
\end{eqnarray}
The spectra and times of exposures are usually transformed into
the heliocentric system. If not, the higher orbit can be used to make
the corresponding correction. The secondary component of this `solar'
orbit can be used to remove the telluric lines (in an approximation)
from the stellar spectra.

 The minimization of $S$ with respect to $p$ is performed by
the simplex method adapted from Kallrath and Linnell
(1987).\footnote{ For details see description of the code FOTEL
 (Hadrava 2004b in this volume), where the same procedure is used.}
Several orbital elements (cf. Table~\ref{tab1}), line
strengths\footnote{ Note that either all line strengths for a
 chosen component can be calculated using the method described
 in Section~\ref{LSV}, or only several of them (e.g. in exposures
 suspicious to be taken during an eclipse) using the simplex
 method.}
or radial velocities\footnote{ For a component which is not tied to
 an orbital motion -- cf. $KEY$ explained in Section~\ref{ContR}.}
can be chosen from all of them
for convergence in one step, the others being fixed.
In each of these `large' steps, there are performed many simplex
transformations. The number of these `small steps' is 10$\times$
the number of the iterated parameters (at maximum 10 parameters
can be converged in one large step). The actual status of
the iteration is displayed on the screen (in order: No., code of
simplex transformation, No. of the worst simplex point, value of
$S$ in this point and its values of parameters, i.e. in the same
way as in FOTEL).
At each simplex step the spectral decomposition is performed
first, and then the line-strengths are calculated (if it is required
by the corresponding key of the star). The self-consistent
solution requires either to repeat these steps, or to store
the values of line-strengths appropriate to the set of orbital
parameters at a particular simplex point. The former approach
is used in the present version of KOREL because the later would
be memory exhausting. To find an exact self-consistent solution
in each step would be very time consuming, hence only up to 5
iterations\label{lsit} of successive spectra decomposition and line-strengths
calculations are performed. The solution with free line-strengths
is usually more sensitive to local minima in the basic parameters.
To prevent this disadvantage it is recommended to hold the line-strengths
fixed during the simplex solution (unless it is only tiny tuning of
an already found solution) and to improve them in a subsequent step.
Another possibility is to include some line-strengths (of chosen
components in chosen exposure -- e.g. very strong telluric
lines or components participating in an eclipse) between the
parameters converged by simplex with other line-strengths held
fixed.

 For the purpose of different numerical tests, there is built-in
into KOREL a possibility to produce synthesized data with chosen
orbital parameters. The profiles of each component can be then
chosen in one of two types, either with broad wings (type 1)
\begin{equation}\label{iin1}
 I_{j}=-c\frac{\Delta^{2}}{(x-x_{0})^{2}+\Delta^{2}}\; ,
\end{equation}
or with continuum (type 2)
\begin{equation}\label{iin2}
 I_{j}=c{\rm min}\left[0,
  \left(\frac{(x-x_{0})^{2}}{\Delta^{2}}-1\right)\right]\; .
\end{equation}
Random noise of a chosen amplitude can be added into each bin.


\cleardoublepage
\setcounter{page}{101}
\chapter{Practice of spectra disentangling}\label{Practice}

\section{Practical conditions for the use of KOREL}
In this part of the text, a guide to practical exercises prepared
for afternoon sessions of the Summer school is provided. These
exercises simulate the future practical use of the code KOREL.

 Each group is provided with an account and a terminal, i.e.
a workstation with Linux Debian and GNOME environment (KDE and
XFCE4 are also installed) in a local net with a central disc-field.
(It means that the users may log in at any terminal.)
The usernames are {\tt korel01}, {\tt korel02}... {\tt korel12}
and the corresponding passwords {\tt 01noh}... {\tt 12noh}.
(The passwords should not be changed and everybody should use
his own account only.) The terminal windows can be opened
e.g. in menu {\tt Application} $\rightarrow$ {\tt Accessories}
$\rightarrow$ {\tt Terminal} or by an icon on the top bar
(FIREFOX, GEDIT, DOSEMU and GEDIT are also on the bar).
The standard applications like SSH, SCP may be used from there
to reach remote computers, editors VIM, GVIM, GEDIT or KEDIT
(also available at  {\tt Application} $\rightarrow$ {\tt Other}
 $\rightarrow$ {\tt KEdit}), or PostScript viewers EVINCE or GV.
A nonstandard but needed is DOS-emulator (available in
{\tt Application} $\rightarrow$ {\tt System tools}
 $\rightarrow$ {\tt DOS emulator} or by the icon on the bar),
which works at the C: directory available in LINUX at directory
{\tt /.dosemu/drive}\_{\tt c}.

 Another account with the same user-name is provided at the
Ond\v{r}ejov cluster of computers\\
\hspace*{15mm} {\tt ***.cz}, IP address {\tt 147.***}.\\
The password {\tt ***01} 
corresponds to username {\tt korel01}
and the last two digits are to be replaced correspondingly for other
users. In the home directories of the users are prepared executable
codes of KOREL and other LINUX-codes. There are also exe-versions of
codes running in DOS only, which we shall download via SCP
to the local C: directory.

 At the cluster the subdirectories {\tt /exe1}... with data for
the exercises are also prepared.
These will be used either directly at the cluster, or they will be
downloded to the local terminal first and results from the
work with them will be sent back, like in future data prepared
from the users own observations.

 The data on real stars which are prepared for the exercises originate
mostly from the Ond\v{r}ejov 2m- telescope and they were observed for
studies either already published in the past or intended in future.
It means that the data should not be used for other purposes than
the training without a special agreement with their authors. The
data are not provided as complete sets and their reduction is not
always the best as required for real scientific use. On the other
hand, some of them were chosen with regards to personal interests of
the participants, so that it is possible that a future cooperation
combining Ond\v{r}ejov and other data could be welcome.

\clearpage
\section{An easy start with KOREL} \label{start}
To perform the disentangling in its simplest form, we have to provide
the KOREL-code with:\\
(1) the spectra to be disentangled,\\
(2) with initial estimates of the parameters of the system and\\
(3) with codes controlling the run of the code.\\
(Other possibilities will be shown later.)
The spectra are to be given in the input file {\tt korel.dat}, the
parameters and codes in {\tt korel.par}, both in the format described
in the Manual in Sections \ref{koreldat} and \ref{ContR}, resp.

 The choice and preparation of the spectra may be a laborious task
for which an auxiliary code PREKOR is written and its use will
explained later. The main demands are that the input spectra should
be sampled in the same equidistant grid of logarithmic wavelength-scale
(with $2^{n}$ grid points, $n$ being integer; a reasonable minimum is
256) and they should be rectified to the continuum. In the chosen spectral 
region it is desirable
to comprise the same lines in all
exposures and to have a continuum at the edges of the region. To
solve for orbital parameters, a maximal resolution around well visible
lines is preferred (e.g. one line only), while for the purposes of
analysis of the disentangled spectra we can choose longer regions with
a smaller resolution (and to keep the orbital parameters fixed from
a previous solution of high-resolution regions of spectra).

 A proper formulation of the questions given to KOREL in the file
{\tt korel.par} is the core of the art of disentangling. KOREL is not
an expert system analyzing the input in an automated way and the user
has to involve his own intuition to estimate which questions could be
answered from which data in which order and to judge if the results
are reasonable and enable one to ask more.
Usually we know or at least we suspect what
the orbital parameters of the studied object may be (e.g. the period
may be known from a photometry) and their uncertainties.
We have to include these values on input and to decide which of them
may be improved using our data. We have also to decide from both the output
values of parameters and the disentangled spectra if the result may be
correct and if it hit an acceptable solution.

\subsection{Example 1 -- simulted data}
\label{exa1}
Here, we shall start first with disentangling of simulated spectra,
which have the advantage that we know what their chosen properties are
and we may check how well they are restored by the disentangling.
KOREL includes an option to create simulated data, however, they
cannot be saved. In the directory {\tt EX1} we have data
(in the file {\tt kor0801a.dat}) created
by an independent code. They approximately repeat those on which
the function of KOREL was demonstrated in the first publication
(Hadrava 1995). There are simulated 7 exposures covering
uniformly one period of a double-lined binary on a circular orbit
with semiamplitudes of radial velocities $K_{1}=22$ and $K_{2}=44$ km/s.
The calculated spectral region is sampled in 256 bins with a step of
2 km/s, contains one line composed of Lorentzian profiles with central
intensities in ratio 4:1 and semi-widths corresponding to radial
velocities 40 and 70 km/s for the primary and secondary, respectively.
An artificial noise is added to the signal with an amplitude up to $\pm 2\%$ of the
continuum. The data-file begins with a line
giving the time of the exposure (here it runs from .0000 to 1.0000;
we chose the period to be 1.), initial wavelength (in \AA{}, here
chosen arbitrarily) and step of the sampling in km/s of the spectral
region, weight of the exposure (chosen here 1.000 for all exposures)
and the number of bins (here 256) in one exposure. Then follow
the values of the intensity in these bins -- cf. the file
{\tt kor0801a.dat}, which we shall copy now to the file
{\tt korel.dat}:\\
\begin{verbatim}
       .0000 4500.0000  2.000  1.000     256
  .99143  .97534  .98784  .98352  .97205  .98681  .98670  .99434  .98397  .98431
  .96184  .97731  .98497  .99025  .97306  .97125 ...
\end{verbatim}

 The file with corresponding parameters and controlling codes for
decomposition of these profiles is prepared as {\tt kor0801a.par},
from where we shall copy it to {\tt korel.par}. Its complete form
reads:\\
\begin{verbatim}
1 1 0 0 0 1 0 2 2     |key(5), k= Nr. of spectra, filter, plot, print
o 0 1 0 1 1 1. 0.001  = PERIOD(0)
o 0 2 0 1 1 0. 0.1    = PERIASTRON EPOCH
o 0 3 0 1 1 0. 0.1    = ECCENTRICITY
o 0 4 0 1 1 0. 10.    = PERIASTRON LONG.
o 0 5 0 1 1 22. 5.    = K1
o 0 6 0 1 1 .5 0.1    = q = M2/M1,  K2 = 44.
x 0 0 0 0 0 0 0       | end of elements
\end{verbatim}
Here the digits 1 between the first five numbers specify that
the spectra of components 1 and 2 from the hierarchical structure
shown in Fig.~\ref{obr1} are present, while the other three are
not. Subsequent 6 lines give the values of orbital parameters
of this pair (in the seventh column). We do not request their
convergence (as given by zero in the fourth column).
The last line denotes end of the input of parameters.
(See Section~\ref{ContR} for a complete explanation of the file
{\tt korel.par}.)

 Having prepared both files {\tt korel.dat} and {\tt korel.par}, we can
run now the code {\tt korel08}. It produces numerical results in the
file {\tt korel.res} and a graphic output in the file {\tt phg.ps},
which we can see in Fig.~\ref{obr2a}.
\setlength{\unitlength}{1mm}
\begin{figure}[hbtp]
\noindent\begin{picture}(140,100)
 \put(0,0){\epsfxsize=100mm \epsfbox{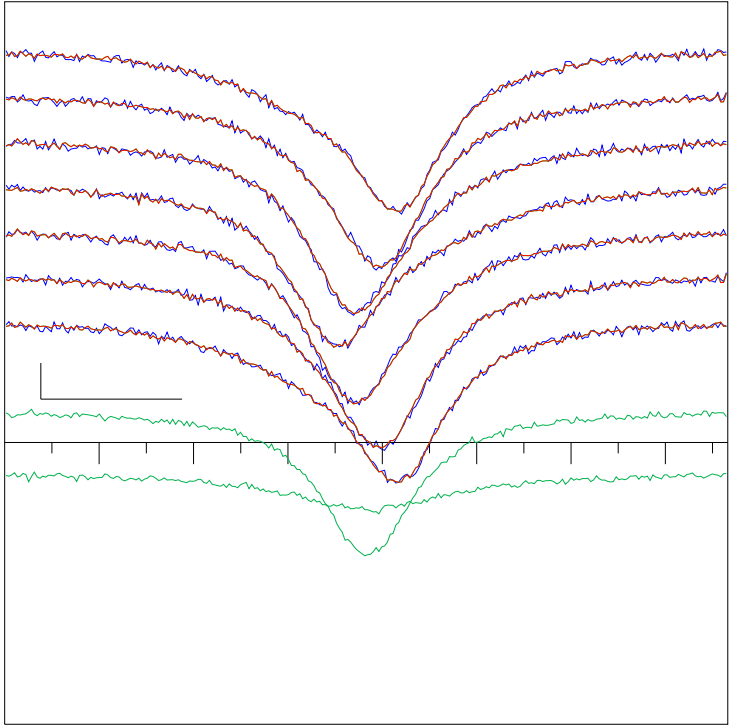}}
\put(110,95){\parbox[t]{50mm}{\caption{
 Simulated composite line profiles (7 blue upper curves)
 of a binary from the file {\tt kor0801a.dat}, the disentangled
 component spectra (2 bottom green curves) and their
 superposition on the input profiles (red lines). \label{obr2a}
}}}
\end{picture}
\end{figure}
We can see here that the disentangling is able to separate the
profile of the secondary component from the phase-locked variations
of asymmetry of the apparent single peak seen as a result of
the blending in the input spectra and to restore also the symmetric
profile of the primary.

 However, this run of KOREL was a simple decomposition and not a
full disentangling, because the orbital parameters were fixed to
their exact values, which we know from the simulation of the data.
It can be seen in Fig.~\ref{obr3} how the $\chi^{2}$ of the residual
noise depends on the amplitudes of radial velocities if we fix them
for the component stars to different values for the file {\tt kor0801a.dat}
and also for {\tt kor0801b.dat}, which is simulated with the same
parameters of the orbit and line profiles, but the random noise is
10-times smaller. It is obvious that the minimum is much deeper and
its position more precisely defined for the data with smaller noise.
\setlength{\unitlength}{1mm}
\begin{figure}[hbtp]
\noindent\begin{picture}(160,90)
 \put(10,05){\epsfxsize=60mm \epsfbox{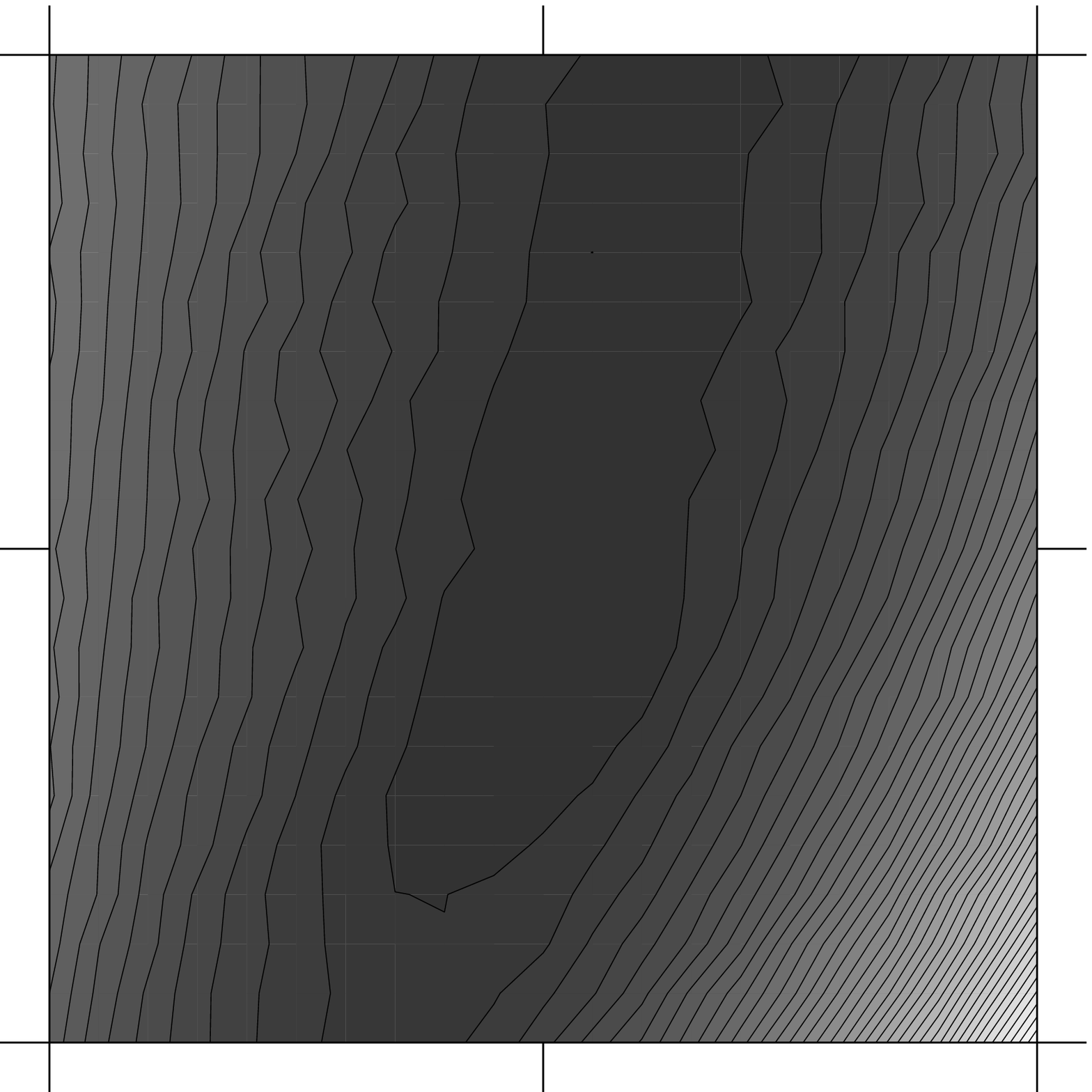}}
 \put(90,05){\epsfxsize=60mm \epsfbox{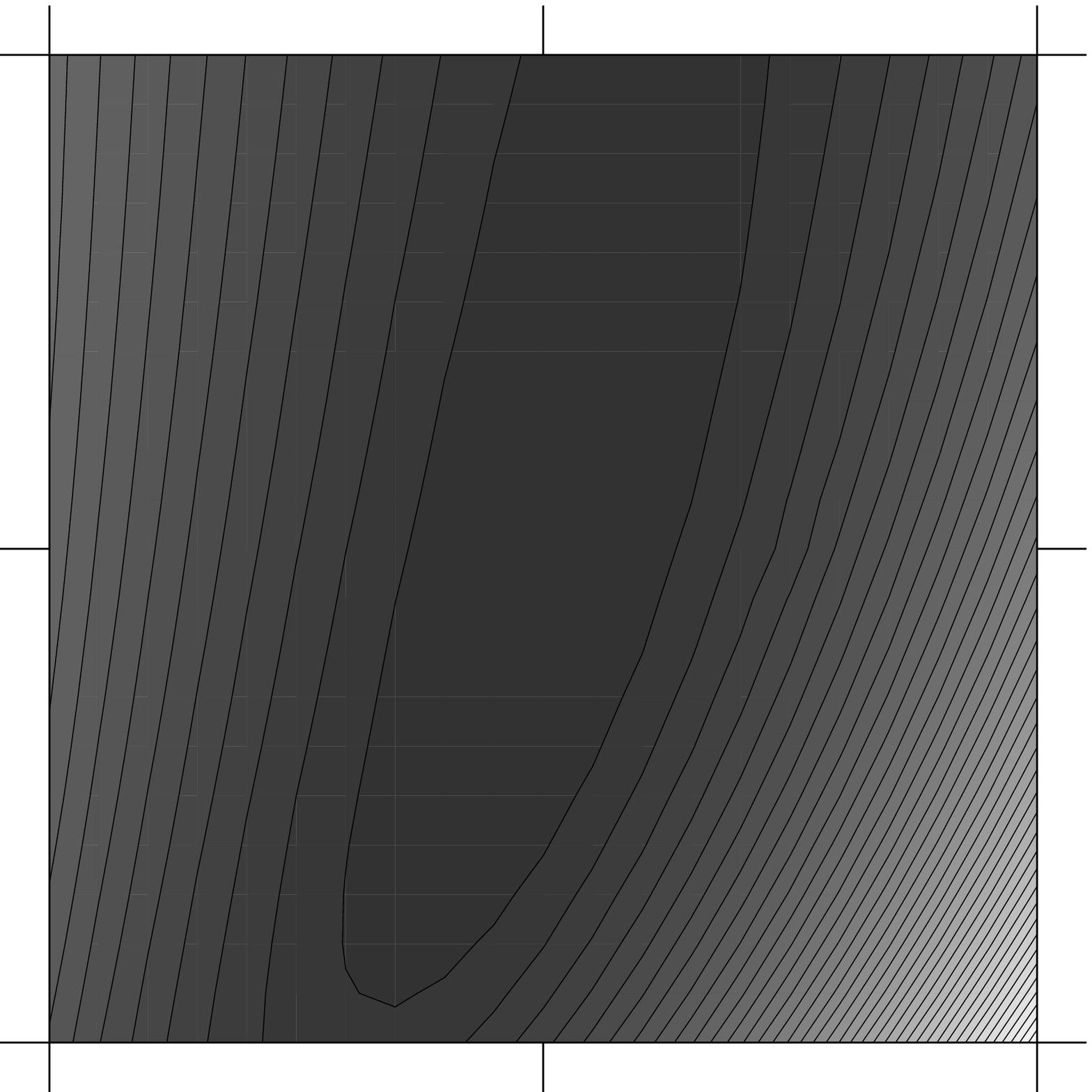}}
\end{picture}
\caption{
 Residual noise as a function of $K_{1}$ and $q$ for data
 from the file {\tt kor0801a.dat} (left) and {\tt kor0801b.dat}
 (right). \label{obr3}
}
\end{figure}

 To simulate a real case of disentangling of spectra for systems
for which the orbital parameters are not well known, we shall change
the initial values of some parameters and then we shall let them
converge. For instance, let us change three lines in the file
{\tt korel.par} as follows:\\
\begin{verbatim}
o 0 2 02 1 1 0.1 0.1    = PERIASTRON EPOCH
...
o 0 5 01 1 1 12. 5.    = K1
o 0 6 01 1 1 .3 0.1    = q = M2/M1,  K2 = 44.
\end{verbatim}
If we shall run now FOTEL again, it will show on the screen and
also in the output-file {\tt korel.res} a protocol about simplex
convergence of the parameters in the form:\\
\begin{verbatim}
   1 C 2 0.36407E+03 0.19428E+00 0.13179E+02 0.32357E+00
   2 A 4 0.21356E+03 0.12357E+00 0.13179E+02 0.39428E+00
   3 A 3 0.21257E+03 0.12357E+00 0.16714E+02 0.32357E+00
   4 B 1 0.17167E+03 0.10000E+00 0.12000E+02 0.30000E+00
   5 B 3 0.10573E+03-0.55232E-01 0.12471E+02 0.30440E+00
   6 A 1 0.72220E+02 0.20518E-01 0.13715E+02 0.30971E+00
   7 B 2 0.59546E+02-0.25708E-01 0.15378E+02 0.36757E+00
...
  30 A 2 0.20125E+02 0.94115E-04 0.21837E+02 0.43925E+00
  31   3 0.20125E+02 0.77977E-05 0.20220E-02 0.52600E-04
  31   4 0.20125E+02 0.98819E-04 0.21835E+02 0.43926E+00
\end{verbatim}
It is natural that the wrong starting values of parameters resulted
in a much higher residual sum -- instead of about 20.76 in the correct
solution to 171.67 in the starting point (see the fourth line) or even
to 364.07 in the first line which has the largest deviation in the
time of periastron. However, the simplex converges finally to values,
which by chance fit the spectra with their noise even better (20.125)
than the true values. The time of periastron is restored with an error
about $10^{-4}$ of the orbital period, $K_{1}$ with error about 1\%,
but the mass ratio (and hence also $K_{2}$) about 10\%. This is obviously
due to the fact, that the the line of the secondary is wide and shallow
and hence the minimum of O--C in Fig.~\ref{obr3} has a shape of a valley
elongated in the $q$-direction with poorly distinguished deepest point.
The same calculation with the less noisy data from {\tt kor0801b.dat}
restore the true values of parameters much better. If we fit $q$ to
some wrong value (e.g. 0.3), we get still quite satisfactory solution.
Another interesting experiment is to disentangle the data a single-line
binary, i.e. to change the first line of {\tt korel.par} to: {\tt 1 0 0 0 0...}
and naturally to skip a convergence of $q$. The $K_{1}$ velocity then
converges to the value about 14.54 km/s. It shows that the disentangling
is sensitive more to the bottom of line-profile, because the radial
velocity defined by moments has amplitude
(22*4*40-44*1*70)/(4*40+1*70)$\simeq$1.91 km/s only. Nevertheless,
the unresolved secondary apparently decreases the amplitude (22 km/s)
of the primary line.


\subsection{Example 2 -- 96 Her}
\label{exa2}
In files {\tt kor0802a.dat}, {\tt kor0802b.dat} and {\tt kor0802c.dat}
in the directory {\tt EX2} we have prepared three regions of the star
96~Her. Corresponding values of parameters with well progressed
solutions are in files {\tt kor0802a.par}, {\tt kor0802b.par} and
{\tt kor0802c.par}.

\setlength{\unitlength}{1mm}
\begin{figure}[hbtp]
\noindent\begin{picture}(160,80)
 \put(5,0){\epsfxsize=70mm \epsfbox{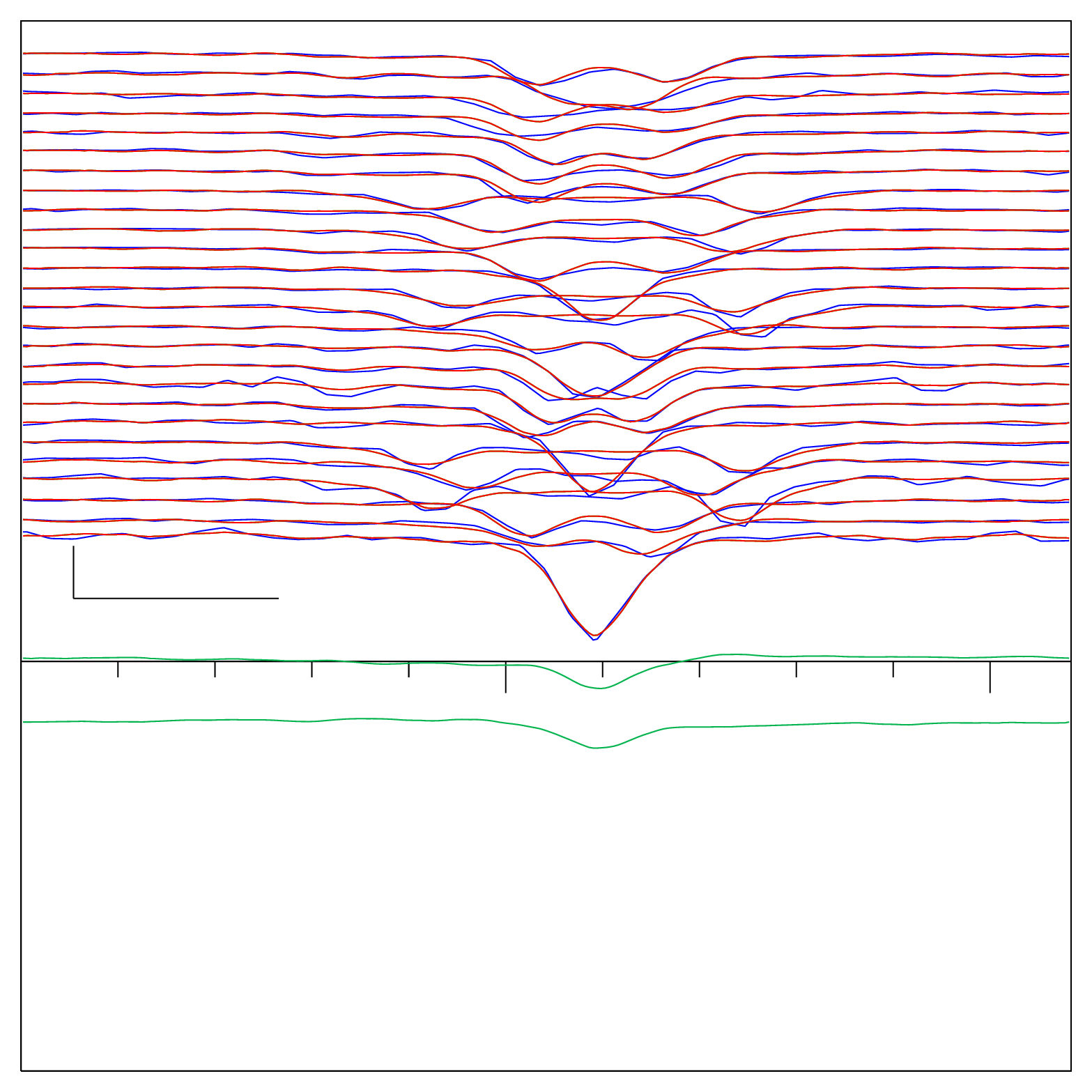}}
 \put(90,10){\epsfxsize=70mm \epsfbox{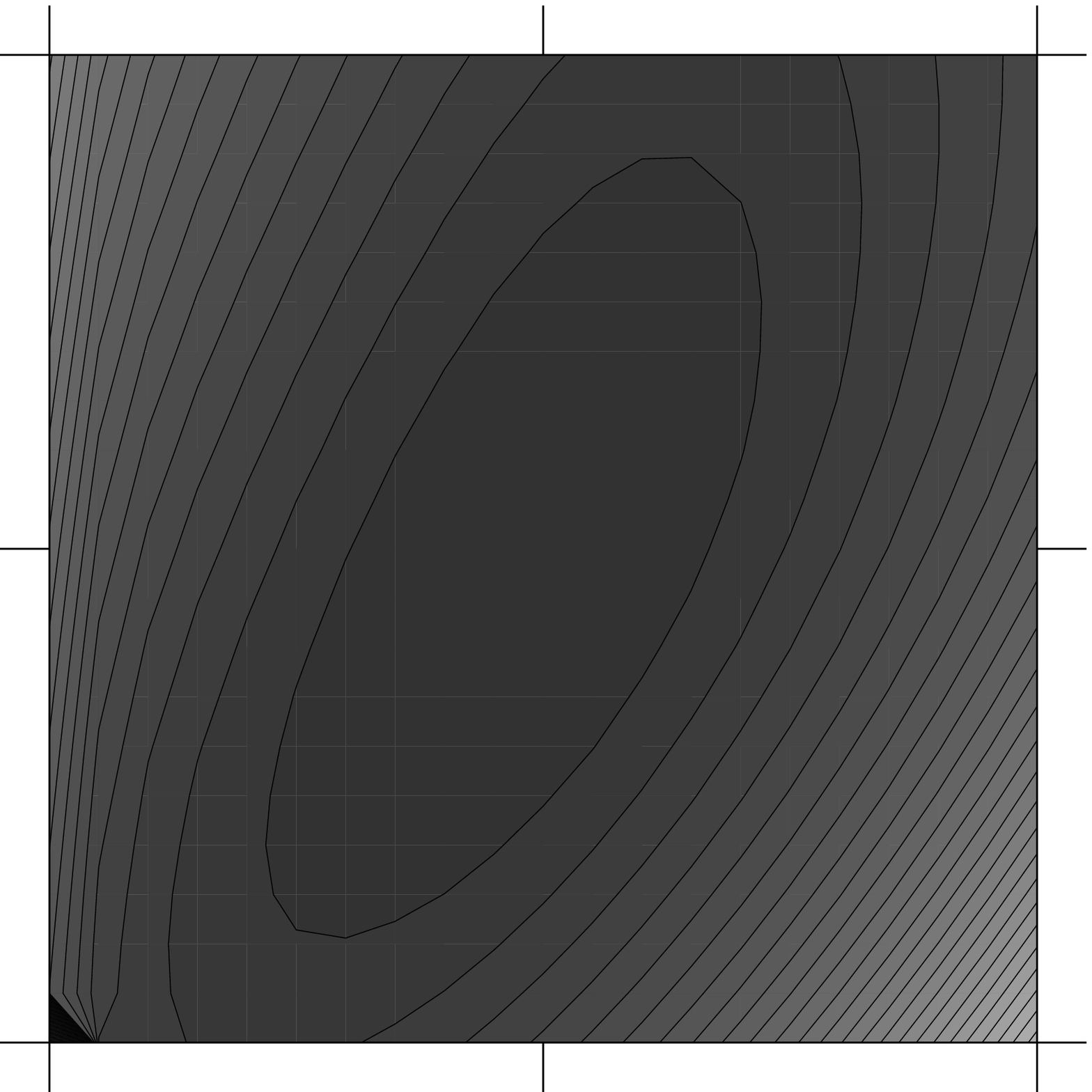}}
\end{picture}
{\caption{
 Disentangling of 96~Her from the file {\tt kor0802b.dat} (left)
 and residuals from {\tt kor0802a.dat} as a function of $K_{1}$
 and $q$ (right). \label{obr4}
}}
\end{figure}
\setlength{\unitlength}{1mm}
\begin{figure}[hbtp]
\noindent\begin{picture}(140,100)
 \put(60,0){\epsfxsize=100mm \epsfbox{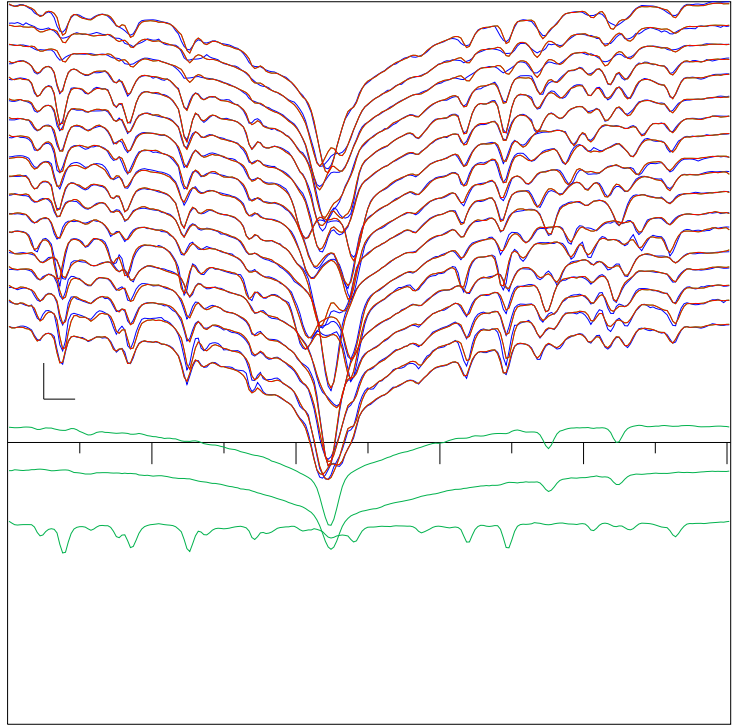}}
\put(0,95){\parbox[t]{50mm}{\caption{
 Disentangled H$_{\alpha}$ from the file {\tt kor0802c.dat}
 (the primary, secondary and the telluric component). \label{obr5}
}}}
\end{picture}
\end{figure}


\subsection{Example 3 -- 55 UMa}
\label{exa3}
In the directory {\tt EX3} we have prepared several spectra of
the triple star 55 UMa in the form of ASCI table. We have to
prepare the input {\tt korel.dat} and {\tt korel.par} using
the code PREKOR. PREKOR runs under the DOS and its executable
version  {\tt prekor14.exe} can be downloaded from the main
directory. We shall download it into the C: directory of the
DOS-emulator, equally as the spectra ({\tt 55u0*.asc}), their
list {\tt prekor.lst}, which also provides the Julian dates
of these exposures, and the file {\tt prekor.par}. In the C:
directory we can run PREKOR, either in the option ``0" to
calculate the orbital parameters for disentangling the telluric
lines (they are contained in the output file {\tt prekor.res}),
or with some higher number, to produce the data {\tt korel.dat}
(in the output {\tt prekor.out}). This output will be sent back
to the cluster. Before it can be run by KOREL, it must be converted
from the DOS-format into the UNIX format. This can be done, in
principle using the utility {\tt dos2unix} (not available here),
or using the trick reccommended by Baju Indradjaja, i.e. by the
command {\tt tr -d "$\backslash$r" < input > output}, or using
the code KORTRANS (cf. Section \ref{KORTRANS}).


\clearpage
\section{Rectification of the input spectra}

\subsection{Example 4 -- $\alpha$ CrB}
\label{exa4}
The FITS-files of unrectified spectra of the binary $\alpha$~CrB
(Gemma) are given in the directory {\tt EX4}. The rectification
may be performed using my code REKTIF (cf.~\ref{REKTIF} on
p.~\pageref{REKTIF}), the DOS exe-version of which can be
downloaded from the main directory. The input spectrum is supposed in
the form of ASCI-table, so we need also to convert it at first.
This is possible using the code FIT2ASC.exe, which can also be
downloaded from the main directory. Each spectrum is to be copied
to the file {\tt fit2asc.in}, then a run of the FIT2ASC-code creates
the output table in the file {\tt fit2asc.out}. The first lines may
contain JD if it is given in the header of the FITS-file (it is
to keep some identification of the exposure. This output should
be copied (or renamed) to the file {\tt rektif.dat}. We also need to
prepare the file {\tt rektif.par} defining the procedure of the
rectification. To specify the window in wavelength and values
of signal, we may either check these quantities from the input-file
{\tt rektif.dat}, or we can run the code with any values and read them from
the output {\tt rektif.res}. In the file {\tt rektif.par} we also have to
specify initial values of the continuum-marks. Provided we do not
have any better estimate, we can chose the signal level at the
maximum and set the acceptable error close to 100\% (1.0).


\subsection{Example 5 -- 68 u Her}
\label{exa5}
In the directory {\tt EX5} data and example of solution are presented
for another system to which I applied KOREL already in its beginning.
Participants of the School may try their own creativity how to treat
these data.



\cleardoublepage
\setcounter{page}{201}
\appendix
\chapter{Appendices}\label{Append}

\section{Historical background}\label{hist}

\begin{wrapfigure}{r}{40mm}
\begin{picture}(37,38)
 \put(5,0){\epsfxsize=35mm \epsfbox{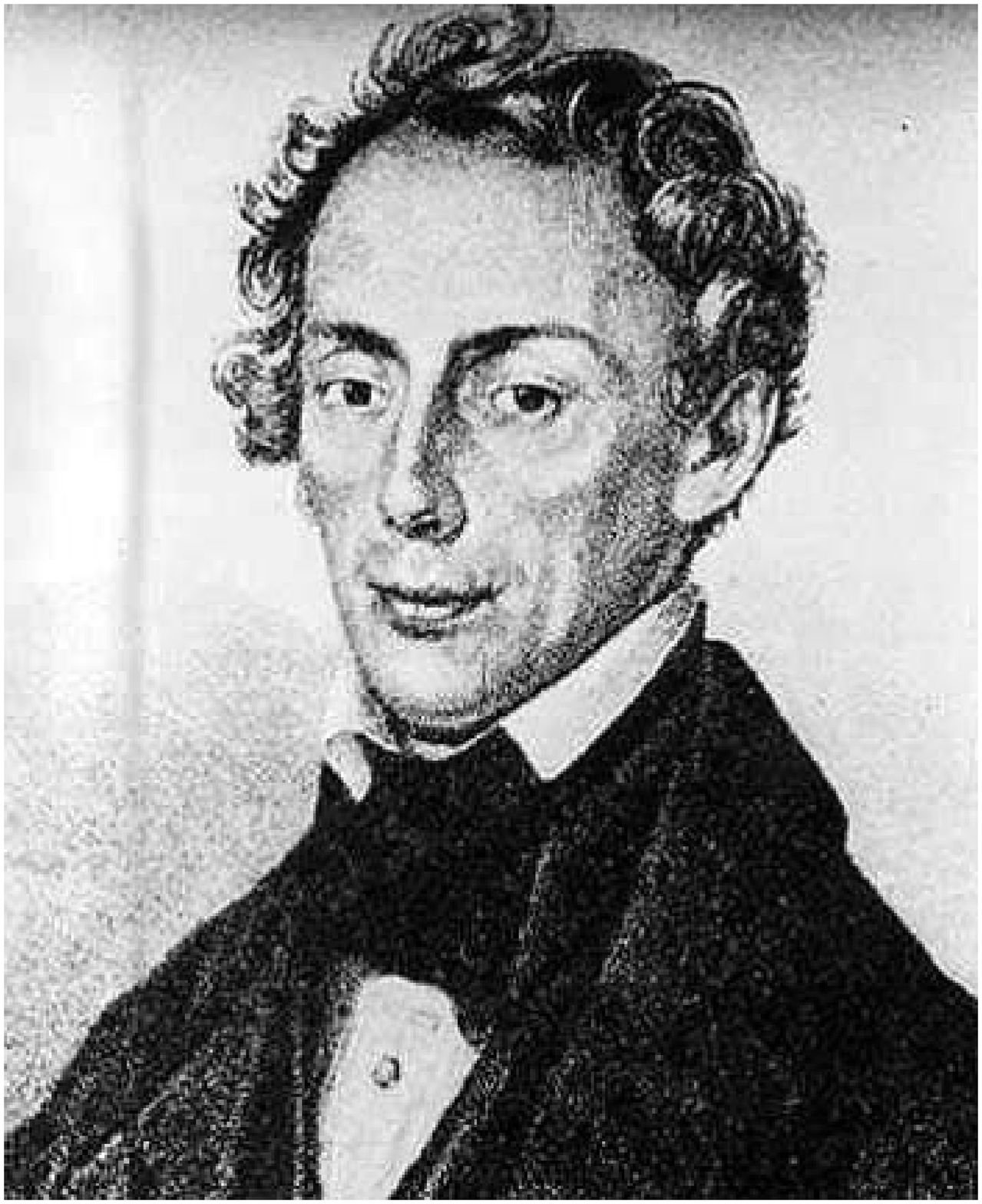}}
\end{picture}
\\[1mm]
\hspace*{2mm} Christian Doppler 
\end{wrapfigure}

\label{Dopplhist}
The method of disentangling, equally as other methods dealing with
observations of spectroscopic binaries, is based on Doppler principle.
Application of this principle is nowadays so widely spread in different
fields not only of science but also of technical or medicine practice,
that it is little known that the original idea which motivated the Austrian
physicist Christian Doppler (1803-1853) to suggest his principle was just
an attempt to explain observations of binary stars. In his seminal work
``Uber farbige Licht der Doppelsterne und einiger anderer Gestirne des
Himmels" (``On the coloured light of the double stars and certain other
stars of heavens", cf. Doppler 1842) presented on 25th of May 1842 to six
listeners only at a session of the Royal Bohemian Society of Sciences in
the Patriotic Hall of Prague Karolinum, he tried to explain differences
between colours of components in some visual binaries by shifts of
frequencies of light due to the different radial velocities of the stars.

\begin{wrapfigure}{l}{40mm}
\begin{picture}(37,41)
 \put(0,0){\epsfxsize=35mm \epsfbox{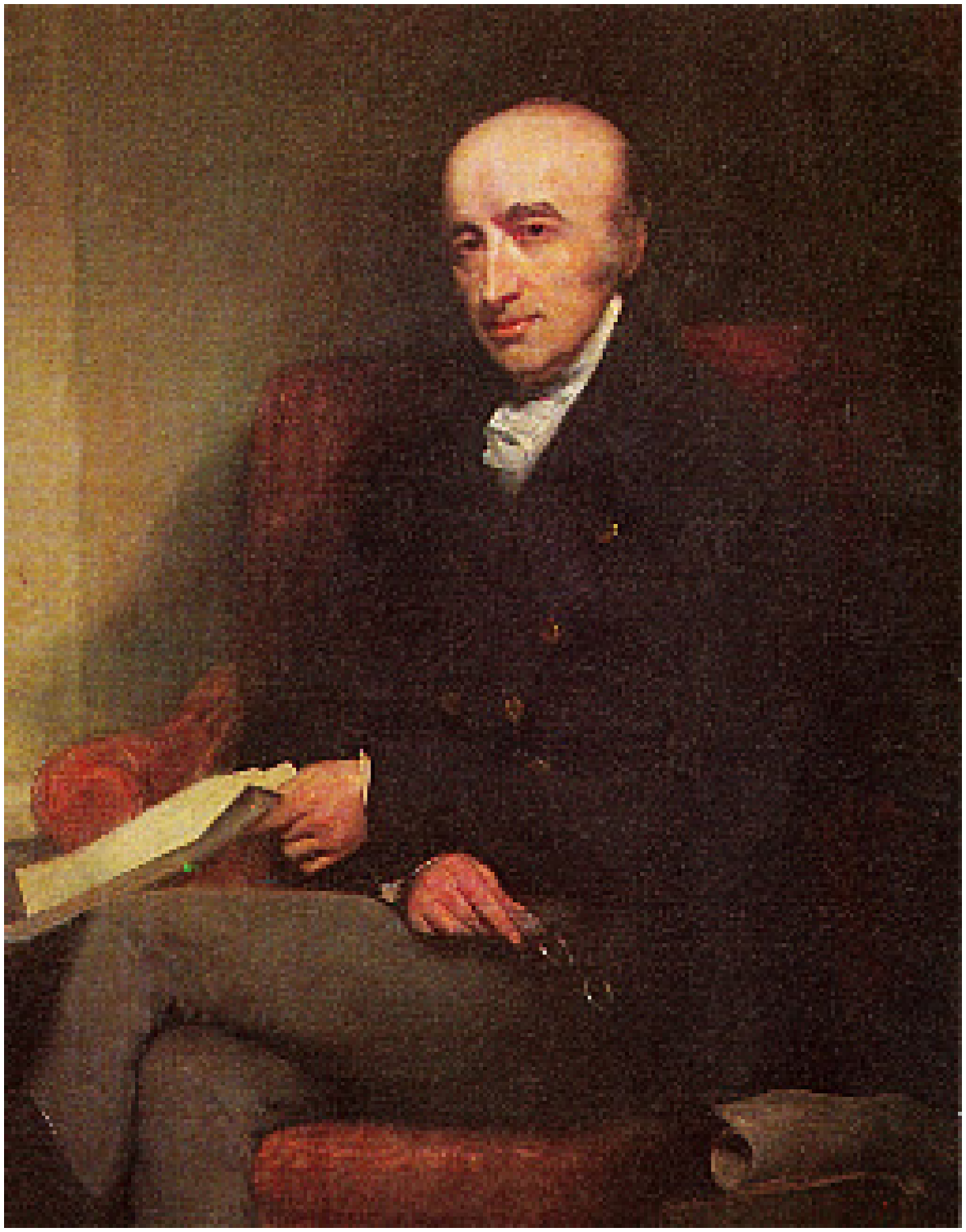}}
\end{picture}
\hspace*{3mm}{W. H. Wollaston}
\end{wrapfigure}

 Doppler's ingenious (but quantitatively exaggerated) idea was modified
in 1848 by French physicist Armand-Hippolyte-Louis Fizeau (1819-1896),
who showed the frequency shift in acoustics. Being unaware of Doppler's
work, Fizeau predicted that the same effect should be observable for light
and that a tiny shifts of absorption lines in stellar spectra could reveal
the velocities of stars. Such absorption lines were observed first in
the solar spectrum by English versatile scientist William Hyde Wollaston
(1766-1828) in 1802 and then by the German optician Joseph von Fraunhofer
(1787-1826) in 1814, but they were explained by German physicists Robert
Wilhelm Eberhard Bunsen (1811-1899) and Gustav Robert Kirchhoff (1824-1887)
in 1859 only.

 In 1863, Italian Jesuit astronomer Pietro Angelo Secchi (1818-1878)
started to investigate systematically the spectra
of stars for the purpose of their spectral classification (introduced
by him) with an intention to measure also their motion, which, however,
was beyond the capability of his spectrograph. The first measurement of
radial velocity was performed in 1868 for $\alpha$~CMa by English
spectroscopist Sir William Huggins (1824-1910), who investigated
(from 1864) spectra of star and nebulae also in order to determine
their chemical composition by identification of their spectral lines.

\begin{wrapfigure}{r}{40mm}
\begin{picture}(37,40)
 \put(5,0){\epsfxsize=35mm \epsfbox{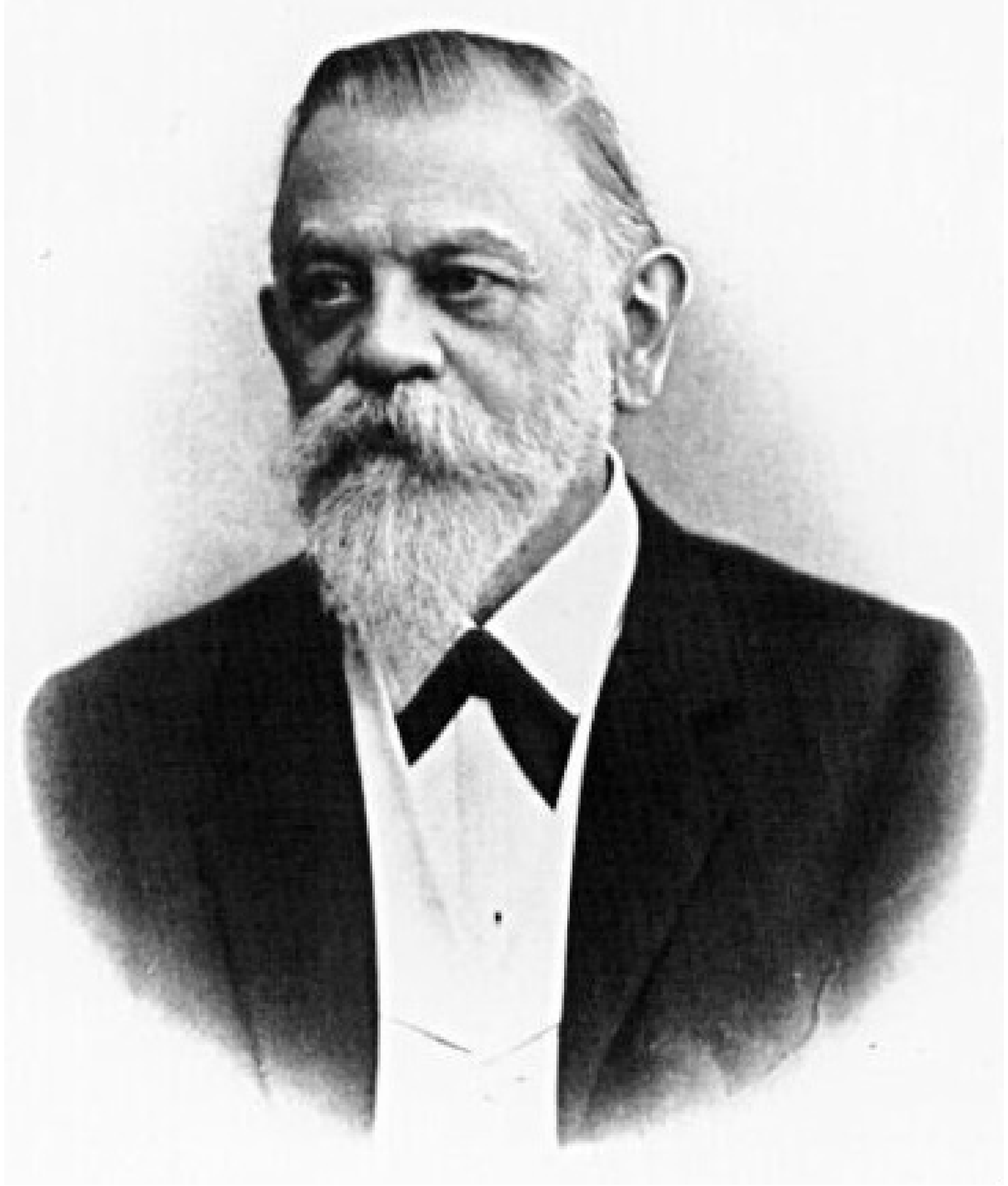}}
\end{picture}
\\[1mm]
\hspace*{10mm} H. C. Vogel 
\end{wrapfigure}

 The German astronomer Hermann Carl Vogel (1841-1907) used the Doppler
principle to measure spectroscopically the rotation of the Sun in 1871,
to compile a catalog of radial velocities of 51 stars between 1871 and
1892 and to identify as spectroscopic binaries the stars $\beta$~Per
(explained already in 1783 as an eclipsing binary by John Goodrick,
1764-1786) and $\alpha$~Vir in 1889.

In the same year 1889, another spectroscopic binaries, $\zeta$~UMa
and $\beta$~Aur were discovered also by American astronomers Edward
Charles Pickering (1846-1919) and Antonia Caetana Maury (1866-1952).\\

\begin{wrapfigure}{l}{41mm}
\begin{picture}(37,40)
 \put(0,0){\epsfxsize=35mm \epsfbox{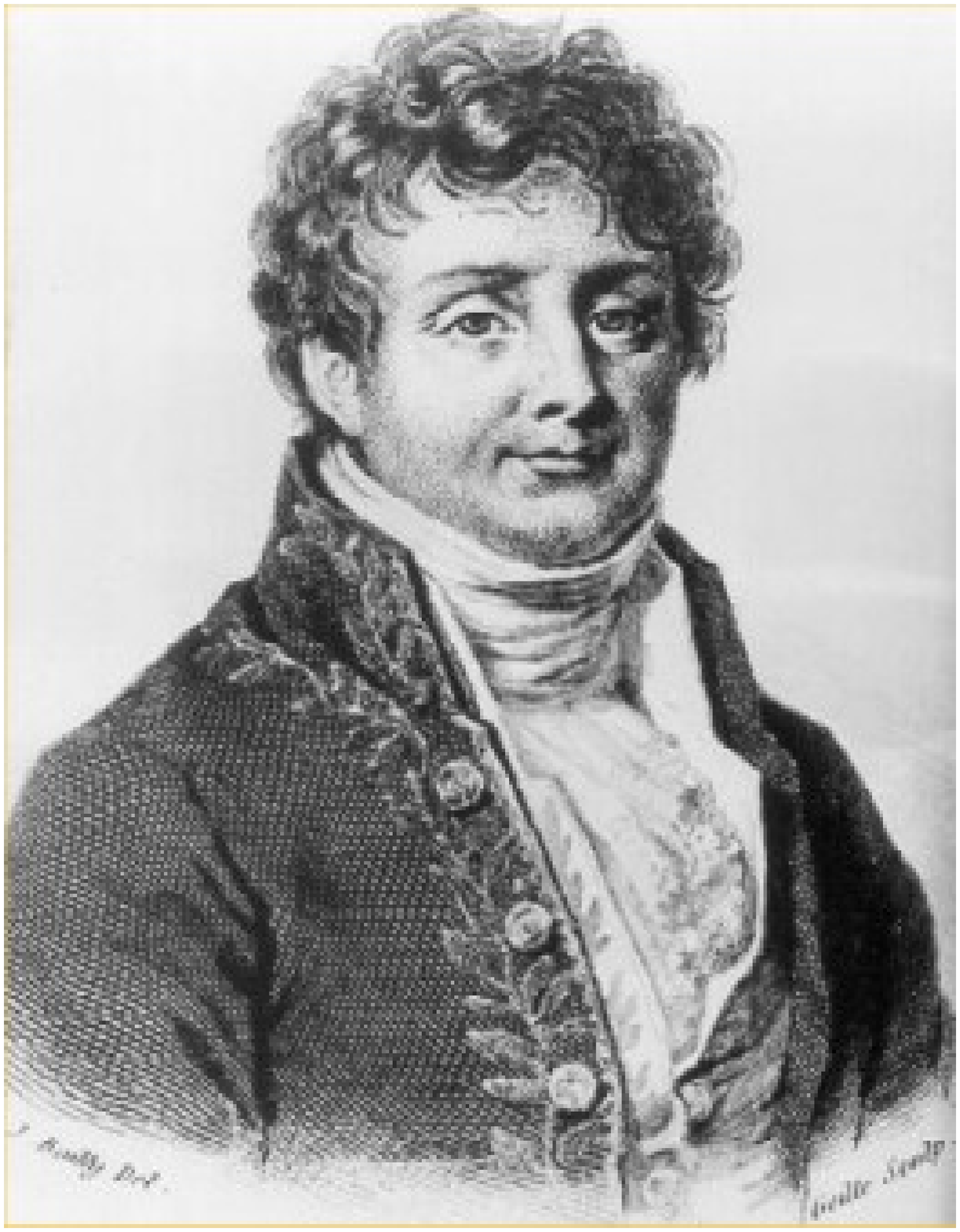}}
\end{picture}
\\[1mm]
\hspace*{3mm} J. B. J. Fourier 
\end{wrapfigure}

 No matter how remarkable is the history of the observational
astrophysics, for a true progress in understanding of Nature are
essential the theoretical conceptions, which enable to rise
appropriate question, to find an efficient way to answers and
to make reasonable conclusions. A backbone of exact sciences is
the mathematics, which is able to yield answers even long time
before the questions arise.

 This is also the case of the Fourier
disentangling of spectra, which is based on the transform
introduced by the French mathematician Jean Baptiste Joseph Fourier
(1768-1830) well before the problem of the Doppler shift could
be settled. Fourier transform has many applications not only in
solving the differential equations (Fourier introduced it to solve
equation of heat conduction) by the so called spectral methods
(or harmonic analysis) but also for computing and processing of signal
in different technical applications.

 One of these applications is also the computer tomography, which
enables to reconstruct spatial distribution of sources from different
projections of their emission. This technique is based on the Radon
transform and its inverse, which was introduced by the Austrian
mathematician Johann Radon (1887-1956). 
In the form of the so called Doppler tomography, it is also applied
in stellar spectroscopy and it is related to the disentangling of spectra
as well.

Some basic conceptions of this mathematics necessary for understanding
of the spectra disentangling will be reminded in the next Section.

\clearpage
\section{Mathematical background}\label{matem}

\subsection{Fourier transform}
Fourier transform of a complex function $f(x)$ of real variable $x$ (to
the complex function $\tilde{f}(y)$ of real variable $y$) is given by
\begin{equation}\label{Fourtr}
 \tilde{f}(y)=\int_{-\infty}^{\infty} f(x)\exp(ixy)dx\; ,
\end{equation}
and its inverse transform
\begin{equation}\label{iFourtr}
 f(x)=\frac{1}{2\pi}\int_{-\infty}^{\infty}\tilde{f}(y)\exp(-ixy)dy\; .
\end{equation}
It should be hold in mind that different normalizations are used in the
literature -- e.g. ${\cal F}[f](\sigma)=\int f(x)\exp(2\pi ix\sigma)dx$
(Gray 2005, p. 26), for which the inverse transform is
$f(x)=\int {\cal F}[f](\sigma)\exp(-2\pi ix\sigma)d\sigma$, or
${\cal F}[f](y)=\tilde{f}(y)/\sqrt{2\pi}$, which has the same
normalization factor at its inverse. Our asymmetric choice corresponds
to the practice in use of the numerically effective algorithm of
Fast Fourier Transform (FFT).

 Obviously, Fourier transform of a real function has an even (symmetric)
real part and odd (antisymmetric) imaginary part
\begin{equation}\label{Foursym}
 \tilde{f}(-y)=\tilde{f}^{\ast}(y) \; ,
\end{equation}
where the asterisk $^{\ast}$ denotes the complex conjugate
($(a+ib)^{\ast}\equiv a-ib$, for $a,b$ real).
Fourier transform is linear (with respect to the addition of functions
and their multiplication by a complex constant).
Rescaling of a function in the $x$ space corresponds to the inverse
rescaling in the $y$ space
\begin{equation}\label{Fourtrs}
 \tilde{f}(\frac{y}{a})=a\int_{-\infty}^{\infty} f(ax)\exp(ixy)dx\; .
\end{equation}

 According to the Parseval theorem, the norm of a function (defined
by $||f||^{2}\equiv\int f^{\ast}(x)f(x)dx$) is proportional to the norm
of its Fourier transform
\begin{eqnarray}\label{Parseval}
 \int|\tilde{f}(y)|^{2}dy&=&
 \int\tilde{f}^{\ast}(y)\int\exp(ixy) f(x)dxdy=
 \int f(x)\left[\int\tilde{f}(y)\exp(-ixy)dy\right]^{\ast}dx=\nonumber\\
 &=& 2\pi\int|f(x)|^{2}dx
\end{eqnarray}
(i.e. the norm $||f||$ is preserved up to the multiplicative constant
$\sqrt{2\pi}$ in our normalization). Consequently, the Fourier transform
also preserves (up to the same constant) the scalar product of functions
defined by $(f,g)=(||f+g||^{2}-||f||^{2}-||g||^{2})/2
=\int(g^{\ast}f+f^{\ast}g)/2dx$.

 It is crucial for the application on the disentangling of spectra
that for a convolution of two functions defined by
\begin{equation}\label{convol}
 h(x)\equiv[f\ast g](x)\equiv\int_{-\infty}^{\infty}f(x-v)g(v)dv \; ,
\end{equation}
the Fourier transform is a simple product of the transforms
\begin{eqnarray}
 \tilde{h}(y) &=&\int\int f(x-v)g(v)dv\exp(ixy)dx= \nonumber\\
   &=& \int f(x-v)\exp(i(x-v)y)d(x-v)\int g(v)\exp(ivy)dv
   =\tilde{f}(y).\tilde{g}(y) \; .\label{Fconvol}
\end{eqnarray}
Another useful property of the Fourier transform concerns the derivatives,
\begin{equation}\label{dFourtr}
 \tilde{f'}(y)=\int \frac{df}{dx}\exp(ixy)dx=-iy\int f(x)\exp(ixy)dx
 =-iy\tilde{f}(y)\; .
\end{equation}

 Fourier transforms for some simple cases can be expressed explicitely
as follows:\\
shifted Dirac delta function $\delta_{v}(x)\equiv\delta(x-v)$
\begin{equation}\label{FourtrDir}
 \tilde{\delta}_{v}(y)=\int \delta(x-v)\exp(ixy)dx=\exp(ivy)\; ,
\end{equation}
Gaussian function $G_{a}(x)\equiv\exp(-x^2/a^2)$
\begin{equation}\label{FourtrGaus}
 \tilde{G}_{a}(y)=\int \exp(-x^2/a^2)\exp(ixy)dx= a\sqrt{\pi}\exp(-a^2y^2/4)\; .
\end{equation}

The Fourier transform can be easily generalized to complex functions
on spaces of higher dimensions $n$
\begin{equation}\label{Fourtrn}
 \tilde{f}(\vec{y})=\int f(\vec{x})\exp(i\vec{x}\vec{y})d^{n}x\; ,
 \hspace{5mm}
 f(\vec{x})=(2\pi)^{-n}\int \tilde{f}(\vec{y})\exp(-i\vec{x}\vec{y})d^{n}y\; .
\end{equation}
Obviously, generalizing Eq.~(\ref{Fourtrs}), a linear transform in
$\vec{x}$ space corresponds to an inverse transform in the $\vec{y}$ space.
In particular a rotation in a plane for an angle $\varphi$ in one space
corresponds to the rotation for $-\varphi$ in the other space. This is
important for the Radon transform and its inverse.

\subsection{Radon transform}
Let us assume a function $f(x_{1},x_{2})$ in Cartesian coordinates in plane.
Its perpendicular projection (in the direction of $t$-axis)
\begin{equation}\label{Radontr}
 p(s,\varphi)\equiv{\cal R}[f](s,\varphi)
 =\int_{-\infty}^{\infty}f(s\cos\varphi-t\sin\varphi,s\sin\varphi+t\cos\varphi)dt
\end{equation}
to a line skewed for angle $\varphi$ and parametrized by its length $s$
is the Radon transform of $f$ to variables $\{s,\varphi\}
|_{s=-\infty}^{\infty}|_{\varphi=0}^{\pi}$. Obviously
\begin{equation}
 p(s,\varphi+\pi)=p(-s,\varphi)\; ,
\end{equation}
which is a condition for existence of $f$ satisfying Eq.~(\ref{Radontr}).
In the case of Doppler tomography it means that the line profiles must
be reversed in the opposite orbital phases if they result from a simple
projection of the same velocity distribution of the emitting matter.
For a fixed $\varphi$, the integral $p(s)$ is the zeroth Fourier mode
in the variable $t$ of the function $f(x_{1},x_{2})$ rotated into the
coordinates $s,t$. This implies the so called Fourier slice theorem
according to which the one-dimensional Fourier transform
$\tilde{p}(u,\varphi)=\int p(s,\varphi)\exp(isu)ds$
of the Radon transform of $f(x_{1},x_{2})$ is a cross-section of
the two-dimensional Fourier transform $\tilde{f}(y_{1},y_{2})$
\begin{equation}\label{sliceteor}
 \tilde{p}(u,\varphi)=\tilde{f}(u\cos\varphi,u\sin\varphi)\; .
\end{equation}
The existence theorem for inverse Radon transform follows from here,
as well as a possibility how to reconstruct the two-dimensional
distribution of the function $f(x_{1},x_{2})$ from a (sufficiently
rich) set of its tomographic projections.

\subsubsection{Back projection}
The so called back-projection can be defined as
\begin{equation}\label{backproj}
 {\cal R}^{\ast}[p](x_{1},x_{2})
 =\int p(x_{1}\cos\varphi+x_{2}\sin\varphi,\varphi)d\varphi\; ,
\end{equation}
which approximates (but it is not) the inverse Radon transform --
obviously if the source is a single point
\begin{equation}
 f=\delta(x_{1}-s_{0}\cos\varphi_{0},x_{2}-s_{0}\sin\varphi_{0})\; ,
\end{equation}
then its projections are $\delta$-functions shifted to follow an S-wave
\begin{equation}
 p(s,\varphi)=\delta(s-s_{0}\cos(\varphi-\varphi_{0}))\; ,
\end{equation}
but its back-projection reads
\begin{equation}
 {\cal R}^{\ast}[p](x_{1},x_{2})
 =((x_{1}-s_{0}\cos\varphi_{0})^{2}
 +(x_{2}-s_{0}\sin\varphi_{0})^{2})^{-\frac{1}{2}}\; .
\end{equation}
It means that it is peaked around the true position of the source
$f$, but with a point-spread function inversly proportional to the
distance from its centre. In a numerical application, where the integral
$\int d\varphi$ in Eq.~(\ref{backproj}) is replaced by a sum over a
finit number of available projections, the back-projection has a shape
of a asterisk with rays in directions of individual projections. Yet
${\cal R}^{\ast}$ is a good approximation to ${\cal R}^{-1}$ for
an iterative tomographic reconstruction.

\subsubsection{Filtered back projection}
Substituting for $\tilde{f}$ from the Fourier slice teorem (\ref{sliceteor})
into the inverse Fourier transform (\ref{Fourtrn}), we can reconstruct
the image of $f$ precisely in the form
\begin{equation}\label{fbackproj}
 f(\vec{x})=(2\pi)^{-2}\int \tilde{p}(u,\varphi)
  \exp(-iu(x_{1}\cos\varphi+x_{2}\sin\varphi))udud\varphi\; .
\end{equation}
This integral in polar coordinates $\{u,\varphi\}$ in the plane of $\vec{y}$
differs from the simple back-projection (\ref{backproj}) by the presence
of the multiplicator $u$ at $du$, which is a filter suppressing the
low-frequency modes of $p$ and enhancing the high-frequency modes.

\subsubsection{Iterative least-square technique}
Suppose now that we have an object consisting of $N=N_{1}\times N_{2}$
bins in coordinates $(x_{1},x_{2})$ (labeled by index $i=(i_{1},i_{2})$),
each one emitting with a rate $f_{i}$.
This emission is registered in $K_{1}$ pixels of the variable $s$
at each of $K_{2}$ tomographic projections in different angles
$\varphi$ (labeled by $k=(k_{1},k_{2})$),
which detect a signal linearly dependent on the emission of source bins
\begin{equation}\label{ILST0}
 p_{k}=\sum_{i}W_{k}^{i}f_{i}
\end{equation}
with some weighting factors $W_{k}^{i}$. Having an estimate of $f_{i}$,
we solve for its correction $\Delta_{i}$, which would minimize the
least-square difference between the observed and calculated signal
\begin{equation}\label{ILST1}
 0=\delta \sum_{k}\left[p_{k}-\sum_{i}W_{k}^{i}(f_{i}+\Delta_{i})\right]^{2}
\end{equation}
with respect to any variation $\delta\Delta_{j}$. This condition
yields
\begin{equation}\label{ILST2}
 \sum_{i}(\sum_{k}W_{k}^{j}W_{k}^{i})\Delta_{i}
 = \sum_{k}W_{k}^{j}\left[p_{k}-\sum_{i}W_{k}^{i}f_{i}\right]\; .
\end{equation}
Regarding Eq.~(\ref{backproj}), the solution of Eq.~(\ref{ILST0})
can be approximated as
\begin{equation}\label{ILST3}
 f_{j} \sim \sum_{k}W_{k}^{j} p_{k}=\sum_{i}\sum_{k}W_{k}^{j} W_{k}^{i}f_{i}\; ,
\end{equation}
which means that
\begin{equation}\label{ILST4}
 \sum_{k}W_{k}^{j} W_{k}^{i}\sim\delta_{ij}\; .
\end{equation}
Consequently, the solution of Eq.~(\ref{ILST2}) with respect to the
correction $\Delta_{i}$ can be estimated as
\begin{equation}\label{ILST5}
 \Delta_{j}\simeq\frac{1}{\sum_{k}(W_{k}^{j})^{2}}
  \sum_{k}W_{k}^{j}\left[p_{k}-\sum_{i}W_{k}^{i}f_{i}\right]\; .
\end{equation}
In practice, this estimate is multiplied by a damping factor $\epsilon<1$
to prevent oscillatios of the iterative procedure $f_{i}\rightarrow
f_{i}+\epsilon\Delta_{i}$.

\subsection{Moments of functions}
For a function $f(x)$ its moments can be defined as
\begin{equation}\label{mom1}
 f_{n}\equiv\int x^{n}f(x)dx\; .
\end{equation}
For instance, for a Gaussian function $G(x)\equiv\exp(-x^2/a^2)$ (centered
around $x=0$)
\begin{equation}\label{mom2}
  G_{2n}=\frac{(2n)!\, a^{2n}}{n!\, 2^{2n}}G_{0}\; ,\hspace*{10mm}G_{2n+1}=0\; .
\end{equation}
For convolution of two functions
\begin{eqnarray}\nonumber
 (f\ast g)_n&=&\int x^{n}f(x-y)g(y)dydx
     =\sum_{k=0}^{n}\frac{n!}{(n-k)!k!}\int (x-y)^{n-k}f(x-y)dx y^{k}g(y)dy\\
    &=&\sum_{k=0}^{n}\frac{n!}{(n-k)!k!}f_{n-k}g_{k}\; .\label{mom3}
\end{eqnarray}
If we define center of a function (representing, e.g., line-profile or
a distribution function)
\begin{equation}\label{mom4}
 \bar{x}_{f}\equiv \frac{f_{1}}{f_{0}}\; ,
\end{equation}
then
\begin{equation}\label{mom5}
 \bar{x}_{f*g}=\bar{x}_{f}+\bar{x}_{g}\; .
\end{equation}
Similarly, for a squared width of $f$ defined by
\begin{equation}\label{mom6}
 \Delta^{2}_{f}\equiv\frac{1}{f_{0}}\int (x-\bar{x})^{2}f(x)dx =
 \frac{f_{2}}{f_{0}}-\bar{x}^{2} \; ,
\end{equation}
it is valid
\begin{equation}\label{mom7}
 \Delta^{2}_{f*g}=\Delta^{2}_{f}+\Delta^{2}_{g}\; .
\end{equation}
Following Eq.~(\ref{mom2}), the fourth moment of the Gaussian function
$G_{4}=3G_{2}^{2}/G_{0}$. If we characterize a non-Gaussianity of
centered (i.e. $f_{1}\equiv 0$) function $f$ by ratio
\begin{equation}\label{mom8}
 R_{f}\equiv\frac{f_{4}f_{0}}{3f_{2}^{2}}
\end{equation}
(which is equal to 1 for Gaussian function), then
\begin{equation}\label{mom9}
 R_{f*g}=
 \frac{R_{f}\Delta^{4}_{f}+R_{g}\Delta^{4}_{g}+2\Delta^{2}_{f}\Delta^{2}_{g}}
 {(\Delta^{2}_{f}+\Delta^{2}_{g})^{2}} \; .
\end{equation}


\clearpage
\section{Manual for users of KOREL}\label{man}

\subsection{Implementation of the code}
The code KOREL is written in FORTRAN 77. It's PC-version
(KOREL.FOR) includes on-line graphics (the package PHG.FOR),
which is written for MicroSoft-fortran. The larger LINUX-version
(KOREL.F) contains analogous package with graphical output to files
(HPGL- or PS-) only and could thus be compiled by any fortran
compiler. Starting from release of May 2004 the LINUX-version
needs also the file KORELPAR.F, where the maximum number of
spectra $nsp$ and pixels (bins) $npx$ are given as parameters.
Note that $npx2=2*npx$ is also required as the array dimension for
the complex representation of Fourier transforms of the spectra.

 The code PREKOR for preparing the data for KOREL exists only in
MS-version, because the use of on-line graphics is crucial for its use.

\subsection{Controlling the run (file {\tt korel.par})}\label{ContR}
KOREL is controlled by the file {\tt korel.par}, from which there are
read (in free format, implicit fortran definition of type is
valid):
\begin{enumerate}
\item Control keys
$$KEY(j)|_{j=1}^{5},K0,IFIL,KR,KPR\; .$$
{\begin{itemize}
\setlength{\itemsep}{-1mm}
\item Key $KEY(j)$ defines if the lines of the star No.~$j$
(according to Fig.~\ref{obr1}) are present in the spectrum
($KEY(j)\ge 1$) or not. If $KEY(j)$ is split into digits
 $KEY(j)=10\times K_{1}+K_{0}$, then $K_{0}=1$ means that the
line-strength of the component $j$ is fixed, while $K_{0}=2$ means
that the intensity of this component is to be solved according
to Eq.~(\ref{sjl}). $K_{1}=0$ means that the radial velocities
of the component $j$ are subjected to the orbital motion
according to Eq.~(\ref{RV}), while $K_{1}=1$ means that its
radial velocities are free parameters, either fixed or
converged.

\item Key $K0$ specifies if there should be read data from the
file {\tt korel.dat} ($K0>0$), or if the calculation should go on with
the previous data ($K0=0$), or (for $K0<0$) the number ($-K0$) of
spectra is to be simulated.

\item $IFIL$ is the number of harmonics to be removed by
filtering.

\item $KR$ is the key of the form of graphic output {\tt phg.out}
($KR=0\Rightarrow$ no output,
$KR=1\Rightarrow$ PCX-format,
$KR=2\Rightarrow$ PostScript-format).

\item $KPR$ controls the level of output.
The value of $KPR$ is to be split into digits,
$KPR= 10\times KPR_{1}+KPR_{0}$.
Then $KPR_{0}>1$ specifies that the information on simplex
convergence has to be printed also into the file
{\tt korel.res}. $KPR_{1}>0$ specifies that the file {\tt korel.o-c} has
to be created, into which the difference spectra O~--~C will be
written in the wavelength scale connected with the star
No.~$KPR_{1}$ (or in the original wavelength scale if $KPR_{1}>5$).
\end{itemize}}

\item Next, there are read lines defining (initial) values
of the free parameters, their (initial) steps for convergence etc.
These lines have the form:
$$c,j,i,Kc,L1,L2,EL(j,i),\Delta(j,i)\; .$$
{\begin{itemize}
\setlength{\itemsep}{-1mm}
\item Here the character $c$ distinguishes the kind of the parameter and
the indices 'j' and 'i' specify which one parameter of the kind is meant.
The character $c$ is equal either to 'o', 's', 'v', 'w' or 'e' (some
other values are to be involved in future versions, other yet non-specified
letters are interpreted as equivalent of 'o').
The letter\\
'o' stands for orbital parameter number $i|_{i=1}^{11}$
(cf. Table~\ref{tab1}) of orbit $j|_{j=0}^{3}$ (cf. Fig.~\ref{obr1}).
The letter\\
's' denotes the strength of lines of component (or telluric lines) $j$
in the exposure number $i$, i.e. the natural logarithm\footnote{ This
 natural-logarithmic scale  of strengths is close to the traditional
 magnitude-scale (because ln$x\simeq 0.434\times\log x$), however, a more
 positive value means a more intensive contribution of the component
 spectrum. These logarithmic values are normalized on output in their
 exponentials, i.e. $\sum_{l}s_{jl}= \sum_{l}1$ (such a normalization
 in intensities seems to be more stable than that in (pseudo-)magnitudes,
 $\sum_{l}{\rm ln}s_{jl}= 0$, used in KOREL up to the year 2002).}
of the line-strengths, $EL(n)={\rm ln}s_{ji}$.
Similarly,\\
'v' denotes the velocity of component $j$ in exposure $i$ (for the case
 $KEY(j)\ge 11$) and\\
'w' enables to change the weight of exposure $i$ ($j$ being ignored).\\
'e' denotes scanning of $S(p)$ given by Eq.~(\ref{FKor3}) in a
two-dimensional cross-section of the $p$-space for calculation of errors
of the orbital parameters specified by $j$ and $i$ like for the letter 'o'.\\
The output of these parameters has the same form (with explanation of
meaning of the parameters at right on each line), so that the output
values can be cut from {\tt korel.res} and copied to the input file
{\tt korel.par} for next run of the code. The input of these lines
is performed in a closed loop until a line with all values equal to zero
is read (the letter may be arbitrary).
\item $Kc>0$ specifies if the above selected element is to be converged
(or, for 'e', the scanning to be performed and saved in the output-file
{\tt korermap.dat}).
The maximum of $Kc$ for all elements gives the number of large
iteration steps.
\item Keys $L1$, $L2>0$ specify if the quantities $EL(n)$ and $\Delta(n)$
are to be read.
\item $EL(n)$ and $\Delta(n)$ are the (initial) value of the element and
its step (some numbers must be present even if they are ignored due to
$L1=0$ and/or $L2=0$). In the case of letter 'e' these values are the
half-widths of value intervals to be scanned on the $x-$ and $y-$ axis, resp.
\end{itemize}}

\begin{table}
\begin{center}
\caption{ Numbering of orbital elements\label{tab1}}
\begin{tabular}{rll}\hline
i&\multicolumn{2}{l}{orbital element}\\ \hline
1&$P$& period [in days]\\
2&$t_{0}$& time of periastron passage [in days]\\
3&$e$& eccentricity\\
4&$\omega$& periastron longitude [in degrees]\\
5&$K$& semiamplitude of radial velocity of the component with
    the lower index [in km/s]\\
6&$q$& the mass ratio of the component with the higher to
    that with the lower index\\
7&$\dot{\omega}$& the rate of periastron advance [in degrees/day]\\ \hline
8&$\dot{P}$& the time derivative of the period\\
9&$\dot{e}$& the time derivative of the eccentricity [in day$^{-1}$]\\
10&$\dot{K}$& the time derivative of $K$-velocity [in km/s/day]\\
11&$\dot{q}$& the time derivative of mass ratio [in day$^{-1}$]\\
\hline
\end{tabular}
\end{center}
\end{table}

\item If $K0<0$ then there are next read $-K0$ values of time,
for which the data should be simulated. Next there must be given
for each visible component its central intensity $c$, its width
$\Delta$ [km/s] and the code 1 or 2 specifying if the profile
should be given by Eq.~(\ref{iin1}) or (\ref{iin2}). Finally
the noise of the spectrum and radial velocity per bin is read
for the simulated data.
\end{enumerate}

 The following example of {\tt korel.par} for the LINUX version:\\
\begin{verbatim}
1 1 0 0 2 1 0 2 2 | key(1,...,5), k= Nr. of sp., filter, plot, print
o 0 1 0 1 1  1.234567 0.000001 |  sum= 9.8765
o 0 2 1 1 1  50001.000 .1 |
o 0 3 0 1 1   0.00 .1 |
o 0 4 0 1 1  90.00 1. |
o 0 5 1 1 1  120.0 .5 |
o 0 6 1 1 1  .5 .1 |
o 3 1 0 1 1    365.256360000     0.100 = PERIOD(3)
o 3 2 0 1 1  51547.520600000    10.000 = PERIASTRON EPOCH
o 3 3 0 1 1      0.016710220     0.001 = ECCENTRICITY
o 3 4 0 1 1    301.795199910    10.000 = PERIASTRON LONG.
o 3 5 0 1 1      0.001000000     0.000 = K1
o 3 6 0 1 1      0.000038185     0.000 = M2/M1, K2 = 26.188293833
o 3 7 0 1 1      0.000009111     0.000 = d omega/dt
o 3 8 0 1 1      0.000000000     0.100 = d P/dt
s 5   1 0 1 1 -0.15890 0.10000
s 5   2 0 1 1  0.07530 0.10000
s 5   3 0 1 1  0.10141 0.10000
s 5   4 0 1 1  0.20137 0.10000
s 5   5 0 1 1  0.18172 0.10000
s 5   6 0 1 1  0.04557 0.10000
s 5   7 0 1 1  0.05994 0.10000
s 5   8 0 1 1 -0.10638 0.10000
s 5   9 0 1 1 -0.17056 0.10000
x 0 0 0 0 0 0 0 | end of elements
\end{verbatim}
will thus converge periastron epoch (identical with the maximum of
radial velocity)
and $K_{1,2}$ of a two-component binary (on circular orbit with
period 1.234567 days) and solve for strengths of telluric lines
(without filtering, with PS- output graphics and a detailed
output protocol).

\subsection{Input data (file {\tt korel.dat})}\label{koreldat}
In the file {\tt korel.dat} there is read the time [in Julian dates],
initial wavelength [in \AA], radial velocity per one bin [in km/s], the
weight of each spectrum and (starting from release of 2003)
also the number $npx$ of pixels (bins) in each exposure.\footnote{
 A frequent mistake done by users is to run older versions of KOREL compiled
 for a fixed value of $npx$ with data prepared for its different value.
 For this reason the value of $npx$ is free in the newer LINUX versions,
 but it must be given in the data itself. The data without this value
 prepared by an older versions of PREKOR can be modified to the new form
 using the code `KORTRANS.F' (cf. Section~\ref{KORTRANS}).}
It must be the same for all exposures and equal to 256 for the PC-version,
while for the Unix-version it can be any multiple of 256 by number of
the form $2^{n},\; n\ge 0$, up to the maximum value of
$npx$ defined by the parameter $npx$ in the file KORELPAR.F.
Next, there follow $npx$ intensities in points with constant step
in radial velocity. There can be read at maximum 27 input spectra
for the PC-version or any number up to $nsp$ defined in KORELPAR.F for
the LINUX-version. These spectra are to be given in up to $mnu$ spectral
regions ($mnu=5$ in the PC-version), each region being characterized
by the initial $\lambda$ and by the step of radial velocity per bin.
Naturally, the number
of input spectra in each region and their phase distribution must
be sufficient for their decomposition, i.e. it must at least higher
(or equal) than the number of calculated components. The number
of decomposed spectra, i.e. the product of number of spectral regions
$\times$ number of components must not exceed the value $mnsu$ given
together with $npx$, $nsp$ and $mnu$ in the file KORELPAR.F (or
fixed to 15 in the PC-version).

 The file {\tt korel.dat} can be prepared by the code PREKOR.

\subsection{The code PREKOR}
The code PREKOR has been developed to facilitate the preparation
of data for KOREL, it means to cut the proper spectral regions
from a set of input spectra, to interpolate them into the
equidistant logarithmic wavelength scale and to write them
in the format required by KOREL in the file {\tt korel.dat}. Because
a visual check of the proper choice of spectral regions is
welcome, the PREKOR is written in MS-FORTRAN only and the
distributed exe-version has to be run under DOS, even if the
data are intended for a LINUX-version of KOREL. A version of
PREKOR with PG-PLOT graphics is in preparation, however some
problems with input of binary data makes it platform-dependent.
Users without access to DOS- (or WINDOWS-) computers have to
produce the {\tt korel.dat} input using some other facilities (e.g.
to select proper regions using MIDAS or IRAF and to rewrite
them into the required format by some user-written code).

 As an input, PREKOR needs a list of input spectra in file
{\tt prekor.lst} and it produces the data for KOREL in its (newly
created) output file {\tt prekor.out}, which has then to be renamed
as {\tt korel.dat}. Versions of PREKOR starting from October 2001
enable also to prepare parameters for disentangling the telluric
lines, in which case an input file {\tt prekor.par} is needed and
the output is written as {\tt prekor.res}.

 When started, PREKOR asks the user at first to choose the mode
of calculation (or the type of KOREL data to be got).
The choice of $mode$ = 0 performs the above mentioned calculation
of parameters for disentangling of telluric lines.
The choice $mode>0$ means to prepare data with the number
$npx=128\times 2^{mode}$ bins in each spectral region. It means that
$mode=1$ is required for the PC-version of KOREL, $mode=2$
for the older LINUX-version and both or some higher integer number
(up to a limit given by the array dimensions in the PREKOR code)
can be chosen for the version with KORELPAR.F- file.

For $mode >0$, PREKOR reads the description of files with
individual exposures from the file {\tt prekor.lst}.
Older versions of PREKOR stopped after finishing the work with
the first 30 spectra from {\tt prekor.lst} and a rearrangement of
this input file was needed to continue with the subsequent
spectra (and to concatenate the corresponding output {\tt prekor.out}
files). Starting from versions of January 2004, PREKOR continues
to cut out from subsequent spectra the region chosen according to
the first displayed spectra and offers them for saving into output.

 The file {\tt prekor.lst} must contain on each line
the name of file with individual exposure, its Julian date,
weight, a code of the type of data file and the value of its
shift in RV's (in the format {\tt a12,f11.4,f8.3,i2,f8.3}).
If the weight is negative, the data file is ignored.\\
The code = 0 denotes ASCI data files with wavelength and
intensity in free format on each line; the first line is a comment.\\
The code =1 refers to files {\tt *.rui} and\\
the code = 2 to {\tt *.uui} of data in format SPEFO used formarly at Ond\v{r}ejov
observatory.\\
The code =3 corresponds to the modified MIDAS output where the first
three lines are a comment and then there follow lines with their sequence
numbers (which are ignored by PREKOR), wavelengths and intensities.\\
The code = 4 means reading of tables produced by IRAF (with headings
consisting of 106 records). Finally, for\\
the code =5, spectra
in FITS- format ($BITPIX=-32$) can be read.\\
The dimensions of arrays in the code limit the length of
the input spectrum to 4100 bins at maximum (the rest is ignored).
The only exception are the input files in the FITS-format (code
=5) where the preview of the whole spectrum is drawn using
averaged pixel-values and the data used for the calculations are
then read only starting from the required wavelength (the limit
of 4100 pixels is thus valid for the chosen spectral region only).
The first spectrum is read from its appropriate input file
and its preview is plotted on the screen. User is asked to insert
the initial wavelength and step in RV per bin.
The corresponding spectral region is marked by a different colour
on the wavelength scale and the chosen parts of spectra for the
first portion of exposures (with non-negative weights) are then
displayed to enable to check if their margins are really in the
continuum. Later versions of PREKOR enable also an approximate
rectification of the chosen spectral region consisting in
normalization by linear function joining the first and the last pixel
of the region.\footnote{ This option can help for quick inspection
 of non-rectified spectra by preventing the jumps between the margins
 of the regions. However, its use may be dangerous because of hiding
 possible spectral lines on margins and because of the influence of
 random noise in the ultimate pixels. A thorough rectification of
 the whole spectrum before the run of PREKOR is always preferable.}
If the result is satisfactory, it can be saved into file {\tt prekor.out}
and subsequent spectral regions can be chosen in an infinite loop
till the end of the run is not required.\footnote{ The commands for
 controlling the run of PREKOR differ depending on version, but they
 are always displayed on the screen and offered to the user.}
The final file {\tt prekor.out} can then be renamed and used as {\tt korel.dat}.
The shift in RV of input spectra can be defined in the last column
of {\tt prekor.lst} either to transform the data from observed to heliocentric
wavelength scale or to compensate for possible errors in wavelength-scale
of the input spectra. These may be measured according to telluric
lines either manually by some other mean or also by KOREL. A spectral
region rich for telluric lines can be chosen first for this purpose
and the column with O--C of RVs of telluric lines can be then copied from
the file {\tt korel.res} into {\tt prekor.lst} to reduce the scatter of RVs of
telluric lines in other spectral regions.

 The theoretical RVs of telluric lines can be predicted from the
coordinates of the observed star. To calculate the corresponding
fictitious orbital parameters of the telluric lines (usually
taken as the component No.~5 on the orbit No.~3), PREKOR can be
run in mod = 0. In this case an additional file {\tt prekor.par} must
be prepared in which are given (on separate lines) the right ascension
and declination of the star (in hours and minutes or degrees and
arc-minutes, resp.). On next line, the equinox of the source coordinates
is read. The following line gives the geographic longitude and latitude of
the telescope in degrees and its altitude above the sea level in meters.
As a result,
a block of lines with orbital parameters of the orbit are written into
the file {\tt prekor.res}, from where they can be directly copied into
{\tt korel.par}.
In addition, radial velocities of the telluric lines are calculated
for each exposure listed in {\tt prekor.lst} with higher precision taking
into account also the planetary perturbations and the rotation of
the Earth.

\subsection{The code REKTIF}\label{REKTIF}
In principle, the spectra of the components could be disentangled
as absolute values of specific intensities $I_{\nu}$ in the whole
range of frequencies $\nu$, if the observed spectra are also
given in the whole range as specific intensity with a sufficient
precision.\footnote{ In fact, this was the original Doppler's idea
 to explain observed colour differences between components of some
 binaries by their frequency shifts due to radial velocities.
 Despite the effect of colour changes is practically negligible
 for usual binaries, it exists and in principle its phase-dependence
 can enable to disentangle individual component's colours.
 Cf.~\ref{Dopplhist} at p.~\pageref{Dopplhist}}
However, in practice, the observed spectra mostly cover only a small
part of the whole spectral range and they are poorly calibrated in
flux.\footnote{ This problem is particularly difficult for echelle
 spectra, where also a smooth connection of subsequent orders is a
 non-trivial task.}
The orbital Doppler shifts, which are of the order $v/c$, i.e.,
typically bellow $10^{-3}$, are thus hardly discernible by
observations of the whole spectra, but they can be better
identified in relative shifts of spectral lines, the widths of
which are often comparable or smaller than the amplitude of
radial velocities. To investigate the Doppler shifts of spectral
lines and to reduce the influence of uncertainties in the
calibration of colour-dependence of spectrograph sensitivity,
the so-called rectified spectra are used, i.e. the observed
spectra normalized with respect to the estimated signal-value
in continuum. From the point of view of physics of stellar
atmospheres, the notion of continuum and hence also the procedure
of rectification is spurious. The idealization of a continuum is
useful if formation of a weak line on atoms immersed in
background photon field is considered. In real atmosphere,
however, the line-profiles often overlap and form the so-called
pseudocontinuum.

For disentangling, the rectification enables to diminish the
low Fourier modes, which can be influenced by the above mentioned
sensitivity effects, and it also removes (or decreases) possible
jumps between the left and right margin of each spectral region,
which could give rise to some unevenness on edges. From this
point of view, it is not much important if the continuum is
chosen for the rectification at a proper level. What matters
is at first the scale of wavelengths, on which the continuum
varies, and also its consistency for all observed spectra. The
scale of continuum variations should be larger compared to
the searched Doppler shifts and characteristic scales of
spectral features (lines) used to reveal the shifts. The
rectification should remove such large-scale variations. If
it fails to do it correctly or even if it introduces such
variations, the disentangling may produce an artifact in form
of long-scale distortions or waves in the continuum of component
spectra. These distortions caused by unreliable rectification
may even prevail in the O--C over the true orbital changes of
the spectra and the solution of orbital parameters may fail
in order to use these degrees of freedom in favour of correcting
the wrong rectification. The danger of this failure can be
escaped or minimized by filtering the low Fourier modes.

A common way of rectification is to draw the continuum in the
form of spline function defined by interactively chosen
points (e.g. in the code SPEFO, cf. \v{S}koda 1996; like in
SPEFO, the splines are calculated in REKTIF according to
G. Hill, 1982).
This `artistic' procedure depends on a skill of user and cannot
be homogeneously repeated in different spectra. To make it a bit
more reliable, the code REKTIF was written by the author for preparing
the data for KOREL. In this code, the marks defining the continuum are
recurrently improved. Starting with a set of marks with chosen
wavelengths and intensities, new mean values of both these
quantities are calculated for all data-points in the observed
spectrum, which fall into a wavelength interval of predefined
width around each mark and the intensities in which fit with the
present spline approximation of the continuum within a precision
limit chosen in a percentage of the continuum. The limit then
shrinks down slowly (by a geometric sequence typically with
quotient about 0.8) in subsequent iterations, so that the
intensity attributed to each mark converges to the mean of the
most populated (rectified) intensity values. These are supposed
to correspond to the (pseudo-)continuum, while the data-points
in lines within the interval are skipped from the mean, when the
limit decreases to the level of signal noise. An experience
shows, that this algorithm gives mostly a result consistent
with the `artistic insight', however, it is recommended to
check the result in a graphic form to improve the cases when
more marks are needed or when some of them converge to a wrong
value of intensity (this can happen due to inconvenient initial
approximation).

The input spectrum to be rectified is read by the code REKTIF
from a file named {\tt rektif.dat}. It is to be given in a form of
table with wavelength in the first and the intensity in the
second column. The number of lines should not exceed the value
of parameter $NPXM$ chosen at the beginning of the code
source-file. The table may contain several lines of comments
at the top, the number $NTXT$ of which is to be given together
with other parameters controlling the run of the code. These
parameters are read from file {\tt rektif.par}. The first line
should give subsequently $\lambda_{min}$, $\lambda_{max}$,
$I_{min}$, $I_{max}$ (i.e. the window to be displayed) and $NTXT$.
Next line gives number $NR$ of marks ($NR\le NRM$ also set as
a parameter in the code), number of iteration steps
and a quotient $q$ by which the range of acceptable errors of
intensity is shrinked in each iteration (value $q\simeq 0.9$
is proved in practice). Then initial values of $\lambda$, $I$,
$\Delta\lambda$ and maximum $\delta I/I_{cont.}$ are given
for each mark on a separate line. The resulting rectified
spectrum is given in the file {\tt rektif.out} and improved values
of marks in the file {\tt rektif.res} (the additional fifth column
gives the number of points taken into account for each mark).

\subsection{The code KORTRANS}\label{KORTRANS}
This code was originally written to transform the file {\tt korel.dat}
from DOS to UNIX format on computers, where the utility DOS2UNIX
is not available. The code PREKOR writes intensities for each
exposure by one command -- always 10 values in one line ended
by mark nonunderstandable to UNIX. The code KORTRANS can read
each line separately from these files renamed to {\tt kortrans.in}
and to write them into {\tt korel.dat} directly readable by KOREL.
Because the older version of KOREL did not require the number
$npx$ of bins in the heading of each exposure, KORTRANS can
update the old data by inserting this value given in file
{\tt kortrans.par}. This file has a single line with values $npx$,
$v_{1}$, $v_{2}$, $\Delta v$ in a free format
(e.g. ``512 0.02 0.98 0.005").
If the value $dv>0$, the spectral region of each input exposure
is transformed (filtered or tapered at edges -- Section~\ref{taper},
p.~\pageref{taper}) according to formula
\begin{equation}
 I_{out}(v)=1+(I_{in}(v)-1)/(1+\exp\frac{v_{1}-v}{\Delta v})
  /(1+\exp\frac{v-v_{2}}{\Delta v})\; ,
\end{equation}
where $v$ is the logarithmic wavelength rescaled to be 0 and 1
on both edges of the spectral regions. This tapering suppresses
smoothly (on a characteristic width $\Delta v$) spectral lines and
deviations of continuum from 1 at edges of the spectral region
(i.e. below $v_{1}$ and above $v_{2}$).

\subsection{Input data (file {\tt korel.tmp})}\label{koreltmp}
The template spectra for constrained disentangling as described in the
Section~\ref{constrai} may be read in the file {\tt korel.tmp}.
The data should have the form:
\begin{verbatim}
6510. 5. 5
1.0014
1.0021
0.9993
1.0006
...
\end{verbatim}
Here the first two numbers in the first line define (in the same way as
in the {\tt korel.dat}) the spectral region for which the template is
prepared and the last digit gives the number of component which should
have this spectrum. Next lines give the values of the intensity (in free
format, it means there may be more of them on the same line) for each
grid point of the input spectra.

\subsection{Outputs of KOREL}\label{outp}
There are several possible forms of output. During the run of the code
a basic information is shown on the screen. In the PC-version this
information includes graphical output on the screen. Simultaneously, the
figure can be written into the file {\tt phg.out}.\footnote{ To facilitate
 the next work with the graphics output, it is named {\tt phg.ps} in the
 case of PostScript format.}
Main information about the run and the results is contained in the
file {\tt korel.res}. Residual spectra can be stored in the file {\tt korel.o-c}.

 After the input of the data and control keys, the basic information on
the task (e.g. the number of data and number of parameters to be solved)
is summarized on the screen. Then the input spectra are drawn in green
colour (in the PC-version) subsequently from the top to the bottom of the
screen. During the iteration of parameters, a protocol on the simplex
procedure is shown on the screen and it can be directed also to
{\tt korel.res}.\footnote{ Cf. the use of key $KPR$ in Section~\ref{ContR}.}
In this protocol, the first number indicates the step (running up to
ten times the number of iterated parameters), letter A, B, C or D shows
the type of simplex operation and the following number the worst point
of the simplex (which is to be improved).\footnote{ See the FOTEL-manual
 (Hadrava 2004b) for a more detailed explanation of the protocol and its
 meaning.}
It is recommended to check if the first point is improved reasonably soon.
The opposite may indicate that there was chosen a too large initial
value of step of some parameter (usually of that, which is improved
first). Next in the line is written the highest value of the minimized
sum. It should be decreasing, however an increase is possible at the
shrinkage operation indicated by D.\footnote{ Obviously, the monotonous
 decrease of the sum can also be violated by an insufficient convergence
 of the line strengths (as explained on page~\ref{lsit}).}
Next there are typed the values of converged parameters, which help to
check the status of the convergence. After the end of the convergence,
the disentangled line profiles are shown at the bottom of the screen in
blue colour and their superpositions with velocities
corresponding to the solution of orbital parameters are plotted in blue
over the green input spectra. Finally, each spectrum is fitted by the
superposition of disentangled profiles with RVs independent of the orbital
parameters and the corresponding fit is drawn in red. The L-shaped line
in the figure indicates the wavelength unit (100km/s) and the unit of
intensity (0.1 of the continuum level).

 The file {\tt korel.res} yields most of the information about the
calculation and its results. First, there are summarized the
values of parameters characterizing the input of the task like
the number of spectra etc. Next, during the iteration, there is
written in each its step the number of the step, the value of the
sum $S$ of squares for the present values of orbital parameters,
its value after recalculation of the strengths of lines and the
mean error of the intensity. If the print-mode is $\ge 2$,
the protocol on the convergence by simplex-method is copied here
as well. After the iteration, the orbital parameters are printed
for each orbit with non-zero period. Next, all non-zero strengths
of lines are printed together with the final values of their
steps (which can be changed with respect to their input values
if the corresponding parameters have been converged by simplex).
Then the values of $S$ are repeated once more.
Next the spectra of individual components are printed (in the order
of columns corresponding to the order of component stars).
It must be kept in mind, that the values of continua of individual
components are unknown, so that the depths of lines are
normalized with respect to the sum all continua. The number 1.0
is added to each component spectrum to prevent the negative
values in absorption profiles. Each spectral region is introduced
by information on the number of exposures contributing to it and
by the mean value of their shifts of continua. Finally, for each
exposure the radial velocities found as the best superposition
of the decomposed spectra are printed. The order of the
components corresponds to the previously used order of the
output spectra and, in addition, it is indicated by the number
of the component. Each radial velocity is followed by the value
of O--C, where C corresponds to the value according to the
final values of orbital parameters.

 The file {\tt korermap.dat} provides a matrix of $20\times20$
values of the (O--C)$^2$ scanned at a two-dimensional cross-section
of the space of orbital parameters. This matrix may be read and
the dependence drawn by the code ISOLIN.

\subsection{The code KORNOR}
To facilitate the normalization of spectra disentangled by KOREL
a simple code named KORNOR has been written (in MS-Fortran).
This code reads the parameters $par$, $mi$, $npx$, $nsp$ and
$kr$ from the first line and $C_{j}|_{j=1}^{nsp}$ from the second
line of the file {\tt kornor.par} and the disentangled spectra cut out
from the file {\tt kornor.dat} in the form of table
$\lambda_{i},I'_{1,i},...,I'_{nsp,i}|_{i=1}^{npx}$ (for each
spectral region separately).
KORNOR draws first these input spectra on the screen in pale
colours and with the unknown continua shifts as they are
received from KOREL (i.e. the mean intensity shifted to the
level = 1). It also draws (in yellow colour) an example composed
spectrum with zero Doppler shifts of components and the proper
shift of the continuum given by the parameter $mi$ which is the
mean intensity given for each spectral region in the output of
KOREL. The continuum is then identified in this spectrum and
marked by red line at the level = 1. It is defined
to be in frequencies where the spectrum differs from 1 for less
than a multiple $par$ (which has to be established by the method
of trials and errors, but typically can be chosen around 0.2) of
the noise of the spectrum. At these points the integrals at continua
of each component are calculated and the continuum shifts are
found according to Eq.~(\ref{E12}). The rectified component
spectra are finally rescaled, i.e. calculated by Eq.~(\ref{E9})
using continua values $C_{j}$ given at the file {\tt kornor.par} and
they are drawn on the screen and written on the output.

\subsection{Problems with KOREL}
In this Section some hints, explanations to problems and answers
to often asked questions of users will be given.

{\it Incomprehensible list input in korel.dat}: This error
message and abort at the beginning of the run of KOREL appears
when the file {\tt korel.dat} is prepared by PREKOR on a DOS operating
computer and it is transmitted to a LINUX operating computer for
the run of KOREL. It is due to the incompatibility of the ends
of lines between these systems. It can be corrected using the
utility DOS2UNIX on the file {\tt korel.dat} or using the code
KORTRANS, which can yield some additional corrections to
the input data.

{\it Array bounds exceeded in subroutine CSWAP}: This subroutine
is called to exchange two columns of a matrix in the solution
of a set of linear equations by the subroutine CGSEV accepted
from the LAPACK package of FORTRAN routines. It uses a standard
trick to speed up matrix operations avoiding the lengthy
straightforward calculation of position in computer memory for
each element of a multidimensional array by a use of its
equivalence with a vector. The dimension of this vector is declared
only formally in this subroutine (and it is underestimated) and
the proper bound of the array is ensured by the correct demands
of the higher subroutine. Array bounds checking of the FORTRAN
compiler should thus be switch off.

{\it Undulation of disentangled spectra}: Sometimes it happens
that the disentangled spectra have large amplitude and long wavelength
wavy perturbations in antiphase (which cancel in the sum of all
components). Such errors appear quite often at the beginning of
the spectra decomposition and in favourable case (but not always)
they are suppressed in the course of convergence of the orbital
and other free parameters. This feature is due to a bad definition
of lower Fourier modes by Eq.~(\ref{FKor3}); the continua of components
cannot be distinguished, because they are not affected by the
orbital motion at all, and the lower modes are affected only slightly,
hence they are determined only poorly --- the corresponding matrix
is nearly singular and a small error (e.g. in rectification of input
spectra) can result in exaggerated values of amplitudes of the modes.
This danger increases if the disentangled regions are long compared
to the real RV- amplitudes. The contributions of these poorly determined
modes to the total minimized sum $S$ can be still small, so that the
solution of higher Fourier modes and the free orbital and others
parameters can be still correct. In such a case, it is sufficient
to rectify the disentangled spectra once more.\footnote{ A simultaneous
 rectification of all components at once with the condition of mutual
 cancelling of the corrections would be desirable.}
However, there is a real danger that the RVs or line-strengths will
converge to a false minimum in which the source of the low modes will
be fitted best and the true spectral lines encrypted in higher modes
will be ignored. Several actions can be undertaken to prevent this
failure. Firstly, user should check if all exposures are correctly
rectified\footnote{ Problems may arise also from overlaps of orders
 in echelle spectra.}
and either to improve or to remove suspicious exposures. The use
of filtering out the lower modes (by the key $IFIL$ described in
Section~\ref{ContR}) may help to hit the proper solution, in which
finally also the undulation may be suppressed. A better initial
estimate of orbital parameters or line-strengths (e.g. from other
spectral regions, from some other sources or simply found by chance
from numerous trials) may also be helpful.

\clearpage
\section{List of objects studied with Fourier disentangling}
The following list gives observational papers, in which the Fourier
disentangling is cited, which mostly means that their results were
achieved with KOREL, but in some cases also that the authors failed
in its use. In both cases these papers may be of some interest for
future studies of the same or similar systems.\\[3mm]
\begin{tabular}{ll}
A\&A 309, 521 (96)& 55 UMa\\
A\&A 315, L401 (96)& $\beta$ Cep\\
A\&A 319, 867 (97)& V436 Per \\
A\&A 322, 565 (97)& $\beta$ Sco \\
A\&A 327, 551 (97)& 4 Her \\
ARep 41, 630 (97)& $o$ Per \\
CoSka 27, 41 (97)& AR Aur \\ 
ARep 42, 312 (98)& V373 Cas\\ 
A\&A 331,  550 (98)& $\chi^2$ Aur \\
A\&A 332,  909 (98)& SZ Cam  \\
A\&A 345,  531 (99)& V606 Mon \\
A\&A 345,  855 (99)& AR Cas \\
A\&A 353, 1009 (00)& $\beta$ Lyr \\
A\&A 358,  553 (00)& V578 Mon \\
ASCP 214, 697 (00)& $\beta$ Lyr \\
IAUS 200, 109 (00)& V578 Mon\\
A\&A 366,  558 (01)& QZ Car \\
A\&A 377,  104 (01)& $\psi^{2}$ Ori \\
AJ 123, 988 (02)& $o$ Leo \\
ApJ 564, 260 (02)& HV 982 (LMC)\\
ApJ 574, 771 (02)& EROS 1044 (LMC)\\
MNRAS 330, 288 (02)& $\alpha$ Equ\\
A\&A 405, 1087 (03)& V497 Cep \\
A\&A 408, 611 (03)& V436 Per \\
ApJ 587, 685 (03)& HV 5936 (LMC)\\
A\&A 419, 607 (04)& $\kappa$ Dra \\
A\&A 422, 1013 (04)& $\kappa$ Sco \\ 
A\&A 427,  581 (04)& $\lambda$ Sco\\
A\&A 427,  593 (04)& $\lambda$ Sco\\
ASCP 318, 103 (04)& b Per\\
ASCP 318, 114 (04)& RV Crt\\
\end{tabular}\hfill
\begin{tabular}{ll}
ASCP 318, 338 (04)& HD 140873, HD 123515\\
ASCP 318, 342 (04)& DG Leo \\
IAUS 215, 166 (04)& 66 Oph \\
IAUS 224, 923 (04)& $\alpha$ Dra, HD 116656 \\
MNRAS 349, 547 (04)& V615 Per, V618 Per \\
A\&A 432,  955 (05)& $\kappa$ Sco \\
A\&A 439,  309 (05)& V578 Mon \\
A\&A 440,  249 (05)& $\varepsilon$ Lup \\
ApJ 623, 411 (05)& TT Hyd \\
ApSpSc 296, 169 (05)& o And \\
ApSpSc 296, 173 (05)& $\kappa$ Dra \\
ASCP 333, 211 (05)& 55 UMa \\
MNRAS 356, 545 (05)& DG Leo \\
A\&A 446, 583 (06)& $\varepsilon$ Per \\
A\&A 455, 1037 (06)& V360 Lac \\
A\&A 455,  259 (06)& $\beta$ Cen \\
CoAst 148, 65 (06)& $\theta^2$ Tau\\
MNRAS 370, 884 (06)& $\lambda$ Sco\\
MNRAS 370, 1935 (06)& $\delta$ Lib\\
A\&A 463, 1061 (07)& V379 Cep \\
ApJ 463, 579 (07)& HD 23642 \\
ApJ 464, 263 (07)& HD 110555 \\
ASCP 361, 482 (07)& $\beta$ Per \\
MNRAS 382, 609 (07)& $\eta$ Mus \\
arXiv 0710.0758 (07)& Cyg X-1 \\
A\&A 481, 183 (08)& V1007 Sco \\
AJ 136, 631 (08)& Cyg X-1 \\
ApJ 678, (08)& Cyg X-1 \\
ESO AS 67 (08)& HD 208905\\
MNRAS 385, 381 (08)& V716 Cen \\
\\
\end{tabular}


\clearpage

\label{eDruhy}

\vfill


\clearpage
\addcontentsline{toc}{section}{Program of the Summer School}
{\Large \bf Program of the Summer School}\\[1mm]
\begin{tabular}{rl}
Monday 15. 9. & \\
 9:00 - 12:00 & P. Hadrava: Introduction into disentangling\\
12:00 - 14:00 & lunch \\
14:00 - 18:00 & Exercises \\
18:00 - \hphantom{19:00} & Wellcome party\\
&\\
Tuesday 16. 9.&\\
 9:00 - 10:20 & P. Hadrava: Related spectroscopic methods \\
10:40 - 12:00 & P. Hadrava: Advanced disentangling \\
12:00 - 14:00 & lunch \\
14:00 - 14:30 & \parbox[t]{100mm}{H. Lehmann: TW Dra -- Spectral disentangling of
                an unusual triple system} \\
14:30 - 15:00 & \parbox[t]{100mm}{H. Ak: KOREL application
                on spectroscopic binary HD 10308} \\
15:00 - 18:00 & Exercises \\
&\\
Wednesday 17. 9.&\\
 9:00 - 10:20 & P. Hadrava: Preparation of spectra for disentangling\\
10:40 - 12:00 & \parbox[t]{100mm}{P. \v{S}koda: Spectra reduction related problems of
 Fourier disentangling}\\
12:00 - 14:00 & lunch \\
14:00 - 14:30 & \parbox[t]{100mm}{N. F. Ak: KOREL application
                on spectroscopic binary IX Per}\\
14:30 - 15:00 & H. V. Senavci et al.: The Echelle Spectra of SW Lac\\
15:00 - 18:00 & Exercises \\
\\
Thursday 18. 9.\\
 9:00 - 10:20 & P. Hadrava: Results from KOREL\\
10:40 - 12:00 & \parbox[t]{100mm}{J. Kub\'{a}t: Model atmospheres and synthetic spectra}\\
12:00 - 14:00 & lunch \\
14:00 - 14:30 & \parbox[t]{100mm}{O. Creevey: Spectroscopic observations
 of a red-edge delta Scuti star in an eclipsing binary}\\
14:30 - 18:00 & Exercises \\
\\
Friday 19. 9.\\
 9:00 - 10:20 & P. Hadrava: Future of disentangling\\
10:40 - 12:00 & \parbox[t]{100mm}{P. \v{S}koda: Futuristic vision of spectra disentangling
 in the new milenium (focus on GRID computing, Virtual Observatory,
 Workflows, Web services)}\\
12:00 - 14:00 & lunch \\
14:00 - 16:00 & Closing discussion\\
\end{tabular}

\pagebreak
\addcontentsline{toc}{section}{List of participants}
{\Large \bf List of participants}\\[1mm]
\begin{tabular}{ll}
Hasan        Ak      &$\langle$hasanak@erciyes.edu.tr$\rangle$ (Kayseri Turkey) \\%
Nurten Filiz Ak      &(Kayseri Turkey) \\%
Eva          Arazimov\'{a}&$\langle$arazimova@sunstel.asu.cas.cz$\rangle$ (Ondrejov, Czech Rep.) \\
Omur         Cakirli &$\langle$omur.cakirli@gmail.com$\rangle$ (Izmir, Turkey)\\%
Orlagh       Creevey &$\langle$orlagh@iac.es$\rangle$ (Spain)\\
Attila       Cseki   &$\langle$attila@aob.bg.ac.yu$\rangle$ (Serbia)\\
Dominik      Drobek  &$\langle$drobek@astro.uni.wroc.pl$\rangle$ (Wroclav, Poland)\\%
Jan          Elner   &$\langle$janelner@centrum.cz$\rangle$ (Ondrejov, CR)\\%
Pedro Amado  Gonzalez&$\langle$pja@iaa.es$\rangle$ (Granada, Spain)     \\%
Lubomir      Hambalek&$\langle$lhambalak@ta3.sk$\rangle$ (Slovakia)\\%
Saskia       Hekker  &$\langle$Saskia.Hekker@oma.be$\rangle$ (Belgium)\\
Baju         Indradjaja&$\langle$bindra@physics.muni.cz$\rangle$ (Indonesia)\\%
Emil         Kundra  &$\langle$ekundra@ta3.sk$\rangle$ (Slovakia)\\
Patricia     Lampens &$\langle$patricia.lampens@oma.be$\rangle$ (Brussel, Belgium)\\%
Olivera      Latkovic&$\langle$olatkovic@aob.bg.ac.yu$\rangle$ (Serbia)\\%
Holger       Lehmann &$\langle$lehm@tls-tautenburg.de$\rangle$ (Jena, Germany)\\%
Joanna       Molenda-Zakowicz&$\langle$molenda@astro.uni.wroc.pl$\rangle$ (Wroclav, Poland)\\%
Andrzej      Pigulski&$\langle$pigulski@astro.uni.wroc.pl$\rangle$ (Wroclav, Poland)\\%
Jan          Polster &$\langle$hans@algieba.asu.cas.cz$\rangle$ (Brno, Czech Republic)\\%
Volkan       Senavci &$\langle$volkan@astro1.science.ankara.edu.tr$\rangle$ (Ankara, Turkey)\\
Tamas        Szalai  &$\langle$szaszi@titan.physx.u-szeged.hu$\rangle$ (Szeged, Hungary)\\
Staszek      Zola    &$\langle$zola@astro1.as.ap.krakow.pl$\rangle$ (Krakow, Poland)\\%
\hline
Melike       Afsar   &(Izmir, Turkey)\\%
Zeynep       Bozkurt &(Izmir, Turkey)       \\%
David        Holmgren&holmgrenm@shaw.ca$\rangle$ (Kanada)\\
Somaya       Saad    &$\langle$somaya111@yahoo.com$\rangle$ (Egypt)\\
Faruk        Soydugan& $\langle$fsoydugan@comu.edu.tr$\rangle$ (Canakkale, Turkey)\\
\end{tabular}

\begin{center}
\begin{picture}(119,60)
 \put(0,0){\epsfxsize=79mm \epsfbox{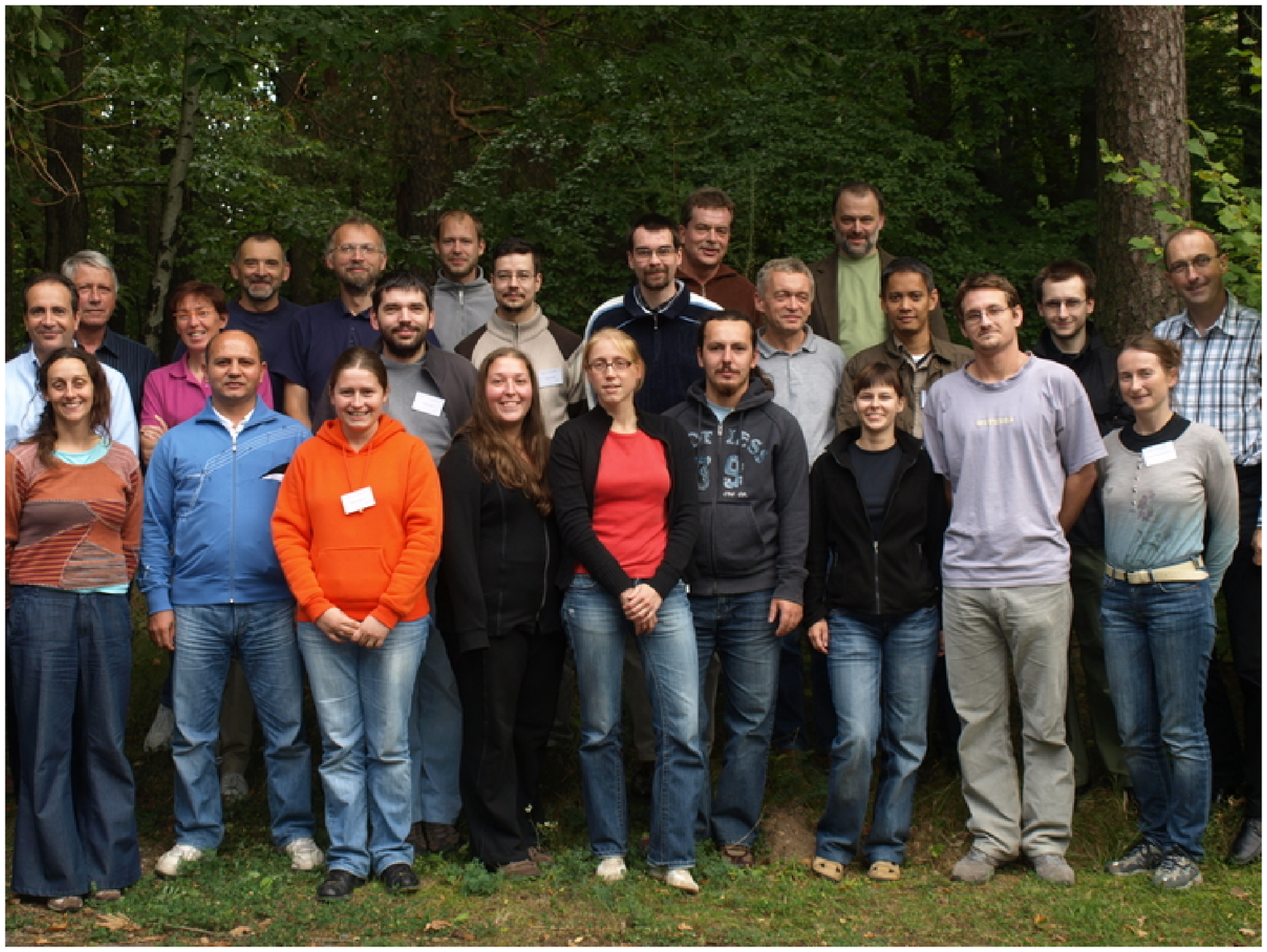}}
 \put(80,30){\epsfxsize=39mm \epsfbox{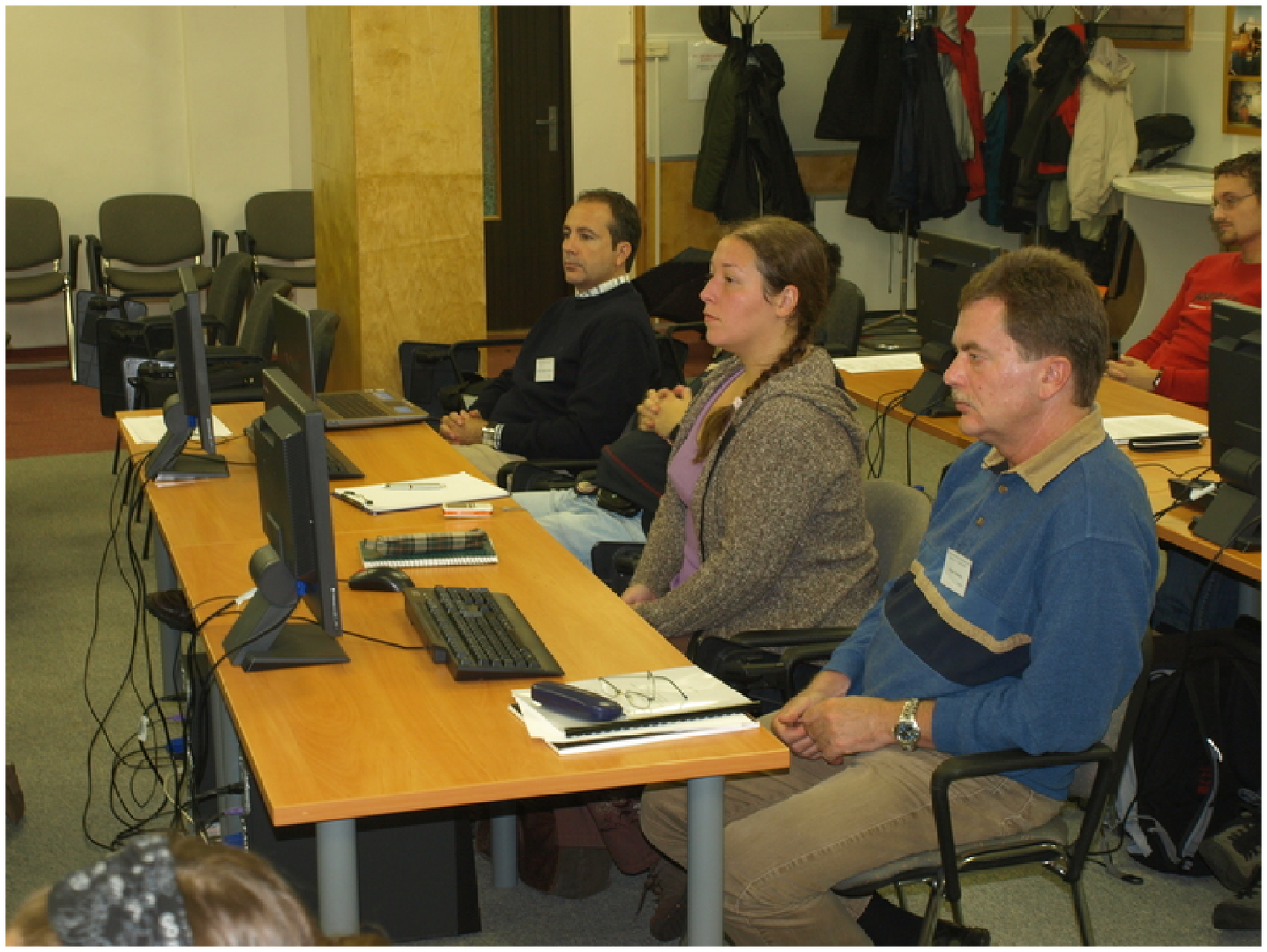}}
 \put(80,0){\epsfxsize=39mm \epsfbox{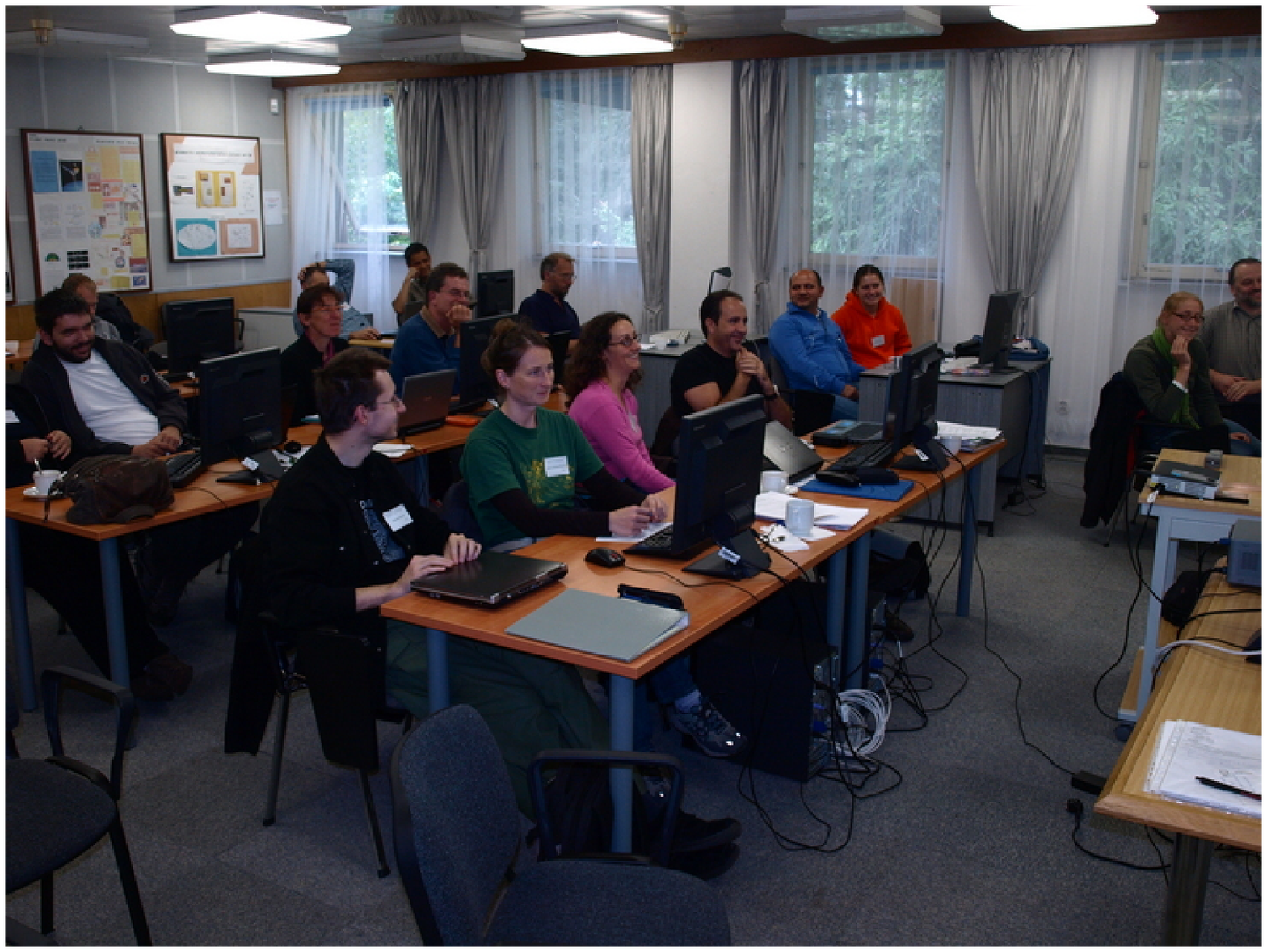}}
\end{picture}
\end{center}


\clearpage

\end{document}